\documentclass[12pt]{article}
\usepackage[top=1.5in,bottom=1.5in]{geometry}
\pdfoutput=1
\usepackage{mathrsfs}
\usepackage{amssymb}
\usepackage{graphicx}
\usepackage[font=scriptsize,labelfont=bf]{caption}
\usepackage{amsmath,amsfonts,amsthm}
\usepackage{multirow}
\usepackage{extarrows}
\usepackage{url}
\usepackage{natbib}
\usepackage{tikz}
\usepackage{listings}
\usepackage{caption}
\usepackage{subfig}

\newtheorem*{theorem*}{Theorem}

\newcommand{\be}{\begin{equation}} \newcommand{\ee}{\end{equation}}
\newcommand{\bd}{\begin{displaymath}} \newcommand{\ed}{\end{displaymath}}
\newcommand{\ba}{\begin{align}} \newcommand{\ea}{\end{align}}
\newcommand{\baa}{\begin{align*}} \newcommand{\eaa}{\end{align*}}
\newcommand{\ben}{\begin{enumerate}} \newcommand{\een}{\end{enumerate}}
\newcommand{\bi}{\begin{itemize}} \newcommand{\ei}{\end{itemize}}

\newcommand{\pkg}[1]{\textbf{#1}}
\newcommand{\proglang}[1]{\textsf{#1}}
\newcommand{\code}[1]{\texttt{#1}}

\newcommand{\footremember}[2]{%
    \footnote{#2}
    \newcounter{#1}
    \setcounter{#1}{\value{footnote}}%
}

\providecommand{\keywords}[1]
{
  \small	
  \textbf{\textit{Keywords---}} #1
}
\usepackage{geometry}
\begin{document}

\title{
Are official confirmed cases and fatalities counts good enough to study the COVID--$19$ pandemic dynamics? A critical assessment through the case of Italy.}

\author{Krzysztof Bartoszek\footremember{Bartoszek}{Department of Computer and Information Science, Link{\"o}ping University, Link{\"o}ping SE--$581$ $83$, Sweden} \and Emanuele Guidotti\footremember{Guidotti}{Institut d'analyse financi{\`e}re, University of Neuch{{\^a}}tel, Switzerland.} \and Stefano Maria Iacus\footremember{Iacus}{European Commission, Joint Research Centre, Via E. Fermi $2749$, I--$21027$ Ispra (VA), Italy.} \and Marcin Okr{\'o}j\footremember{Okroj}{Department of Cell Biology and Immunology, Intercollegiate Faculty of Biotechnology, University of Gda{\'n}sk and Medical University of Gda{\'n}sk, Poland.}}

\maketitle
\begin{abstract}
As the COVID--$19$ outbreak is developing the two most frequently reported statistics seem
to be the raw confirmed case and case fatalities counts.
Focusing on Italy, one of the hardest hit countries, we look at how these two values
could be put in perspective to reflect the dynamics of the virus spread. In particular, we
find that merely considering the confirmed case counts would be very misleading.
The number of daily tests grows, while the daily fraction of confirmed cases to total tests 
has a change point. It (depending on region) generally increases with strong fluctuations
till (around, depending on region) $15^{\mathrm{th}}$--$22^{\mathrm{nd}}$ March and then decreases linearly after.
Combined with the increasing trend of daily performed tests,
the raw confirmed case counts are not representative of the situation and are confounded with the sampling effort. This we observe when regressing on time the logged fraction of positive tests and
for comparison the logged raw confirmed count.
Hence, calibrating model parameters for this virus's dynamics should not be done based only
on confirmed case counts (without rescaling by the number of tests), but take also fatalities and hospitalization count under consideration as variables not prone to be distorted by testing efforts.
Furthermore, reporting
statistics on the national level does not say much about the dynamics of the disease, which
are taking place at the regional level. 
These findings are based on the official data of total death counts up to $15^{\mathrm{th}}$ April $2020$ released by ISTAT and up to $10^{\mathrm{th}}$ May $2020$ for the number of cases.
In this work we do not fit models but we rather investigate whether this task is possible at all.

This work also informs about a new tool to collect and harmonize
official statistics coming from different sources in the form of a package for the \proglang{R} statistical environment and presents the ``\texttt{COVID-19 Data Hub}''.

\section*{Highlights}
\begin{itemize}

\item confirmed cases are related to the total tests and time
   \item confirmed case counts without number of tests would likely misinform epidemiological models
   \item  national level statistics do not say much about the dynamics of the disease
   \item a new \proglang{R} package and a COVID--$19$ web hub with harmonized data is made available
\end{itemize}
\end{abstract}

\keywords{COVID-19, coronavirus, R language, data}

\pagebreak

\tableofcontents

\pagebreak

\section{Introduction}
In December $2019$ the first cases of pneumonia of unknown etiology were reported in Wuhan
city, People's Republic of China. Analyses of patients' samples collected from their respiratory tract revealed
that a novel coronavirus, later named as severe acute respiratory syndrome coronavirus $2$ (SARS--CoV--$2$) 
is the pathogen responsible for infection \citep{CHuaetal2020}. The disease, officially called COVID--$19$ by 
World Health Organization (WHO) is
characterized by higher transmissibility and infectivity but lower mortality than Middle East
Respiratory Syndrome (MERS) and Severe Acute Respiratory Syndrome (SARS) caused by other
coronaviruses \citep{LWabYWanDYeQLiu2020}. 

Apart of the source of infection, the spread of the virus depends on the transmission route and general
susceptibility of the population. SARS--CoV--$2$ is believed to be transmitted mostly by close contact (and
further carry--over to the mucous surfaces of the body) and inhalation of aerosol produced by an infected
person. The presence of the virus was also reported in samples from the gastrointestinal tract \citep{FXiaetal2020} but the
potential role of the oral--fecal route of infection is unknown. The evidence of asymptomatic carriers who
may unintentionally transmit the virus together with relatively long incubation period up to $24$ days
\citep{YBaietal2020} increase the risk of viral spread worldwide and make prevention measures difficult. On the other
hand, separation of identified cases, prior immunity to SARS--CoV--$2$ or cross--reactivity of human
antibodies naturally risen against other viruses would act as a barrier for virus transmission. The latter
is probable as RNA sequences of SARS--CoV--$2$ are in $79\%$ identical to the sequences of 
SARS--CoV responsible for the previous pandemic in Far East countries in $2002$ and $50\%$ identical to MERS--CoV
\citep{RLuetal2020}. All above mentioned issues would act as confounding factors for
any 
modelling of pandemic progression.

Except of the city of Wuhan where the first reports of COVID--$19$ were announced in December $2019$,
there was another outbreak of disease, which took place in January--February $2020$ on the 
\textit{Diamond Princess} cruise ship with more than $3700$ people onboard. As such a great number of people were
locked in a confined space using common facilities, air--condition systems, restaurants etc. and once
the chronology of infections, symptoms and undertaken health measures are known \citep{ENakHInoAAka2020,JRocHSjoAWil2020,SZhaetal2020}, one can
consider this as a unique, naturally--occurring epidemiological study useful for prediction of mortality,
disease spread and other parameters of the COVID--$19$ pandemic. Since the virus has spread across the world
and new pandemic epicenters like Italy, Spain, Iran, South Korea and USA have emerged, a multitude
of new data has appeared. Different countries have applied different strategies of testing people for the coronavirus
(mass testing vs. testing of selected patients), different testing methods (serological vs. PCR--based
assays) and count of case fatalities (solely SARS--CoV--$2$ positive tested cases vs. cases with comorbidities).
Therefore, any direct comparison of pandemic dynamics is difficult but still, comparison to a ``golden standard'', which the Diamond Princess case could be considered as, may be useful. 

\begin{figure}[!ht]
\begin{center}
\includegraphics[width=0.98\textwidth]{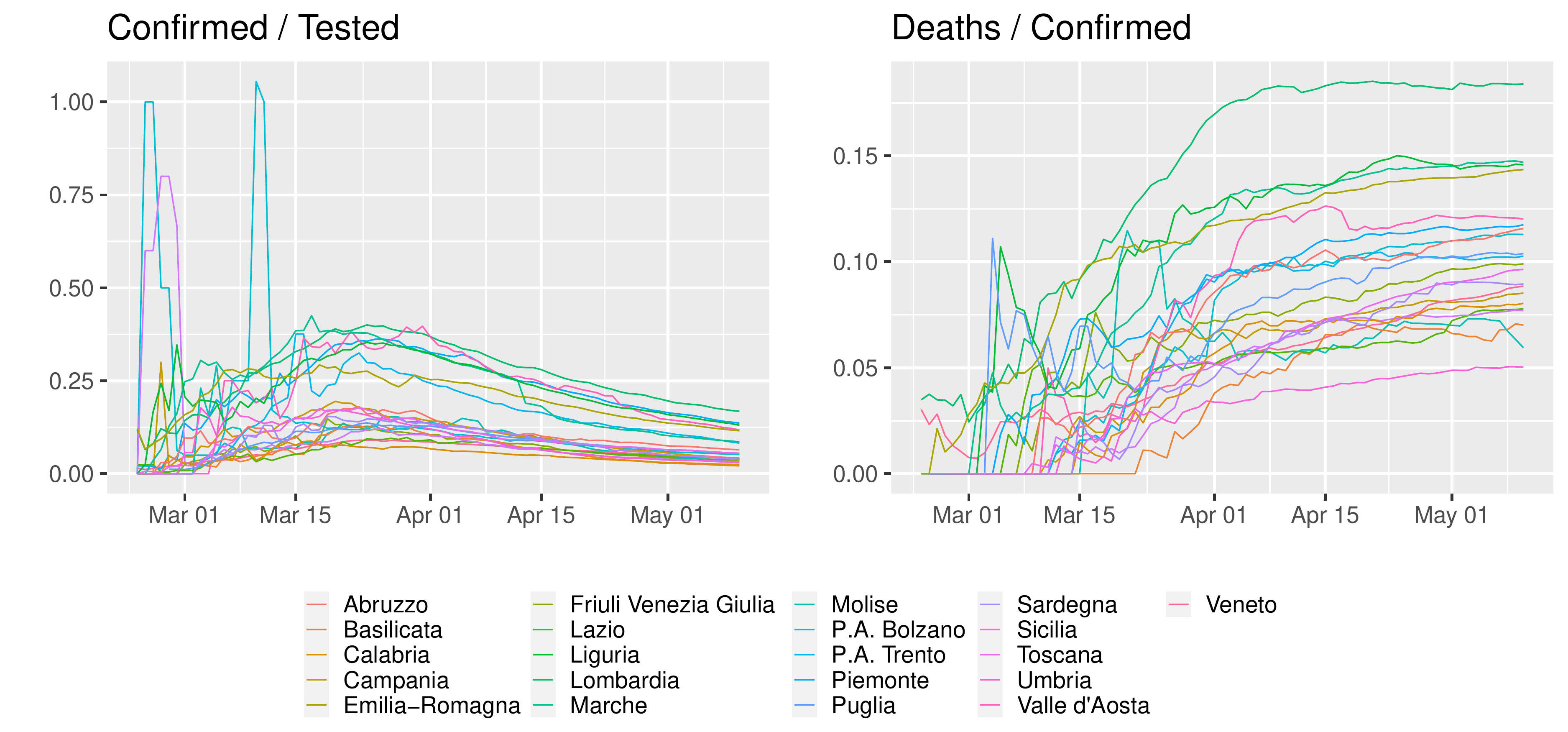} 
\end{center}
\caption{Cumulative confirmed cases and case fatalities for all the regions of Italy.
Right: Cumulative case fatalities divided by confirmed cases, left:
cumulative confirmed cases divided by the cumulative number of tests.
}\label{figITregionsTestConfirmed}
\end{figure}

Since the outbreak of the disease a multitude of papers modelling the dynamics of the infection
have appeared, especially on the \texttt{arXiv} preprint server. They are usually concerned with 
connecting the pandemic with 
various epidemiological models (e.g. \citet{JKumKHem2020arXiv,AMor2020arXiv,HSin2020arXiv,FZul2020arXiv} following 
a brief survey of arXiv at the start of April $2020$).
However, such models of course require data concerning the infected individuals. 
Furthermore, the media are bombarding today with two basic numbers 
(for each country)---the number of confirmed cases and the number of
case fatalities. Given that supposedly the vast majority of people are asymptomatic and
testing is not done as random sampling of the population but due to particular
protocols these values by themselves might be misleading. 
We can only second \citet{SWoo2020arXiv} in 
``\emph{Despite millions of tests having been
performed, there are still no results from statistically well founded sampling based testing programmes
to establish basic epidemic quantities such as infection fatality rate and infection rates.  In the absence
of such direct data, epidemic management has to proceed on the basis of data produced largely as a side
effect of the clinical response to the disease.}''
As a motivating example
we present Figure~\ref{figITregionsTestConfirmed} from which we can see that in Italy the
case fatality to confirmed ratio is constant while the confirmed cases to 
number of tests has been decreasing since around March $22^{\mathrm{nd}}$. Indeed, the time period since March $22^{\mathrm{nd}}$ is longer than the median time of $19.5$ days of infection till death \citep{ZhouetAl2020}, so 
one should already start observing some drop in the
case fatality to confirmed ratio. 

Through the case of Italy, this paper tries to investigates the following issues: 
\begin{itemize}
    \item With each country having their own reporting standard and testing strategy are these raw numbers comparable across countries?
\item Do these data actually mean what they are being said to be and are they appropriate for model fitting at all? 
\end{itemize}

Clearly, the curves presented in Figure~\ref{figITregionsTestConfirmed}
suggest that a more in--depth look at the raw numbers is required and that there is a need to put the data in a correct perspective
before trying to fit any epidemiological model to them, especially because 
the viral dynamics are starting to be inferred from reported case fatalities
\citep{TBri202004medRxiv,APugSSot2020arXiv,GVat2020arXiv}. 

In this work we approach these issues by looking in detail at the available infection
data for individual Italian regions (Section \ref{secItreg}) and present 
the \proglang{R} \citep{RCoreTeam} package \pkg{COVID19} (Section \ref{secCOVID19}) that 
unifies COVID--$19$ datasets across different sources in order to simplify the data acquisition process and 
the subsequent analysis. Section~\ref{sec:discussion} contains a discussion on what other
data would be useful (if of course possible to collect for the already overworked public services),
in understanding the dynamics of the pandemic. Most regional analyses are contained in the Appendices.

\section{Italian regional data analysis}\label{secItreg}
Italy is a country which is being very extremely hard--hit with the COVID--$19$
pandemic. It is currently (as of $13^{\mathrm{th}}$ May $2020$) as a whole in lockdown and the medical services 
are extremely strained. However, due to this situation it has also 
very detailed epidemiological data that has been made publicly available.
Its constantly increasing infected and
case fatality count has lead us looking in greater detail into this data,
especially as it is used for curve--fitting of epidemiological models 
(e.g. \citet{JKumKHem2020arXiv,AMor2020arXiv,HSin2020arXiv,FZul2020arXiv} following 
brief survey of arXiv) and presented in public media. 

The first hurdle that one comes across is what do the presented counts 
actually represent. This seems to be region dependent\footnote{Initially the Veneto region 
blanked tested a significant part of the population, while Lombardy did not (private communication
with Marco Picariello and Paola Aliani).}.
Furthermore, any deceased
whose test result is found positive is classified as a COVID--$19$ case fatality, 
regardless of any past or underlying diseases,
and this methodology has been consistently applied in Italy since the beginning \citep{MPicPAli2020arXiv}. 
It is important to point out that different countries seem to have different 
testing strategies and classification systems of deaths---hence raw
counts between countries might not be comparable.
Given the huge amount of tests performed in Italy \citep[$2735628$
as of $13^{\mathrm{th}}$ May $2020$ (\pkg{COVD19} package),  ][]{COVID19pkg}
an important question is:
``what fraction of them were serological tests?'' as  
there is no official data on this. 
A serological test may not distinguish between a person actively infected with the virus and a person that was exposed to the virus in the past. Alternatively, serological test may not detect person actively infected with still low viral titer of anti-virus antibodies.
On the other hand, if the protocol is to test only people exhibiting
symptoms and medical personnel, then given that it is hypothesised that the vast majority of cases are asymptomatic, such a raw count might not be representative of the scale of the epidemic.

Given the above uncertainties we set out to see how the Italian
regional data could be presented in a standardized manner.
Furthermore, we see how the data of each region compares to the
\textit{Diamond Princess'} data. We focus on the two values that
are being presented everywhere---the confirmed case count and the case fatalities
count. However these should be scaled. We scale the confirmed case count
by the total number of tests performed. Scaling the case fatalities
is more problematic. A common way is to present them as
the case fatality ratio but these may be misleading when estimated
during an epidemic \citep{LBotMXiaTCho2020arXiv}. Furthermore, 
assuming that the vast majority of cases are asymptomatic---hence
not tested and not inside the case count, we are uncertain to what
the fatalities would actually be compared to. 

Given, the lack
of hard data another objective approach would be to compare
the daily count of case fatalities to the total
deceased count for the day. To the best of our knowledge such
statistics are not centrally reported in Italy in real--time.
Daily deceased counts (from nearly all of the Italian 
municipalities---see Discussion) are available though for the period $1^{\mathrm{st}}$ January--$15^{\mathrm{th}}$ April $2020$\footnote{\url{https://www.istat.it/it/archivio/240401}}. Hence, for this time 
period we are able to plot  the weakly  ``nearly''-desired ratios (see Section~\ref{sec:discussion}). 
We aggregate per week to remove daily fluctuations, which obscure the picture.
Furthermore, the same data source
provides deceased counts for the years $2015$--$2019$
(for the same time period). This allows
us to also visualize the excess mortality (with respect to the per week
average from the past five years). Beyond this time interval, 
it is impossible to provide such curves. However, having daily case fatalities
counts and past mortality (this is taken as a constant value
equalling the average number of deceased for $15^{\mathrm{th}}$ April)
we are able to plot the (per week) ratio of 
case fatalities to previous average mortality. This provides some 
indication of the magnitude of excess mortality\footnote{A similar graphical analysis for  appeared in \citet{TheEconomist, NYT, FT}.}. 
However, it is worth noticing that when  looking at the current excess mortality it could be appropriate to compare with past mortality peaks 
\citep[e.g. for UK death toll, the
$2014/2015$ and 
$1999/2000$ peaks\footnote{\url{https://www.ons.gov.uk/peoplepopulationandcommunity/birthsdeathsandmarriages/deaths/articles/highestnumberofexcesswinterdeathssince19992000/2015-11-25}}, Figs. $1$, $5$ and $6$ of][]{DTho2020arXiv}, taking into consideration the causes of death. Here for Italy and its regions, in Figs. 
\ref{figRawDeathsLombardyVeneto}, \ref{figRawDeathsItaly} and Fig. \ref{figRawDeathsRegion} in Appendix B we compare the current deceased peak with the seasonal start of the January one.

We should remark that perhaps more focus should be on the 
cumulative positive test fraction instead of the daily positive test fraction.
This is because the daily fraction is extremely noisy and furthermore it sometimes happens
that this fraction, in the official data source for Italy, exceeds $1$. 
For similar reasons we plot the weekly scaled deaths and cumulative scaled
deaths. The daily counts are extremely noisy as well. 

We plot the scaled daily and cumulative positive test count and
scaled case fatalities next to the cumulative 
positive tested fraction of passengers on the 
\textit{Diamond Princess}. Here we present the graphs
from two special regions in Italy---Lombardy and Veneto. 
The remaining regions are presented in the Appendix A.
Lombardia is the center of the epidemic, where the cases and 
deaths counts are the highest. Veneto seems to be a region
where the pandemic's dynamics are special---it was a region
that very early on undertook population--wide testing and drastic lockdown measures\footnote{Private communication
with Marco Picariello and Paola Aliani.}.

\begin{figure}[!ht]
\begin{center}
\includegraphics[width=0.98\textwidth]{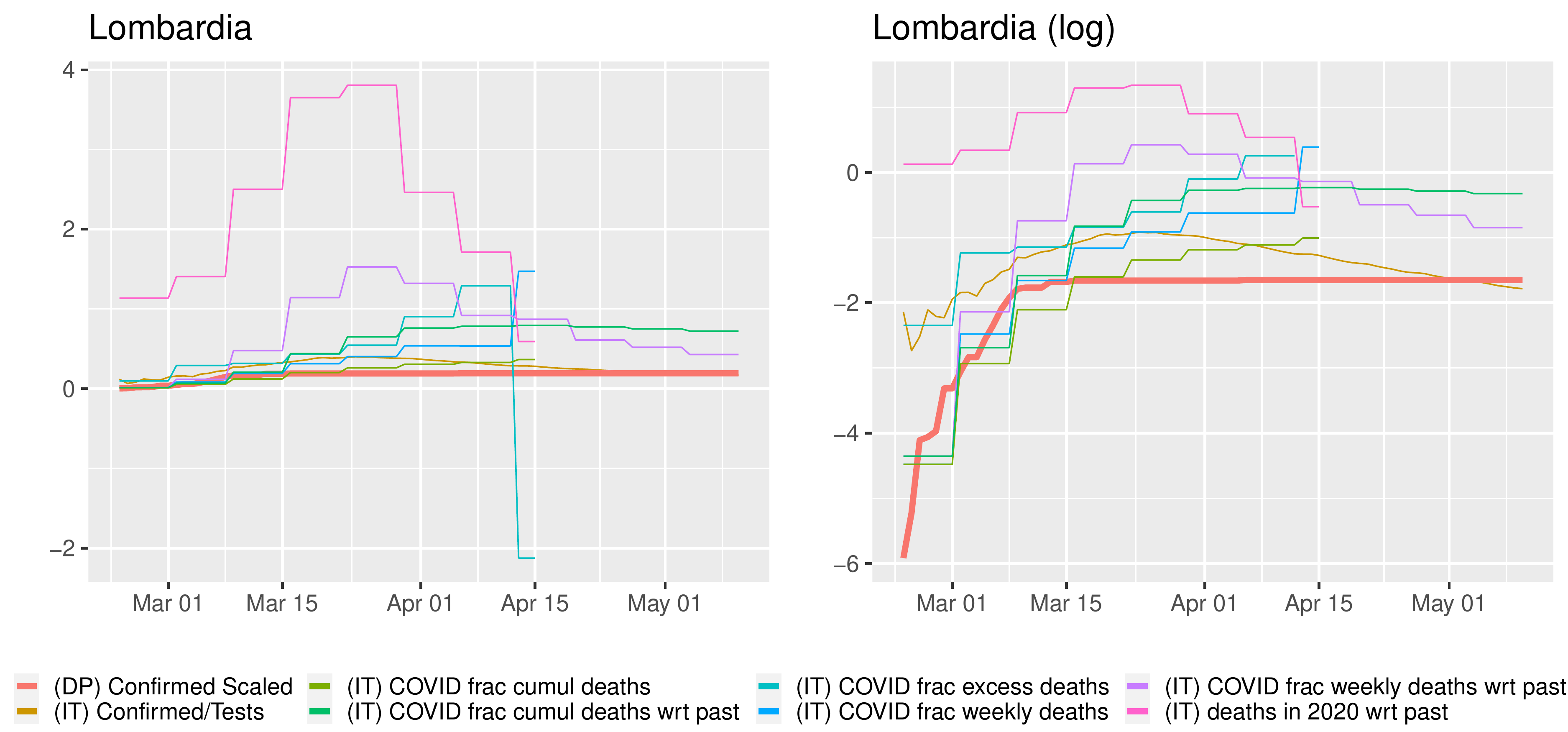}
\\
\includegraphics[width=0.98\textwidth]{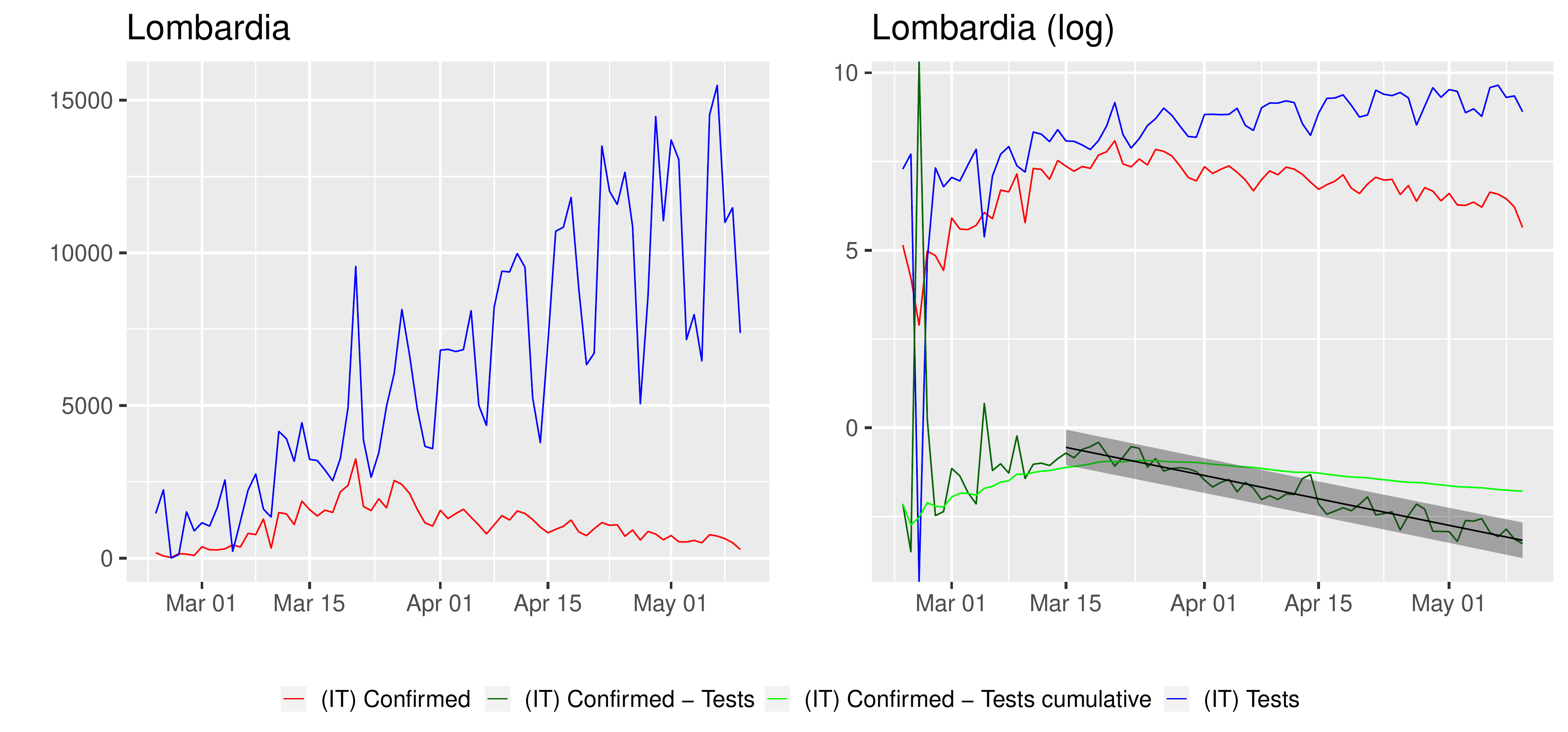}
\end{center}
\caption{Comparison of curves for Lombardy region. Left: $y$--axis on normal scale, right: on logarithmic
scale.
Regression line shown for $\log(\mathrm{daily~confirmed})-\log(\mathrm{daily~tested})\sim time$
with $95\%$ prediction band. Slope of regression with $95\%$ confidence interval: 
$a_{\mathrm{f}}=-0.047 (-0.051,-0.043)$,
this corresponds to a half--life (in days) of $14.829 (13.712,16.144)$.
The slope of the regression
$\log(\mathrm{daily~confirmed})\sim time$ is 
$a_{\mathrm{raw}}=-0.025 (-0.029,-0.021)$ corresponding to a half--life (in days) of $27.656 (23.961,32.698)$.
The slope of the regression
$\log(\mathrm{daily~tests})\sim time$ is 
$0.022 (0.016,0.030)$ corresponding to a doubling time (in days) of $31.970 (23.280,42.506)$.
Ratio of slopes for $a_{\mathrm{f}}/a_{\mathrm{raw}}=1.865$, with corresponding half--lives' ratio: $0.536$.
The slope of the regression
$\log(\mathrm{cumulative~confirmed})-\log(\mathrm{cumulative~tested})\sim time$ is 
$-0.016 (-0.018,-0.015)$ corresponding to a half--life (in days) of $42.525 (39.386,46.208)$.
}\label{figLombardy}
\end{figure}

\begin{figure}[!ht]
\begin{center}
\includegraphics[width=0.98\textwidth]{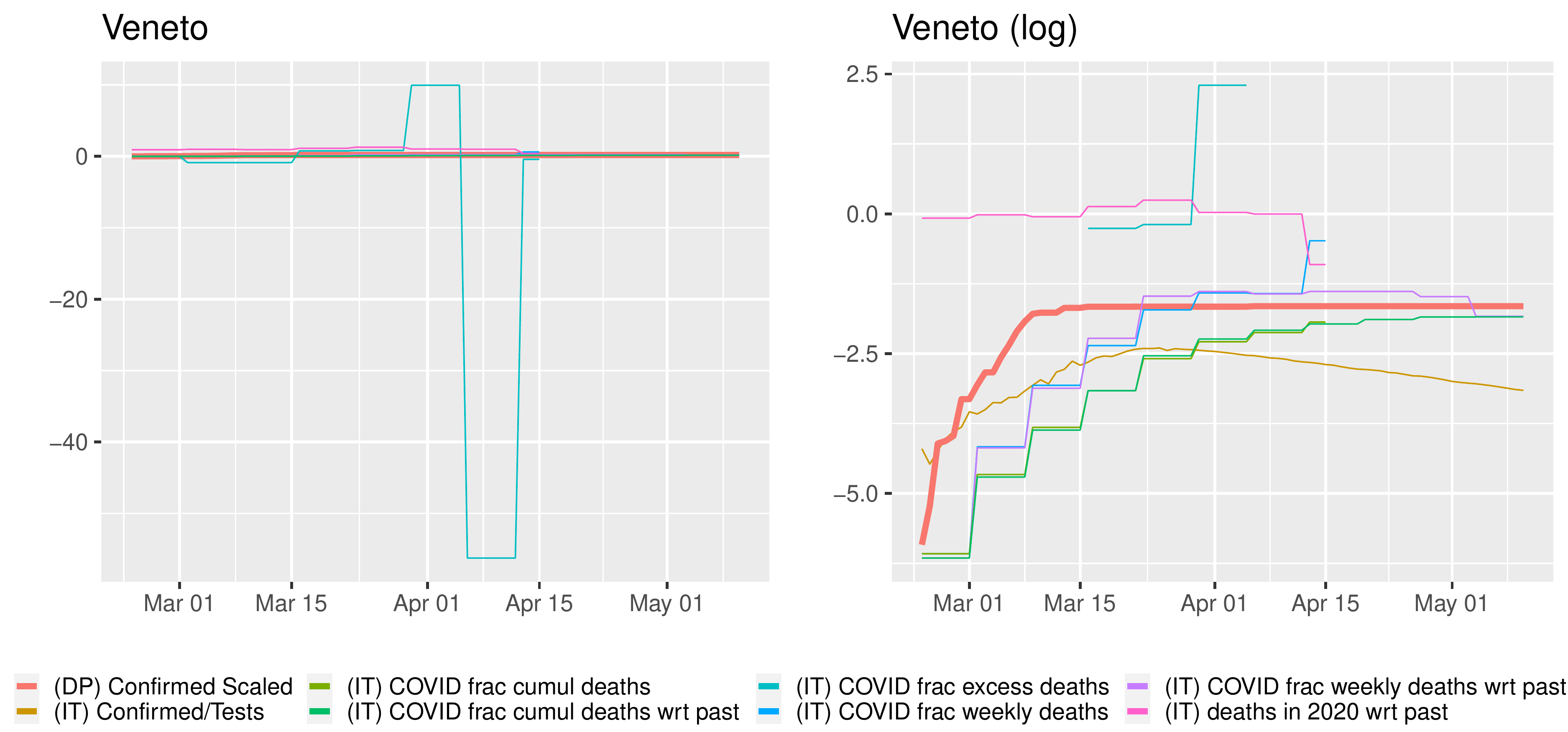}
\\
\includegraphics[width=0.98\textwidth]{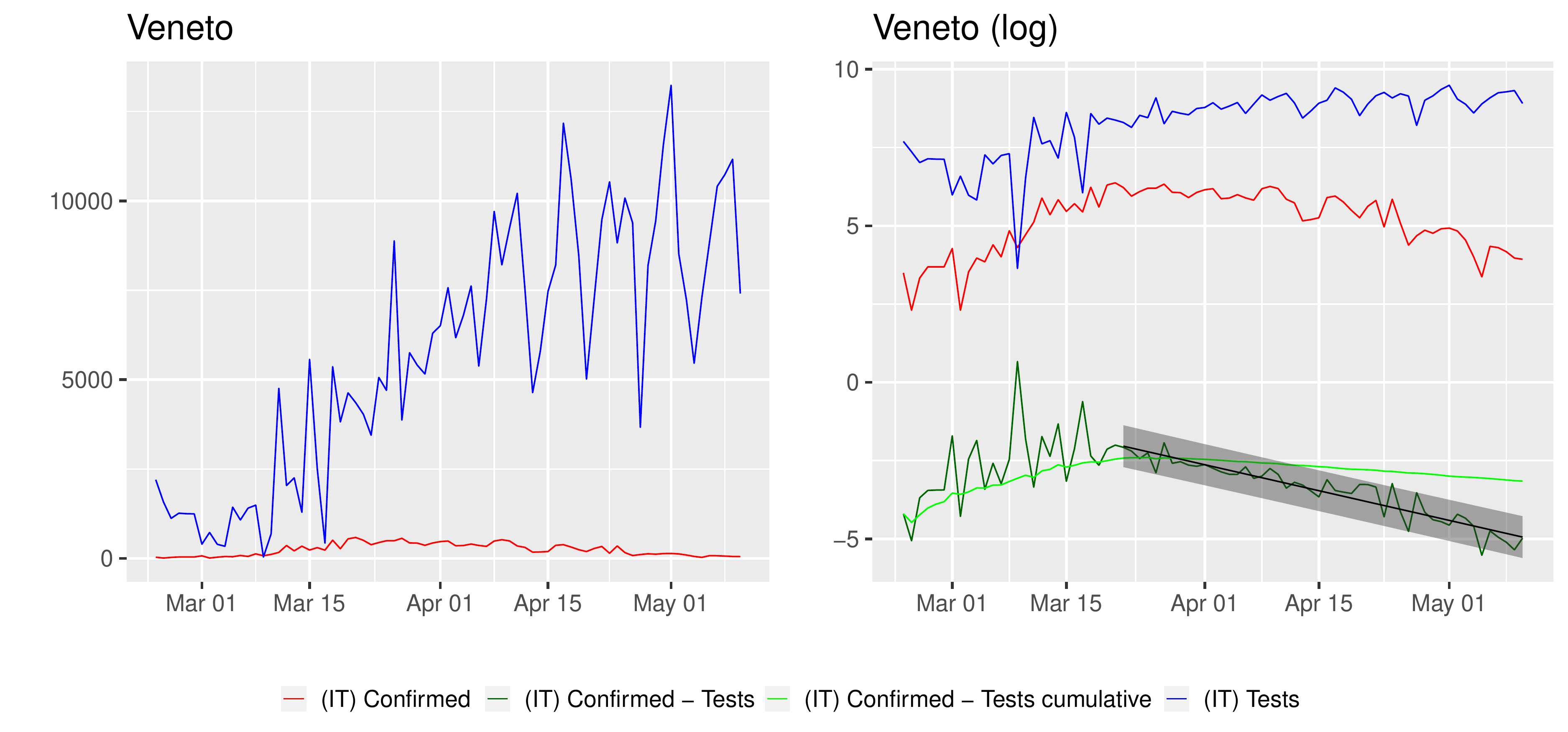}
\end{center}
\caption{Comparison of curves for Veneto region. Left: $y$--axis on normal scale, right: on logarithmic
scale. 
Regression line shown for $\log(\mathrm{daily~confirmed})-\log(\mathrm{daily~tested})\sim time$
with $95\%$ prediction band. Slope of regression with $95\%$ confidence interval: 
$a_{\mathrm{f}}=-0.059 (-0.065,-0.053)$ corresponding to a half--life (in days) of $11.693 (10.592,13.049)$.
The slope of the regression
$\log(\mathrm{daily~confirmed})\sim time$ is 
$a_{\mathrm{raw}}=-0.047 (-0.054,-0.040)$ corresponding to a half--life (in days) of $14.843 (12.879,17.513)$.
The slope of the regression
$\log(\mathrm{daily~tests})\sim time$ is 
$0.013 (0.007,0.026)$ corresponding to a doubling time (in days) of $55.100 (26.467,95.637)$.
Ratio of slopes for $a_{\mathrm{f}}/a_{\mathrm{raw}}=1.269$, with corresponding half--lives' ratio: $0.788$.
 The slope of the regression
$\log(\mathrm{cumulative~confirmed})-\log(\mathrm{cumulative~tested})\sim time$ is 
$-0.016 (-0.017,-0.016)$ corresponding to a half--life (in days) of $42.325 (40.947,43.799)$.
}\label{figVeneto}
\end{figure}  

\begin{figure}[!ht]
\begin{center}
\includegraphics[width=0.98\textwidth]{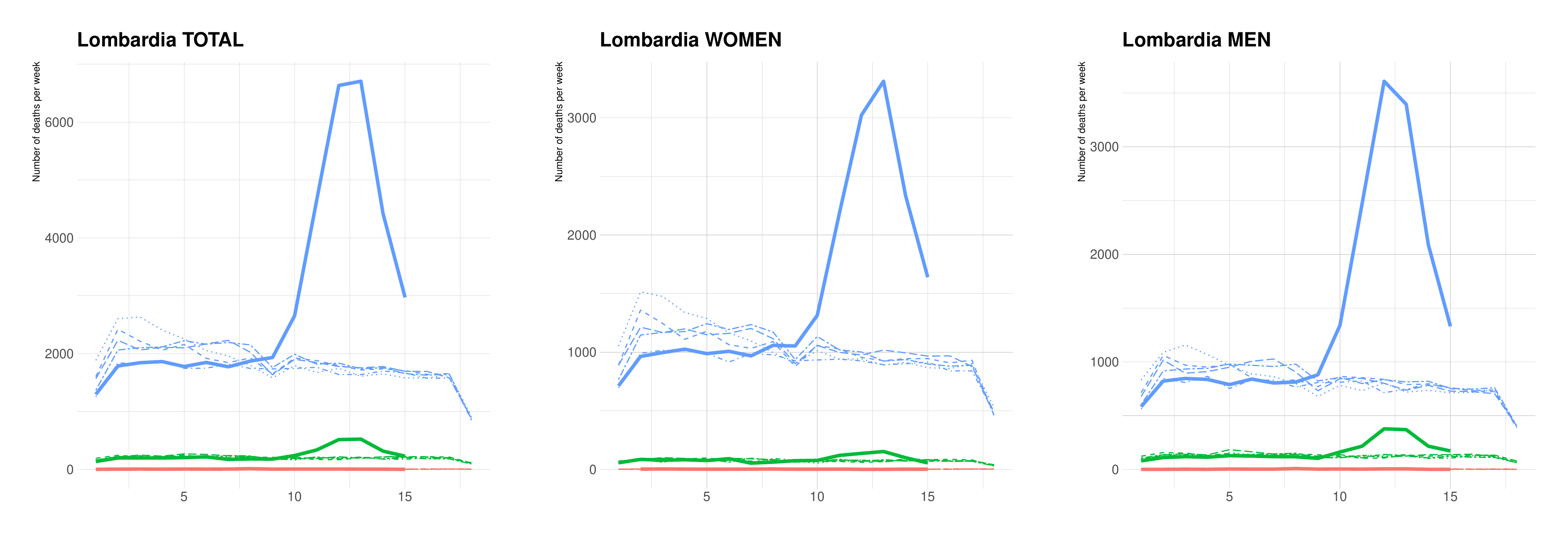} \\
\includegraphics[width=0.98\textwidth]{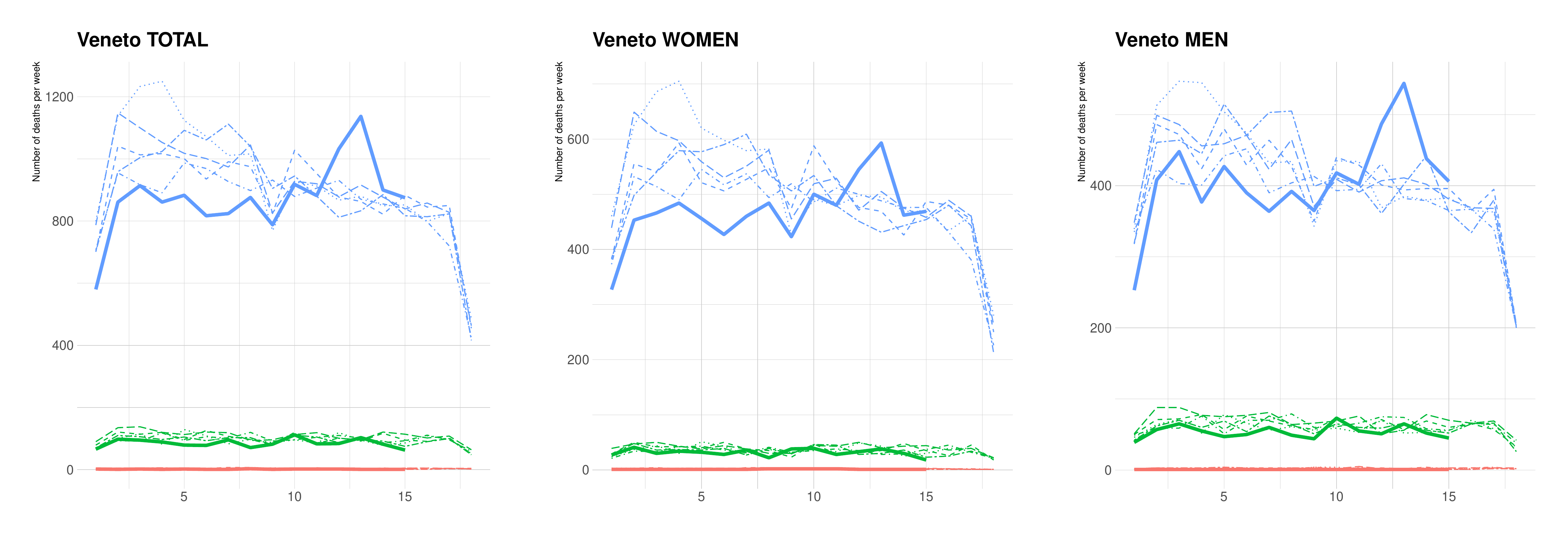} \\
\includegraphics[width=0.98\textwidth]{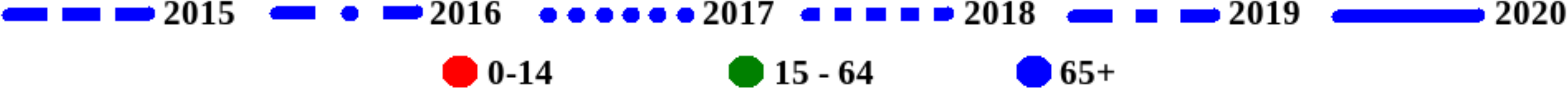}
\end{center}
\caption{Weekly raw death toll comparison in different age groups between $2020$ and $2015$--$2019$
for Lombardy and Veneto. 
}\label{figRawDeathsLombardyVeneto}
\end{figure}  

On all of the graphs the curve labels have the following meaning.
\begin{enumerate}
    \item (DP) Confirmed Scaled: cumulative number of cases on the Diamond Princess divided by $3711$, the number of passengers and crew onboard
    \item (IT) Confirmed/Tests: cumulative confirmed case to cumulative number of tests ratio for Italy or region
    \item (IT) COVID frac cumul deaths: cumulative number of case fatalities to cumulative number of deceased in $2020$ ratio for Italy or region
    \item (IT) COVID frac cumul deaths wrt past:  cumulative number of case fatalities to cumulative number of average from $2015$--$2019$ number of deceased ratio for Italy or region
    \item (IT) COVID frac excess deaths: number of case fatalities for given week to
    difference between deaths in $2020$ and average from $2015$--$2019$ number of deceased for given week ratio for Italy or region
    \item (IT) COVID frac weekly deaths: number of case fatalities for given week to
    number of deceased in $2020$ for given week ratio for Italy or region
    \item (IT) COVID frac weekly deaths wrt past: number of case fatalities for given week to average from $2015$--$2019$ number of deceased for given week ratio for Italy or region
    \item (IT) deaths in 2020 wrt past: number of deceased in $2020$ for given week to average from $2015$--$2019$ number of deceased for given week ratio for Italy or region
    \item (IT) Confirmed: daily number of confirmed cases for Italy or region
    \item (IT) Confirmed - Tests: $\log(\mathrm{(IT)~Confirmed}) - \log(\mathrm{(IT)~Tests})$
    \item (IT) Confirmed - Tests cumulative: $\log(\mathrm{cumulative~number~of~confirmed~cases~in~Italy~or~region}) - \log(\mathrm{cumulative~number~of~tests~performed~in~Italy~or~region})$
    \item (IT) Tests: daily number of tests performed for Italy or region
\end{enumerate}

We obtain data for the period $24^{\mathrm{th}}$ February--$10^{\mathrm{th}}$ May $2020$ and
we plot the curves from the moment of the first death. 
From both Figure~\ref{figLombardy} and \ref{figVeneto} (and those present in the Appendix A) we can
notice a number of facts. Firstly the daily fraction of infected cases fluctuates very wildly
and sometimes can be greater than $1$. This can only be due to some changes in protocols
or reporting. Similarly, such an explanation seems plausible for the fluctuations in the fractions.
In fact, \citet{MPicPAli2020arXiv} report a change in the way positive cases and deaths are calculated
on $10^{\mathrm{th}}$ March. The cumulative case fraction on the other hand does not exhibit such fluctuations. 
For most regions it is flat and then decreasing. 
In a number of regions (e.g. Abruzzo, Basilicata, Campania, Friuli Venezia Giulia, 
Lazio, Molise, Puglia, Sardegna, Sicily, Toscana, Umbria, Veneto)
on the log--scale graphs the cumulative case to tests ratio curve seems the peak around or below the 
\textit{Diamond Princess}' cumulative case curve and then start dropping. The scaled death curves exceed this curve.

When looking at the graphs of the number of tests per day two things can be seen.
Firstly, the number of positive cases closely follows the number of tests 
(this is clearly visible on the log--scale graphs and supported by the regression study). 
We look at this issue in detail and present for each region and Italy the confirmed cases
with respect to the total tests carried out. We also regress the 
\mbox{
$\log(\mathrm{daily~confirmed~cases})-\log(\mathrm{daily~total~tests})$} on time
(in days) and \\
\mbox{
$\log(\mathrm{cumulative~confirmed~cases})-\log(\mathrm{cumulative~total~tests})$} on time
(in days). The slope of such a regression can be presented in terms of the 
half--lives, if it is negative.
Such a presentation in terms
of effect sizes is important, otherwise it is difficult to assess if the raw slope
is big or small. The linear model approach means that the proportion of infected behaves exponentially
$$
\left(\mathrm{daily(cumulative)~confirmed~cases})\right)/\left(\mathrm{daily(cumulative)~total~tests}\right))
=: p(t) = b e^{at},
$$
then to get the half--life (for $a$ negative, $t_{2}>t_{1}$) one takes
$$
2=\frac{p(t_{2})}{p(t_{1})}=e^{a(t_{2}-t_{1})} 
$$
obtaining $(t_{2}-t_{1})=\log(2)/(-a)$. For $a>0$ one will obtain the doubling time in the same way as $(t_{2}-t_{1})=\log(2)/a$. It is important to point out that this is a rather
rule--of--thumb approach---our aim is \emph{not} to model the dynamics of infections, but
rather to visualize and understand what the data in front of us is. 
These regressions were not performed from the first day, as initially there seems to be a lot of noise in the tests, the starting time considered is visible in each graph---where the fitted line with prediction confidence band is fitted. We performed a regression for both the daily and cumulative counts. For some regions (Molise, Valle d'Aosta) no regression is performed as the daily counts seem to noisy. Secondly, one can very clearly
identify days when something must have changed due to the testing methodology in the Emilia Romagna region---there are huge dips in the numbers of tests performed. Hence, for this region the dates $28^{\mathrm{th}}$--$30^{\mathrm{th}}$ March were removed for the regression estimation. 
In the Basilicata and Calabria regions spikes to $0$ can also be observed---these
are also removed, as on the log scale would result in infinite values which cannot be handled by the regression
procedure in \code{lm()}. However such dips require careful investigation.

The directly plotted death toll in Figs. \ref{figRawDeathsLombardyVeneto}, \ref{figItaly} and
\ref{figRawDeathsRegion} shows that 
in
the regions Emilia--Romagna, Lombardia, P.A. Bolzano combined with P.A. Trento 
and Valle d'Aosta there is a larger current (spring $2020$) mortality peak than the 
past December/January ($2015$--$2020$ are plotted separately)
maximum one. In the regions 
Liguria, Marche and Piemonte such a larger current peak is present for men only.
In the other regions
for all age groups and both men and women the current ``COVID--$19$ peak'' seems to be approximately of the same height, or lower, 
than past December/January ($2015$--$2019$)
maximum ones. Looking at Italy for men and both sexes combined it is higher, but women seem to have the same peak height.
However, it must be stressed that this is only considering the peak's height, not the total amount of deceased
during the current peak and the December/January ones.

\section{\pkg{COVID-19} \proglang{R} package}\label{secCOVID19}
We used the, available on CRAN, \pkg{COVID19} \proglang{R} package for the purpose
of obtaining the data\footnote{\url{https://cran.r-project.org/web/packages/COVID19/}}.
The package unifies COVID--$19$ datasets across different sources in order to simplify the data acquisition process and the subsequent analysis. 
COVID--$19$ data are pulled in real time and merged with demographic indicators from several 
trusted sources including but not limited to: 
Johns Hopkins University Center for Systems Science and Engineering 
(JHU CSSE)\footnote{\url{https://github.com/CSSEGISandData/COVID-19}}; 
World Bank Open Data\footnote{\url{https://data.worldbank.org/}}; 
World Factbook by CIA\footnote{\url{https://www.cia.gov/library/publications/resources/the-world-factbook/fields/343rank.html}}; 
Ministero della Salute, Dipartimento della Protezione Civile\footnote{\url{https://github.com/pcm-dpc/COVID-19}}; 
Istituto Nazionale di Statistica\footnote{\url{https://www.istat.it/en/population-and-households?data-and-indicators}}; 
Swiss Federal Statistical Office\footnote{\url{https://www.bfs.admin.ch/bfs/en/home/statistics/regional-statistics/regional-portraits-key-figures/cantons/data-explanations.html}}; 
Open Government Data Zurich \footnote{\url{https://github.com/openZH/covid_19}}.
Besides worldwide data, the dataset includes fine--grained data for the \textit{Diamond Princess}, Switzerland and Italy. 
At the time of writing, these include the number of confirmed cases, deaths and tests, total population, 
population ages $0$--$14$, $15$--$64$ and $65+$ (\% of total population), 
median age of population, population density per km$^{2}$, population mortality rate. 
Depending on the data provider, the data are available at the country level, state level, or city level.
For non \proglang{R} users, the combined datasets are available in csv format\footnote{\url{https://covid19datahub.io}}.

\section{Discussion or Should we use these data to calibrate epidemiological models?}
\label{sec:discussion}
In this work we analyzed in depth the two statistics that are commonly
reported for the currently ongoing COVID--$19$ pandemic---the number
of confirmed cases and the number of case fatalities
for the different regions of Italy. 
We found significant variability between regions but also
some common insights. In particular, the number of confirmed
cases is clearly related to the number 
of tests and their ratio seems to be decaying for some time now in all regions.
This is confirmed when looking at the log--scale plot. The difference
between the logarithm of the cumulative number of tests 
and the logarithm of the cumulative number of confirmed seems to 
be (visually) dropping linearly (apart from the below, extremely noisy ones) 
regions and Italy as a whole. Furthermore, for a number of regions
(Molise, Valle d'Aosta), on the log scale, the tests, total, positive and difference
behave very chaotically, suggesting rather various test handling situations, than 
any pattern. Such oscillations can be visible in all regions at the initial stages, but they settle down (apart from the previously mentioned three regions). However, in regions 
with seemingly well--behaved curves
individual huge dips can be observed (Emilia-Romagna, Marche).
Therefore, reports claiming the growth of the epidemic
based only on the increasing number of confirmed individuals
will not be catching its dynamics. 

Furthermore, studying daily positively tested counts could be misleading.
On a number of days we found (for some regions) that this count 
was greater than the number of tests performed. This can certainly be understood,
as the result of reporting procedures, in a crisis situation. However, this 
also implies that any statistical analysis or modelling of such data has to be done very carefully.
We find that the cumulative positively tested fraction behaves much more stably, even 
though in the official cumulative counts decreases can be observed. 

More importantly, using the raw confirmed case counts
one could risk combining the sampling effort with the actual disease spread. 
In our regressions, for the logarithm of the ratio confirmed cases to total tests on time
the fitted slopes are all negative (indicating that the virus is receding and this was observed also by \citet{Rep}). 
 
\begin{figure}[!ht]
\begin{center}
\includegraphics[width=0.98\textwidth]{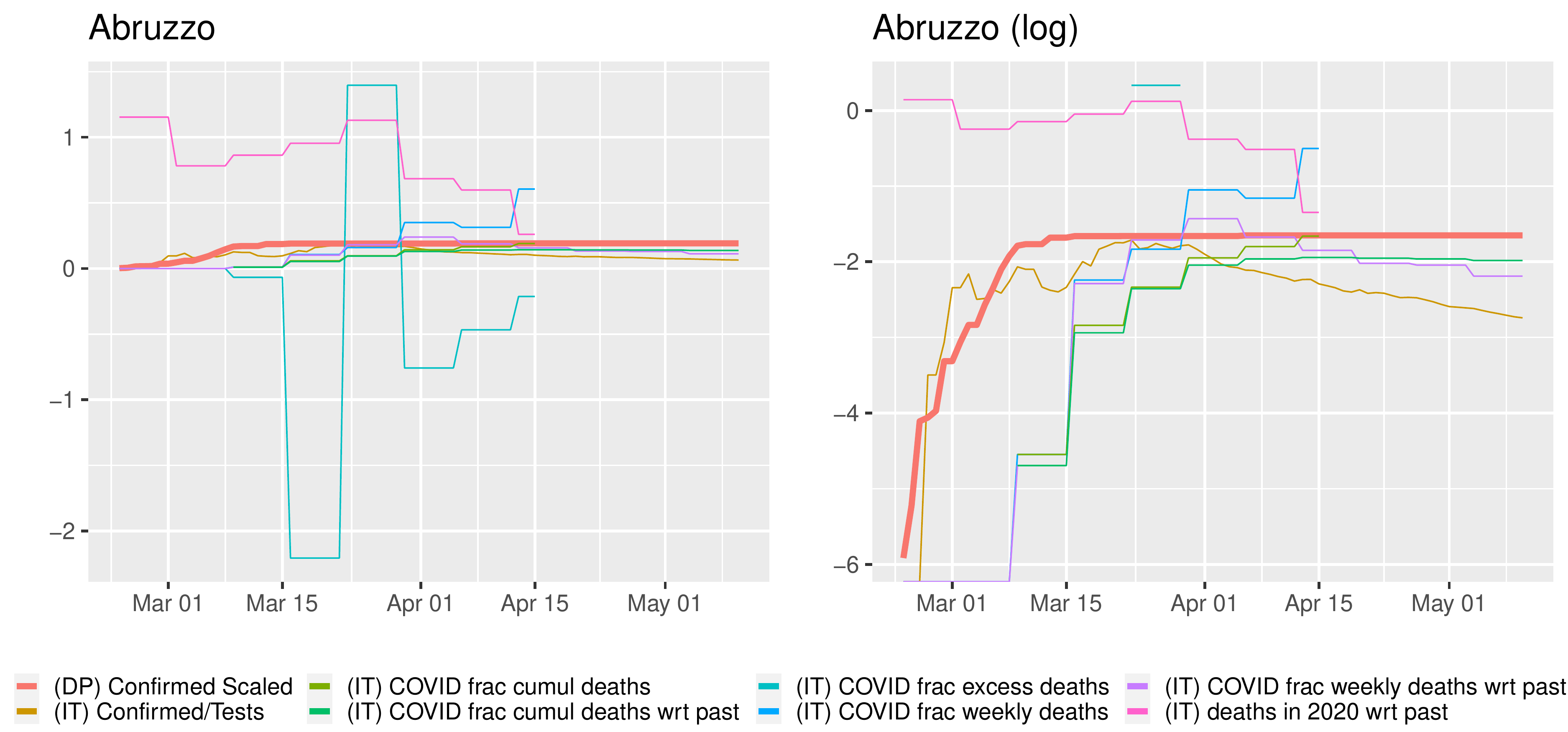}
\\
\includegraphics[width=0.98\textwidth]{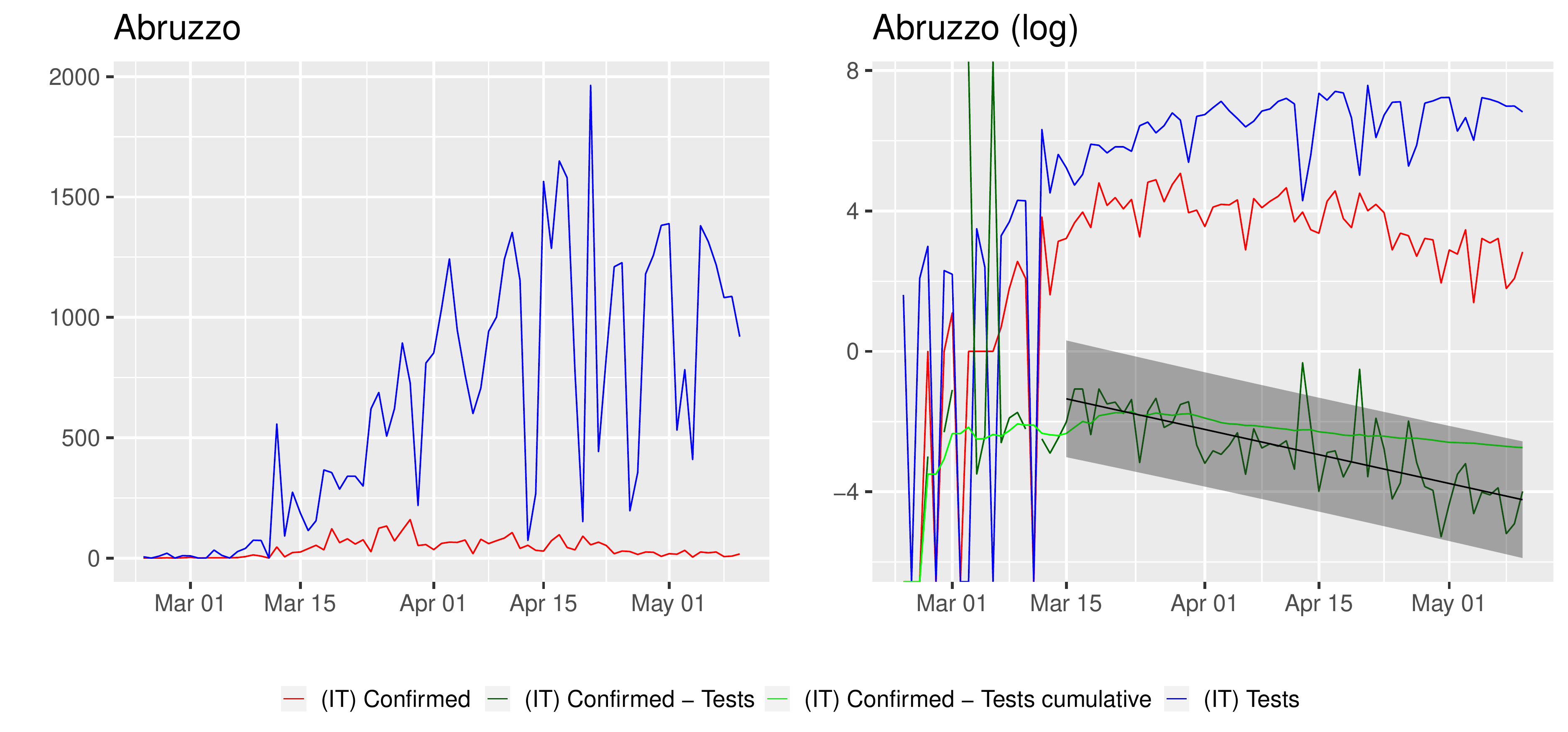}
\end{center}
\caption{Comparison of curves for Abruzzo region. Left: $y$--axis on normal scale, right: on logarithmic
scale. 
Regression line shown for $\log(\mathrm{daily~confirmed})-\log(\mathrm{daily~tested})\sim time$
with $95\%$ prediction band. Slope of regression with $95\%$ confidence interval: 
$a_{\mathrm{f}}=-0.051 (-0.064,-0.039)$, this corresponds to a half--life time (in days) of $13.501 (10.830,17.923)$.
The slope of the regression
$\log(\mathrm{daily~confirmed})\sim time$ is 
$a_{\mathrm{raw}}=-0.030 (-0.040,-0.021)$ corresponding to a half--life (in days) of $22.801 (17.203,33.796)$.
The slope of the regression
$\log(\mathrm{daily~tests})\sim time$ is 
$0.021 (0.010,0.048)$ corresponding to a doubling time (in days) of $33.105 (14.398,68.114)$.
Ratio of slopes for $a_{\mathrm{f}}/a_{\mathrm{raw}}=1.689$, with corresponding half--lives' ratio: $0.592$.
The slope of the regression
$\log(\mathrm{cumulative~confirmed})-\log(\mathrm{cumulative~tested})\sim time$ is 
$-0.017 (-0.019,-0.015)$ corresponding to a half--life (in days) of  $40.259 (35.924,45.784)$.
}\label{figAbruzzo}
\end{figure}

\begin{figure}[!ht]
\begin{center}
\includegraphics[width=0.98\textwidth]{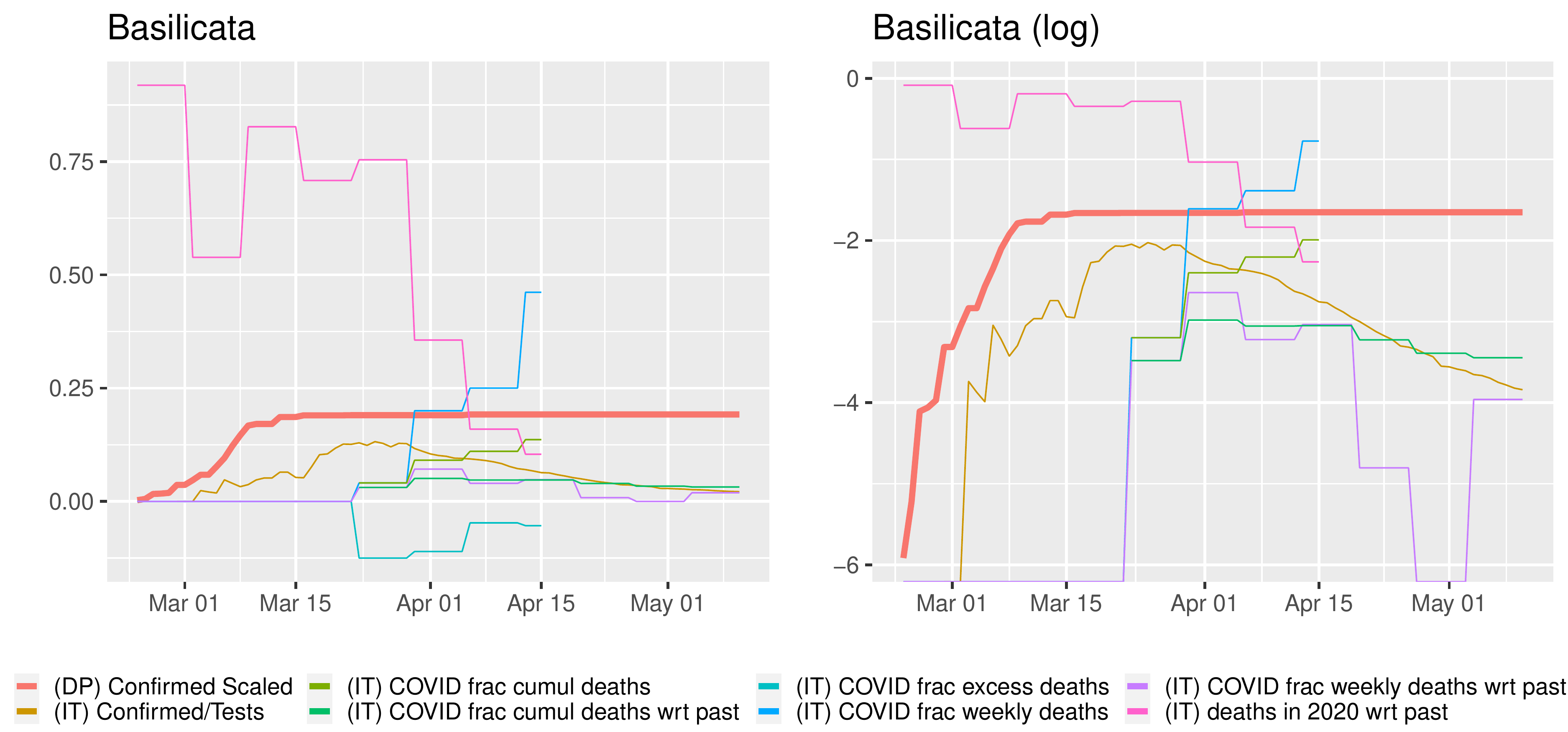}
\\
\includegraphics[width=0.98\textwidth]{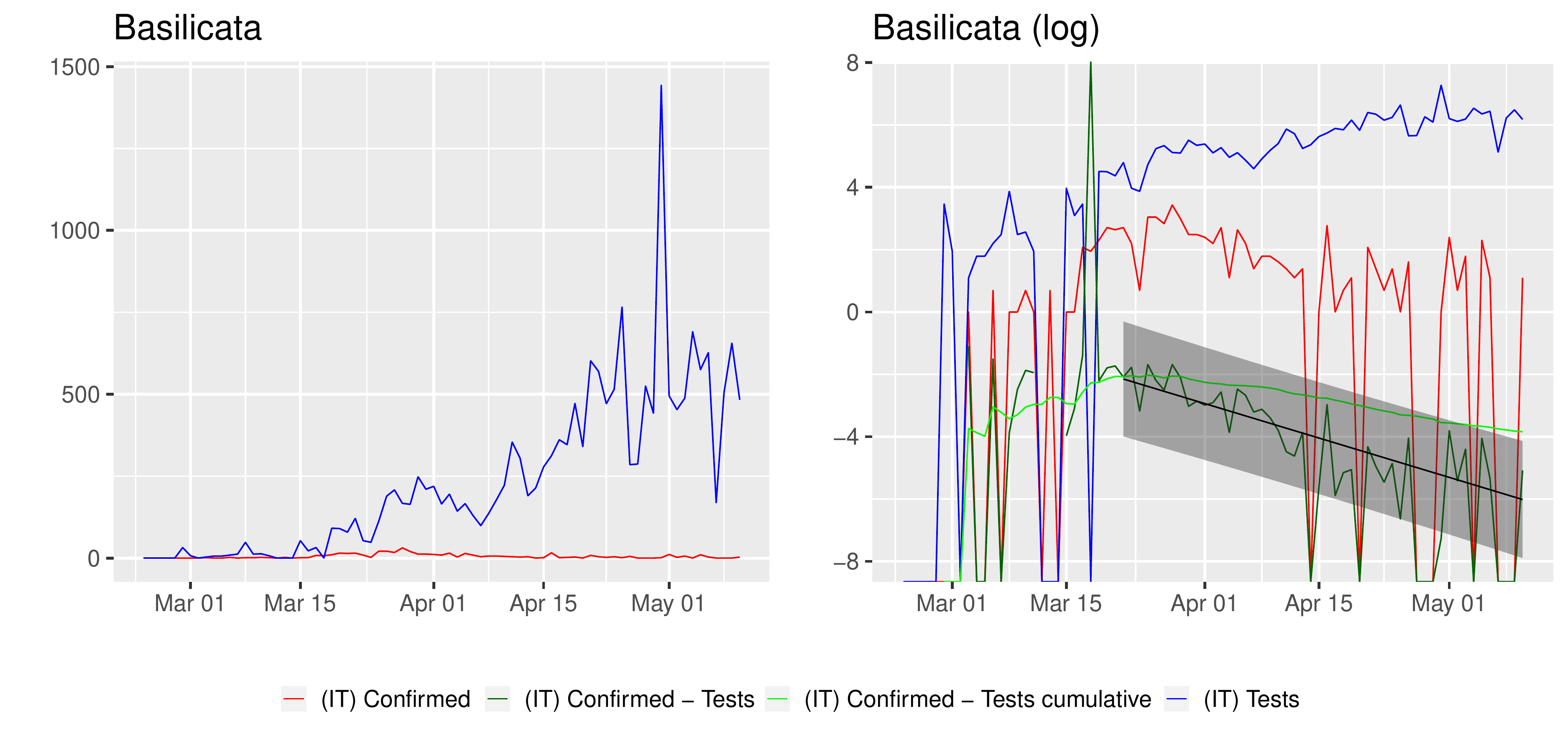}
\end{center}
\caption{Comparison of curves for Basilicata region. Left: $y$--axis on normal scale, right: on logarithmic
scale.
Regression line shown for $\log(\mathrm{daily~confirmed})-\log(\mathrm{daily~tested})\sim time$
with $95\%$ prediction band. Slope of regression with $95\%$ confidence interval: 
$a_{\mathrm{f}}=-0.079 (-0.099,-0.059)$, this corresponds to a half--life (in days) of $8.786 (7.037,11.693)$. 
The slope of the regression
$\log(\mathrm{daily~confirmed})\sim time$ is 
$a_{\mathrm{raw}}=-0.036 (-0.054,-0.018)$ corresponding to a half--life (in days) of $19.170 (12.816,38.025)$.
The slope of the regression
$\log(\mathrm{daily~tests})\sim time$ is 
$0.038 (0.030,0.028)$ corresponding to a doubling time (in days) of $18.446 (24.860,23.491)$.
Ratio of slopes for $a_{\mathrm{f}}/a_{\mathrm{raw}}=2.182$, with corresponding half--lives' ratio: $0.458$.
The slope of the regression
$\log(\mathrm{cumulative~confirmed})-\log(\mathrm{cumulative~tested})\sim time$ is 
$-0.041 (-0.043,-0.040)$ corresponding to a half--life (in days) of $16.866 (16.265,17.513)$.
}\label{figBasilicata}
\end{figure}

\begin{figure}[!ht]
\begin{center}
\includegraphics[width=0.98\textwidth]{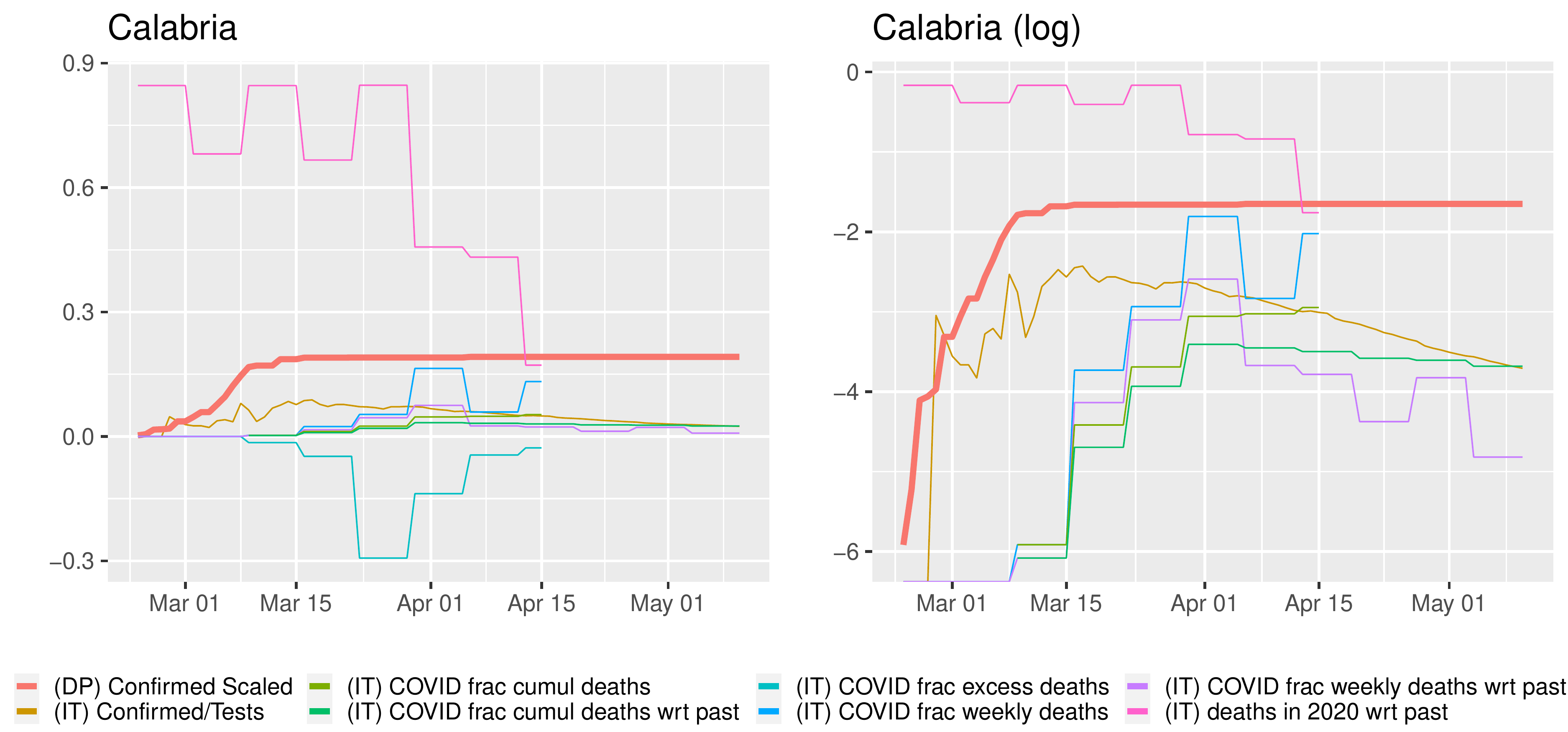}
\\
\includegraphics[width=0.98\textwidth]{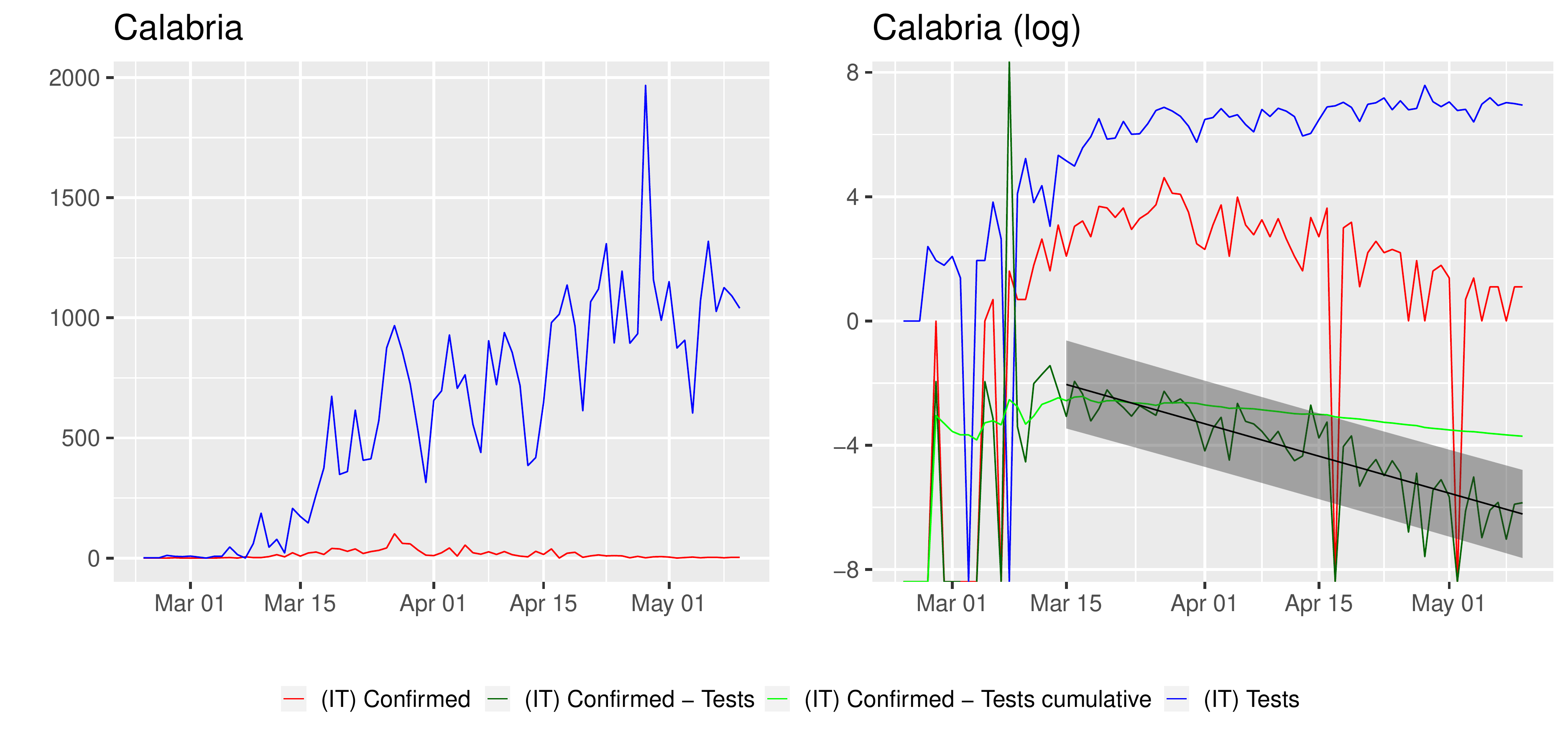}
\end{center}
\caption{Comparison of curves for Calabria region. Left: $y$--axis on normal scale, right: on logarithmic
scale. 
Regression line shown for $\log(\mathrm{daily~confirmed})-\log(\mathrm{daily~tested})\sim time$
with $95\%$ prediction band. Slope of regression with $95\%$ confidence interval: 
$a_{\mathrm{f}}=-0.074 (-0.085,-0.063)$,this corresponds to a half--life (in days) of $9.315 (8.122,10.918)$. 
The slope of the regression
$\log(\mathrm{daily~confirmed})\sim time$ is 
$a_{\mathrm{raw}}=-0.052 (-0.064,-0.040)$ corresponding to a half--life (in days) of $13.249 (10.747,17.271)$.
The slope of the regression
$\log(\mathrm{daily~tests})\sim time$ is 
$0.022 (0.016,0.043)$ corresponding to a doubling time (in days) of $31.709 (16.233,42.761)$.
Ratio of slopes for $a_{\mathrm{f}}/a_{\mathrm{raw}}=1.422$, with corresponding half--lives' ratio: $0.703$.
The slope of the regression
$\log(\mathrm{cumulative~confirmed})-\log(\mathrm{cumulative~tested})\sim time$ is 
$-0.023 (-0.024,-0.022)$ corresponding to a half--life (in days) of $30.582 (29.262,32.028)$.
}\label{figCalabria}
\end{figure}

\begin{figure}[!ht]
\begin{center}
\includegraphics[width=0.98\textwidth]{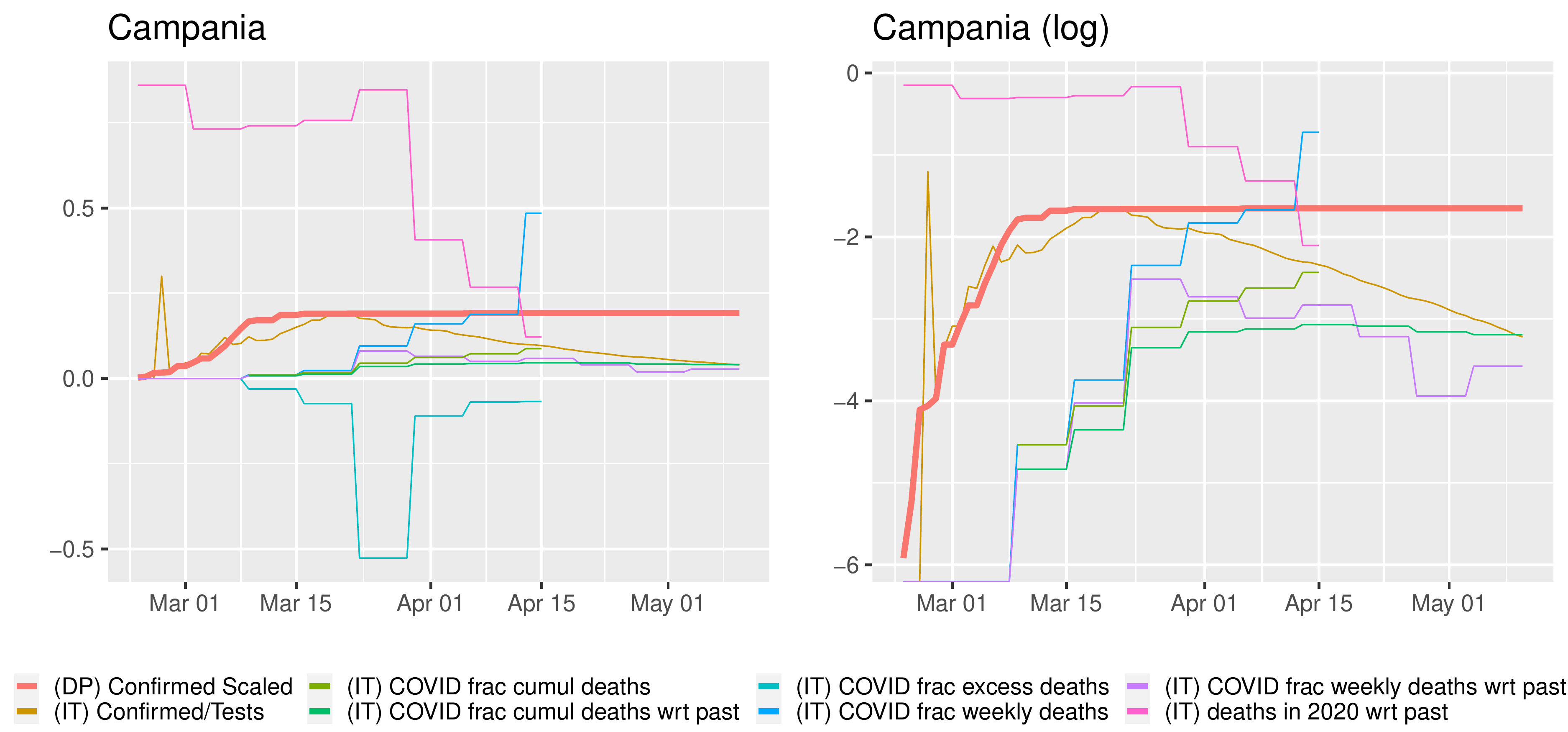}
\\
\includegraphics[width=0.98\textwidth]{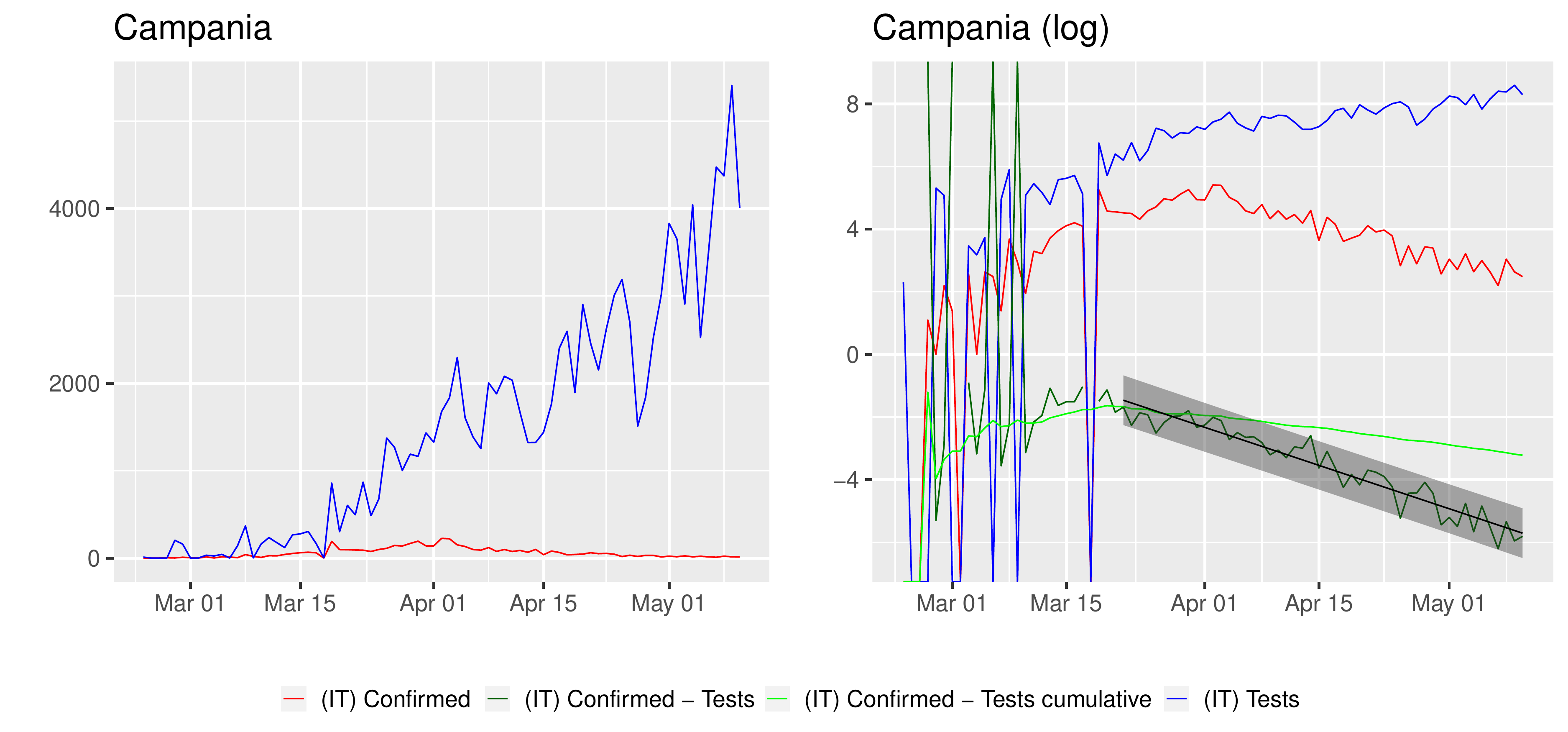}
\end{center}
\caption{Comparison of curves for Campania region. Left: $y$--axis on normal scale, right: on logarithmic
scale.
Regression line shown for $\log(\mathrm{daily~confirmed})-\log(\mathrm{daily~tested})\sim time$
with $95\%$ prediction band. Slope of regression with $95\%$ confidence interval: 
$a_{\mathrm{f}}=-0.087 (-0.094,-0.079)$ ,this corresponds to a half--life (in days) of $7.998 (7.376,8.734)$. 
The slope of the regression
$\log(\mathrm{daily~confirmed})\sim time$ is 
$a_{\mathrm{raw}}=-0.055 (-0.062,-0.047)$ corresponding to a half--life (in days) of $12.707 (11.155,14.762)$.
The slope of the regression
$\log(\mathrm{daily~tests})\sim time$ is 
$0.032 (0.027,0.032)$ corresponding to a doubling time (in days) of $21.580 (21.771,25.624)$.
Ratio of slopes for $a_{\mathrm{f}}/a_{\mathrm{raw}}=1.589$, with corresponding half--lives' ratio: $0.629$.
The slope of the regression
$\log(\mathrm{cumulative~confirmed})-\log(\mathrm{cumulative~tested})\sim time$ is 
$-0.031 (-0.032,-0.030)$ corresponding to a half--life (in days) of $22.438 (21.964,22.933)$.
}\label{figCampania}
\end{figure}

\begin{figure}[!ht]
\begin{center}
\includegraphics[width=0.98\textwidth]{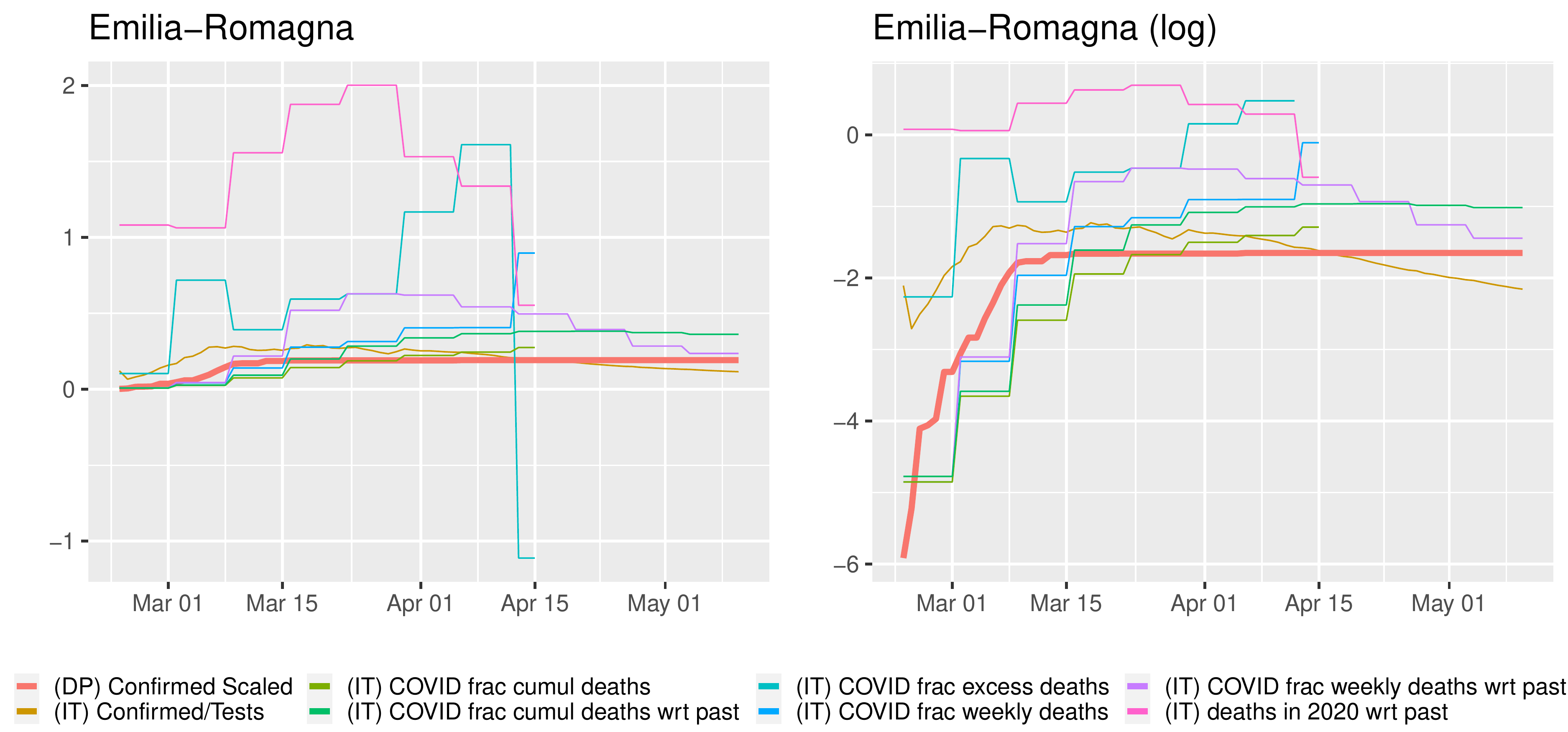}
\\
\includegraphics[width=0.98\textwidth]{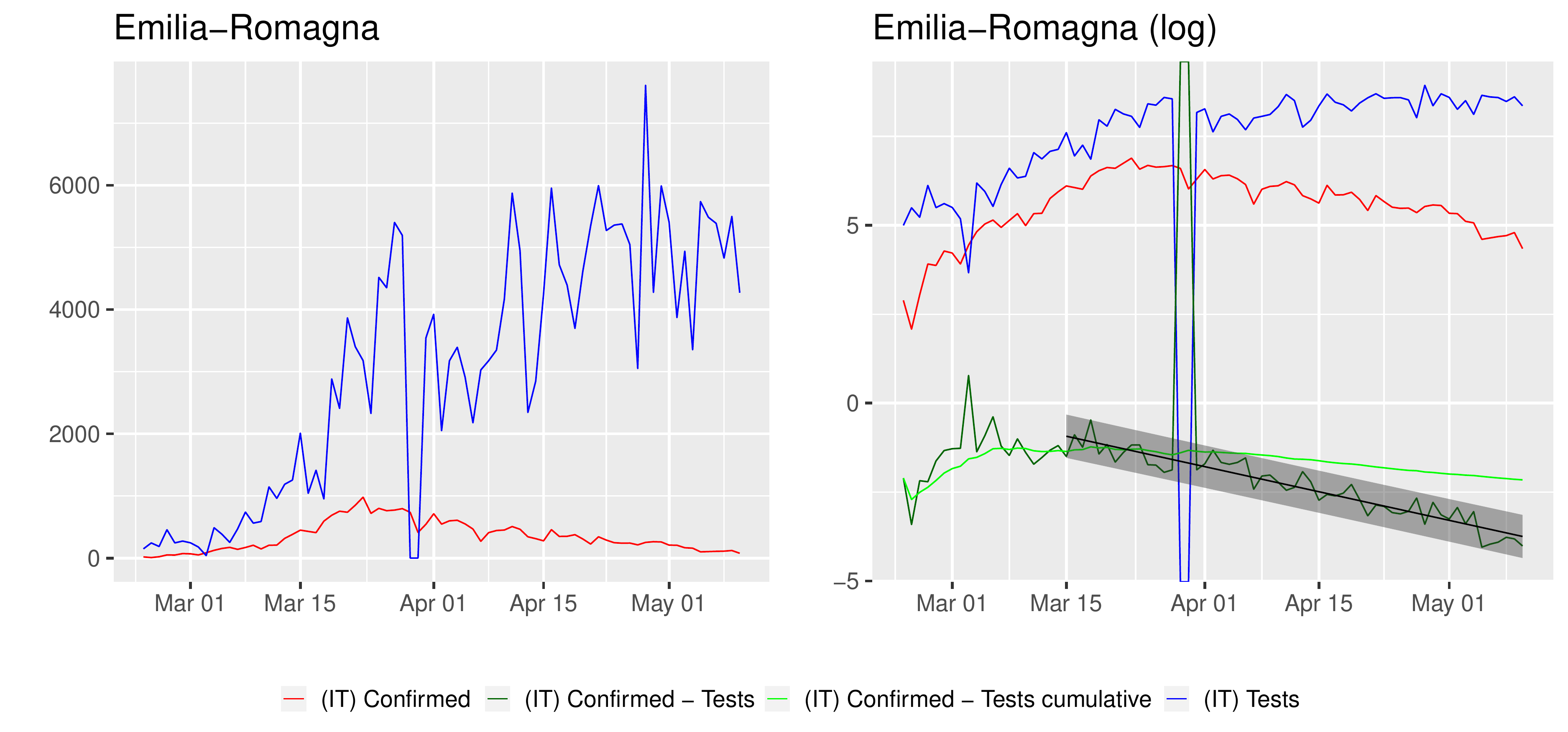}
\end{center}
\caption{Comparison of curves for Emilia Romagna region. Left: $y$--axis on normal scale, right: on logarithmic
scale.
Regression line shown for $\log(\mathrm{daily~confirmed})-\log(\mathrm{daily~tested})\sim time$
with $95\%$ prediction band. Slope of regression with $95\%$ confidence interval:
$a_{\mathrm{f}}=-0.050 (-0.055,-0.046)$, this corresponds to a half--life (in days) of $13.774 (12.594,15.196)$.
The slope of the regression
$\log(\mathrm{daily~confirmed})\sim time$ is 
$a_{\mathrm{raw}}=-0.034 (-0.038,-0.029)$ corresponding to a half--life (in days) of $20.681 (18.199,23.948)$.
The slope of the regression
$\log(\mathrm{daily~tests})\sim time$ is 
$0.016 (0.011,0.035)$ corresponding to a doubling time (in days) of $42.042 (19.760,62.087)$.
Ratio of slopes for $a_{\mathrm{f}}/a_{\mathrm{raw}}=1.502$, with corresponding half--lives' ratio: $0.666$.
The slope of the regression
$\log(\mathrm{cumulative~confirmed})-\log(\mathrm{cumulative~tested})\sim time$ is 
$-0.017 (-0.018,-0.016)$ corresponding to a half--life (in days) of $40.674 (38.280,43.387)$. 
The data for the dates $28^{\mathrm{th}}$--$30^{\mathrm{th}}$ March are removed for the regression analysis.
}\label{figEmiliaRomagna}
\end{figure}

\begin{figure}[!ht]
\begin{center}
\includegraphics[width=0.98\textwidth]{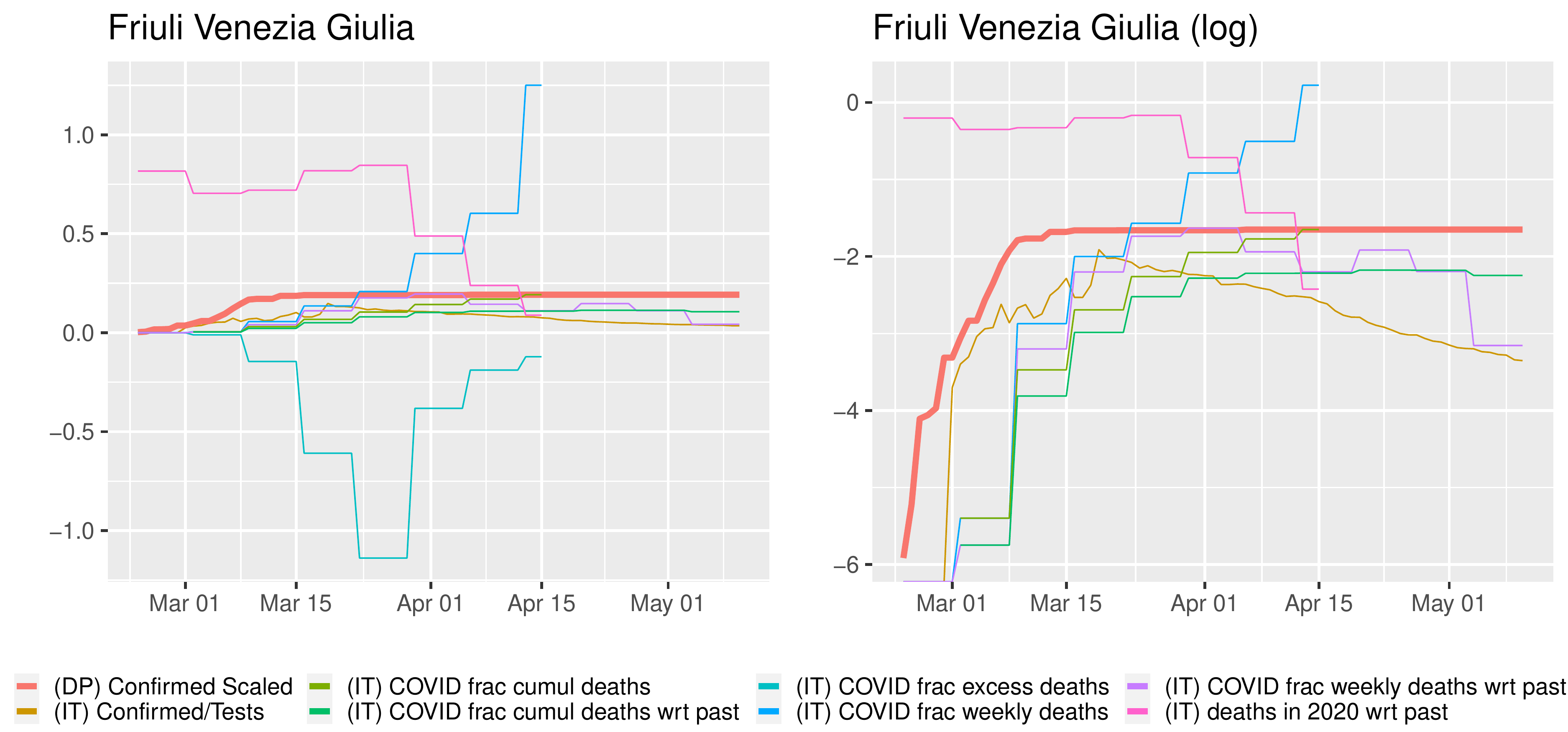}
\\
\includegraphics[width=0.98\textwidth]{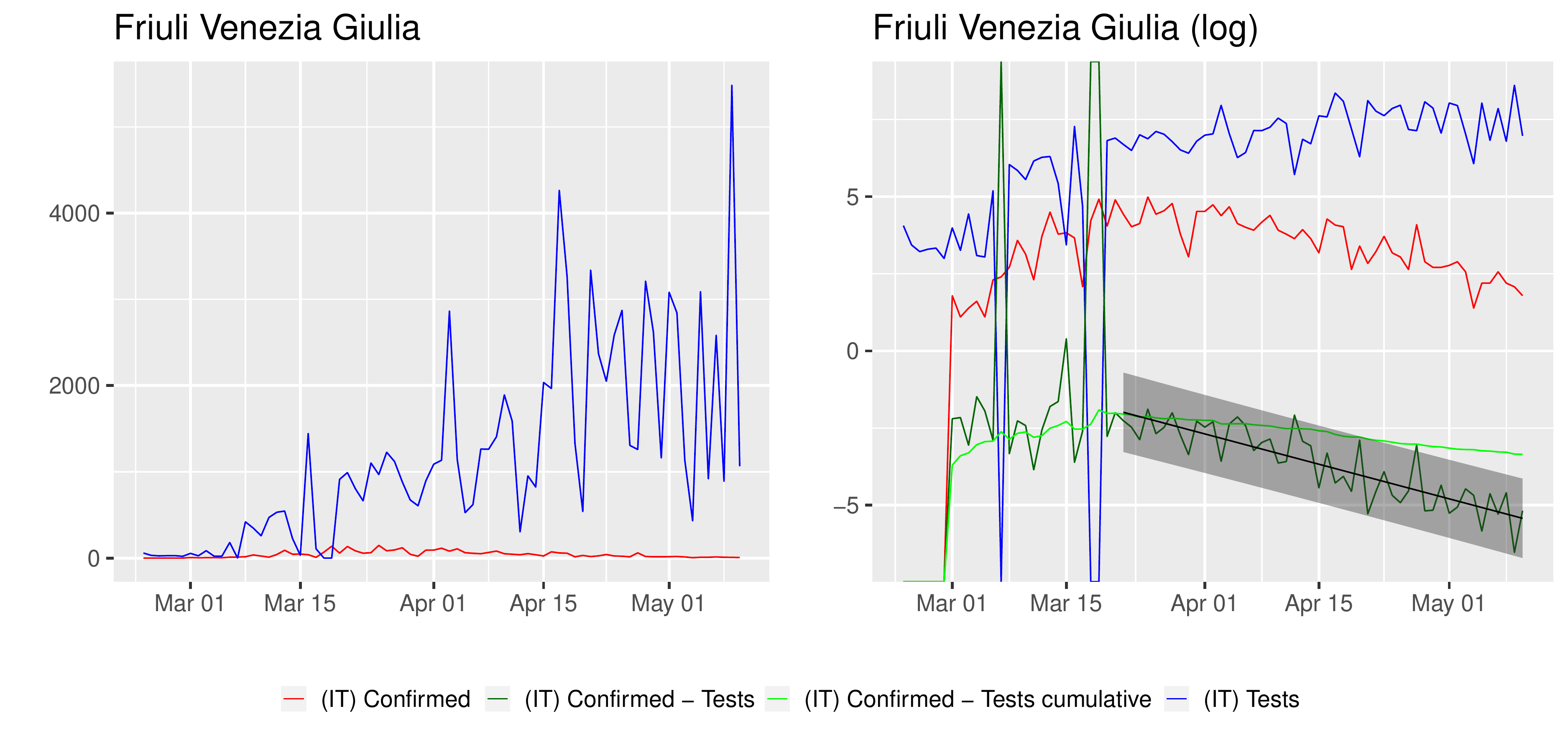}
\end{center}
\caption{Comparison of curves for Friuli Venezia Giulia region. Left: $y$--axis on normal scale, right: on logarithmic scale. 
Regression line shown for $\log(\mathrm{daily~confirmed})-\log(\mathrm{daily~tested})\sim time$
with $95\%$ prediction band. Slope of regression with $95\%$ confidence interval:
$a_{\mathrm{f}}=-0.070 (-0.082,-0.058)$, this corresponds to a half--life (in days) of $9.885 (8.456,11.894)$. 
The slope of the regression
$\log(\mathrm{daily~confirmed})\sim time$ is 
$a_{\mathrm{raw}}=-0.052 (-0.061,-0.043)$ corresponding to a half--life (in days) of $13.315 (11.352,16.098)$.
The slope of the regression
$\log(\mathrm{daily~tests})\sim time$ is 
$0.018 (0.007,0.025)$ corresponding to a doubling time (in days) of $38.365 (27.562,103.079)$.
Ratio of slopes for $a_{\mathrm{f}}/a_{\mathrm{raw}}=1.347$, with corresponding half--lives' ratio: $0.742$.
The slope of the regression
$\log(\mathrm{cumulative~confirmed})-\log(\mathrm{cumulative~tested})\sim time$ is 
$-0.028 (-0.029,-0.027)$ corresponding to a half--life (in days) of $24.866 (24.101,25.680)$.
}\label{figFVG}
\end{figure}

\begin{figure}[!ht]
\begin{center}
\includegraphics[width=0.98\textwidth]{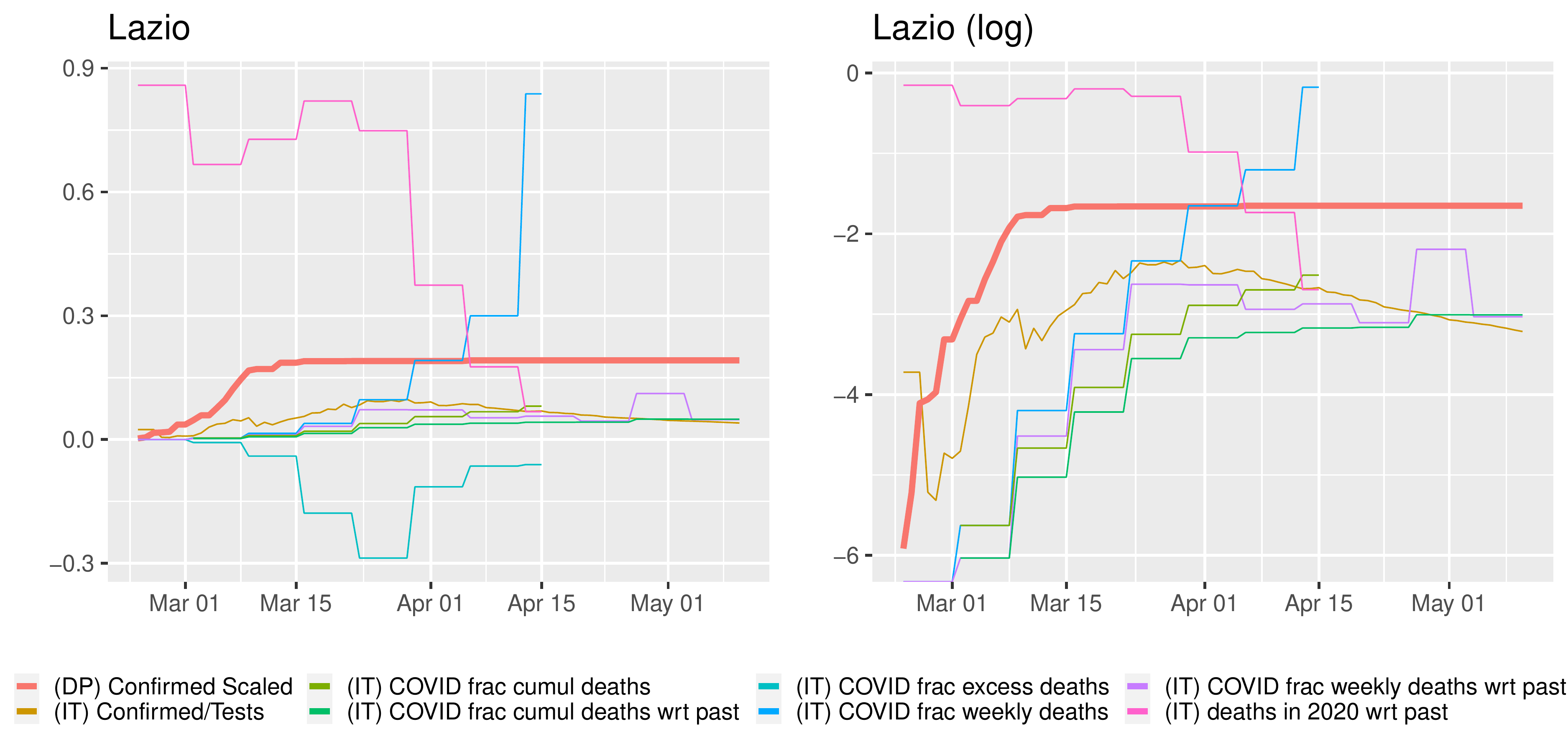}
\\
\includegraphics[width=0.98\textwidth]{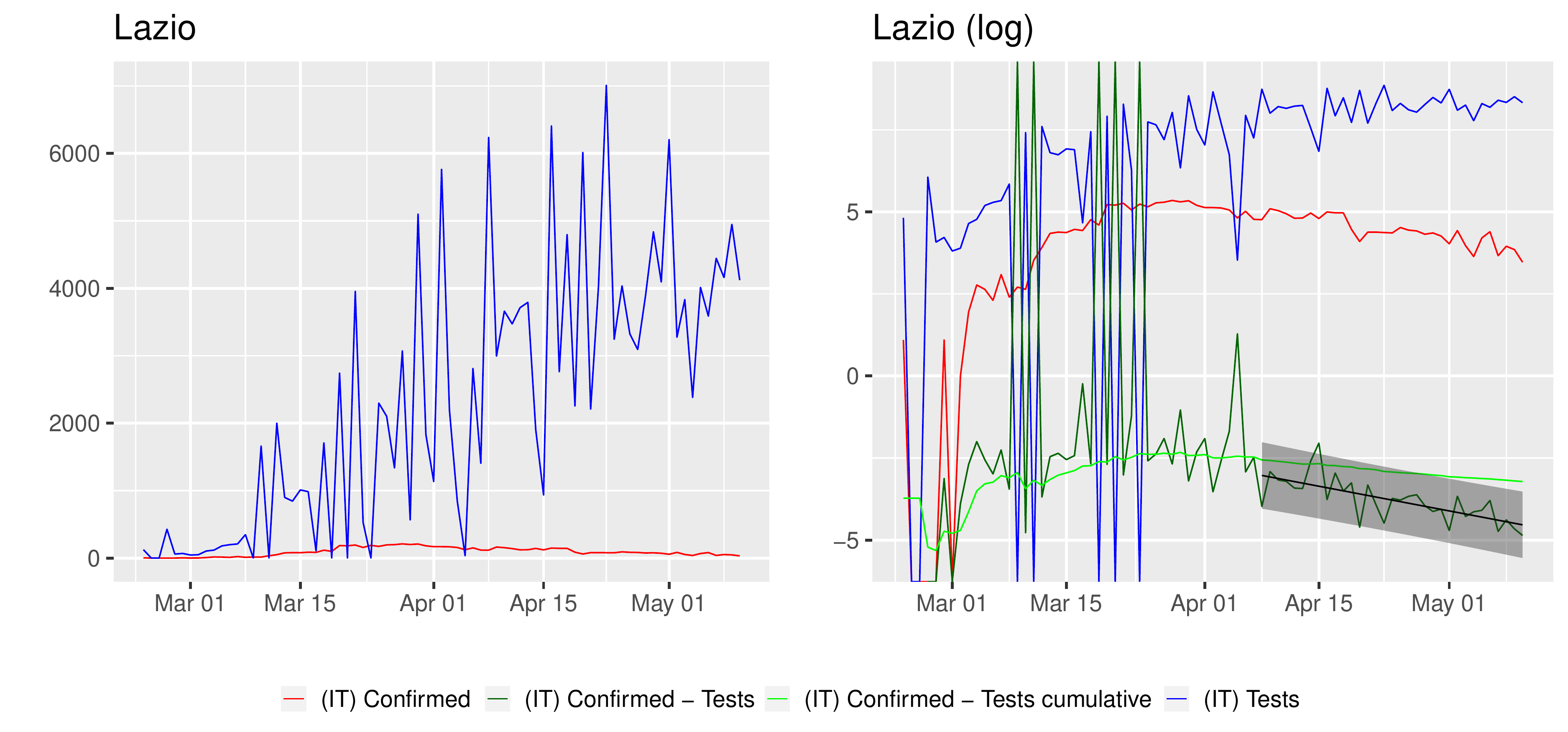}
\end{center}
\caption{Comparison of curves for Lazio region. Left: $y$--axis on normal scale, right: on logarithmic
scale.
Regression line shown for $\log(\mathrm{daily~confirmed})-\log(\mathrm{daily~tested})\sim time$
with $95\%$ prediction band. Slope of regression with $95\%$ confidence interval: 
$a_{\mathrm{f}}= -0.047 (-0.064,-0.030)$, this corresponds to a half--life (in days) of $ 14.740 (10.854,22.960)$. 
The slope of the regression
$\log(\mathrm{daily~confirmed})\sim time$ is 
$a_{\mathrm{raw}}=-0.039 (-0.047,-0.032)$ corresponding to a half--life (in days) of $17.554 (14.679,21.828)$.
The slope of the regression
$\log(\mathrm{daily~tests})\sim time$ is 
$0.008 (-0.007,0.017)$.
Ratio of slopes for $a_{\mathrm{f}}/a_{\mathrm{raw}}=1.191$, with corresponding half--lives' ratio: $0.840$.
The slope of the regression
$\log(\mathrm{cumulative~confirmed})-\log(\mathrm{cumulative~tested})\sim time$ is 
$-0.021 (-0.022,-0.020)$ corresponding to a half--life (in days) of $32.935 (32.015,33.910)$.
}\label{figLazio}
\end{figure}

\begin{figure}[!ht]
\begin{center}
\includegraphics[width=0.98\textwidth]{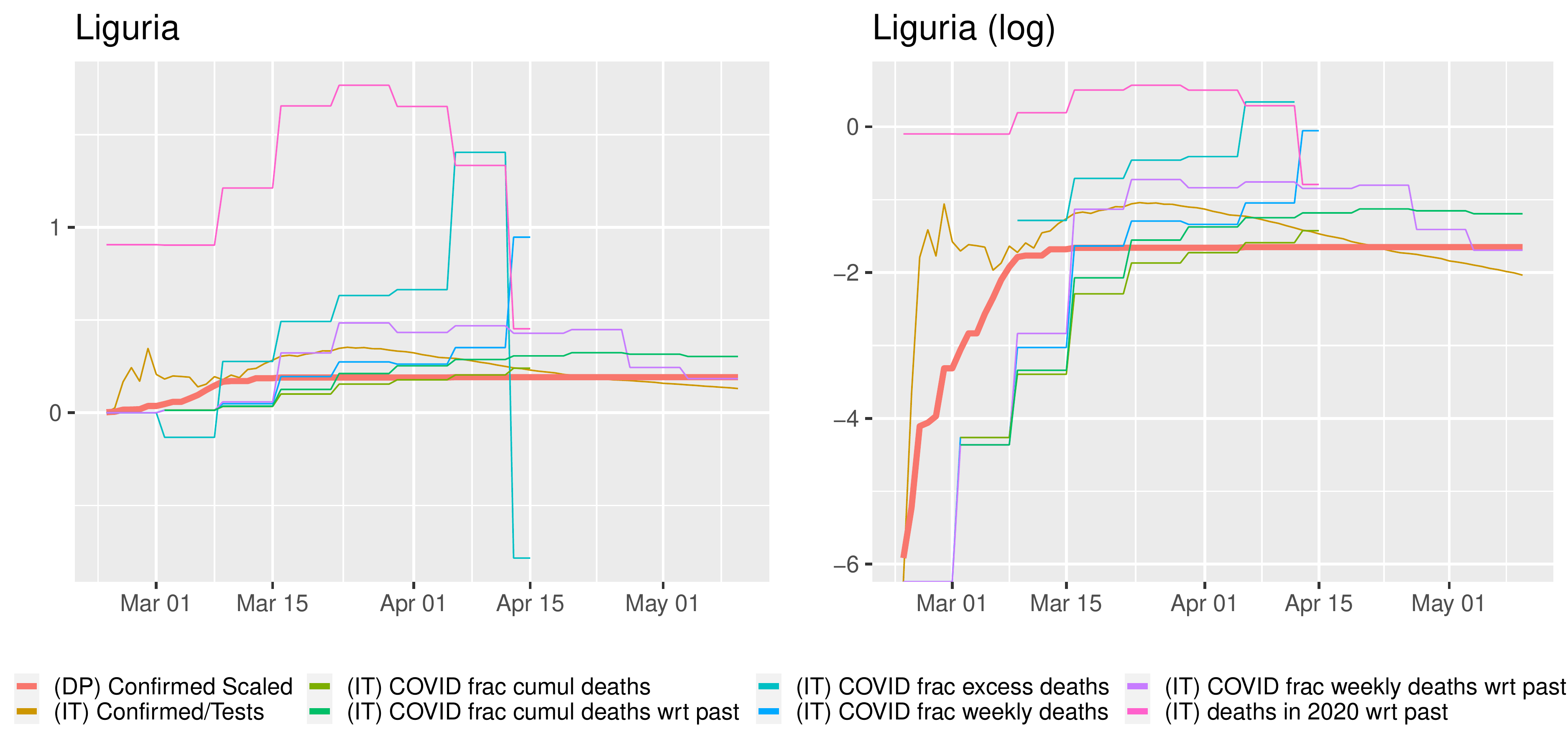}
\\
\includegraphics[width=0.98\textwidth]{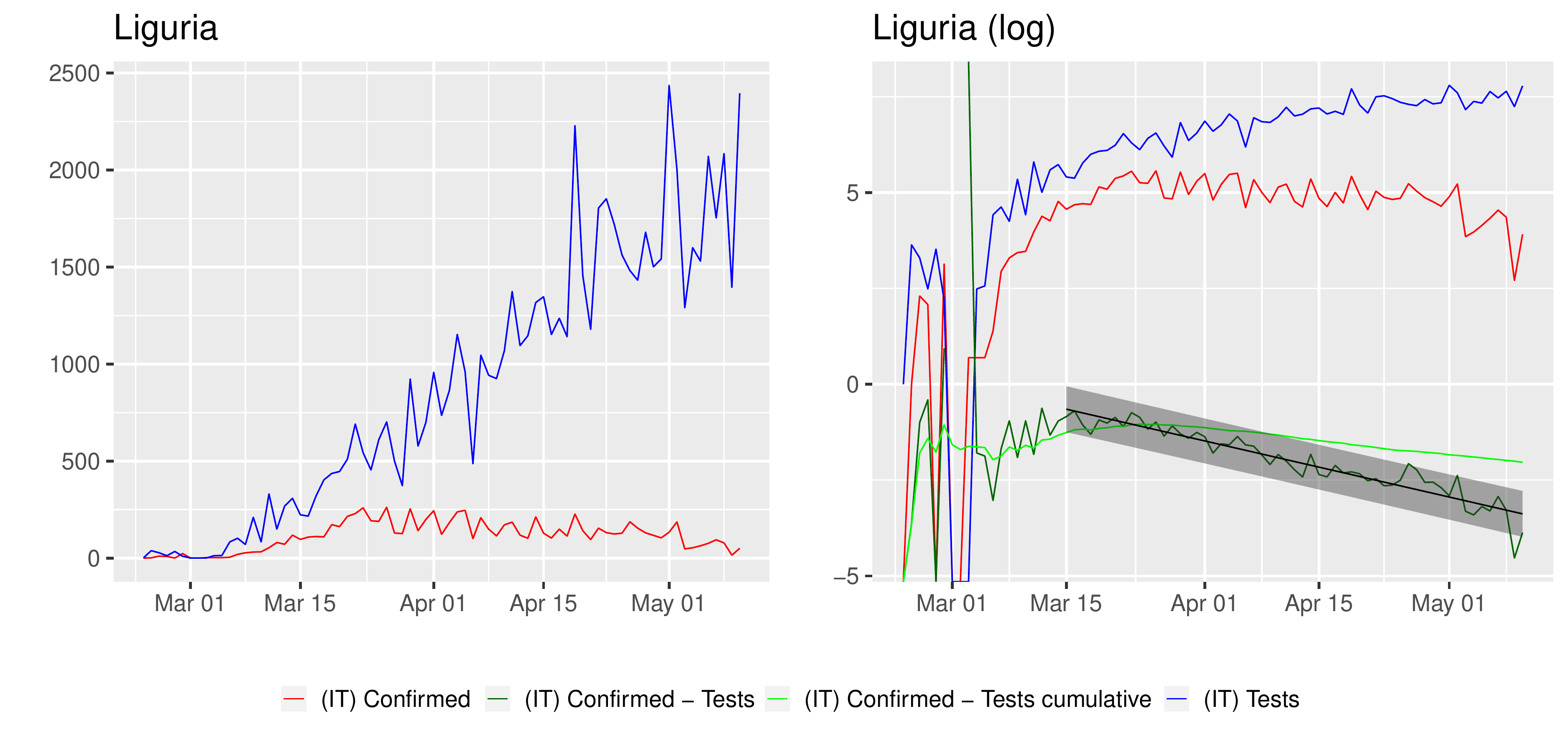}
\end{center}
\caption{Comparison of curves for Liguria region. Left: $y$--axis on normal scale, right: on logarithmic
scale. 
Regression line shown for $\log(\mathrm{daily~confirmed})-\log(\mathrm{daily~tested})\sim time$
with $95\%$ prediction band. Slope of regression with $95\%$ confidence interval: 
$a_{\mathrm{f}}= -0.049 (-0.053,-0.044)$, this corresponds to a half--life (in days) of $14.217 (13.002,15.682)$.
The slope of the regression
$\log(\mathrm{daily~confirmed})\sim time$ is 
$a_{\mathrm{raw}}=-0.016 (-0.023,-0.010)$ corresponding to a half--life (in days) of $42.208 (29.848,72.036)$.
The slope of the regression
$\log(\mathrm{daily~tests})\sim time$ is 
$0.032 (0.028,0.045)$ corresponding to a doubling time (in days) of $21.437 (15.371,24.575)$.
Ratio of slopes for $a_{\mathrm{f}}/a_{\mathrm{raw}}=2.969$, with corresponding half--lives' ratio: $0.337$.
The slope of the regression
$\log(\mathrm{cumulative~confirmed})-\log(\mathrm{cumulative~tested})\sim time$ is 
$ -0.018 (-0.020,-0.017)$ corresponding to a half--life (in days) of $37.641 (34.775,41.024)$.
}\label{figLiguria}
\end{figure}

\begin{figure}[!ht]
\begin{center}
\includegraphics[width=0.98\textwidth]{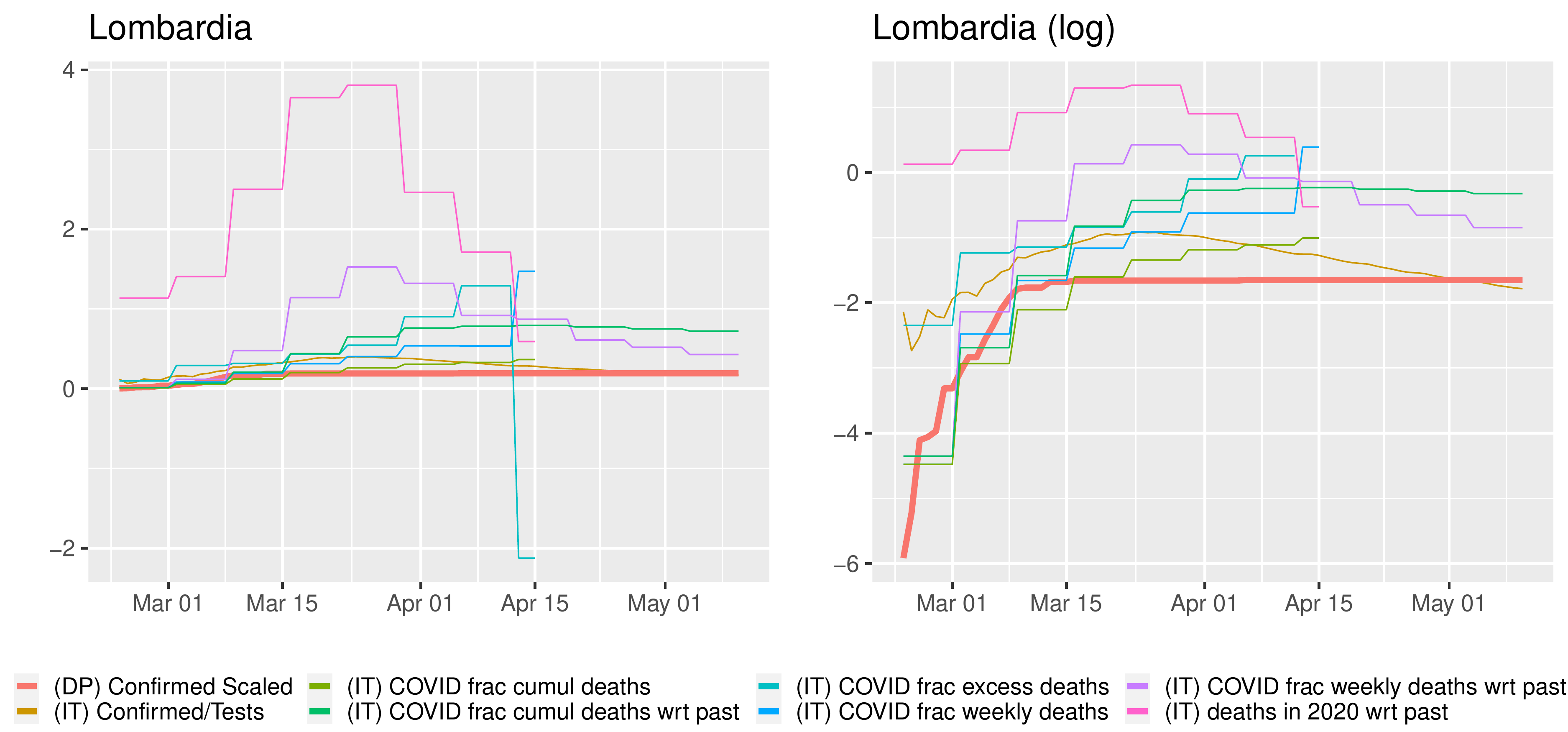}
\\
\includegraphics[width=0.98\textwidth]{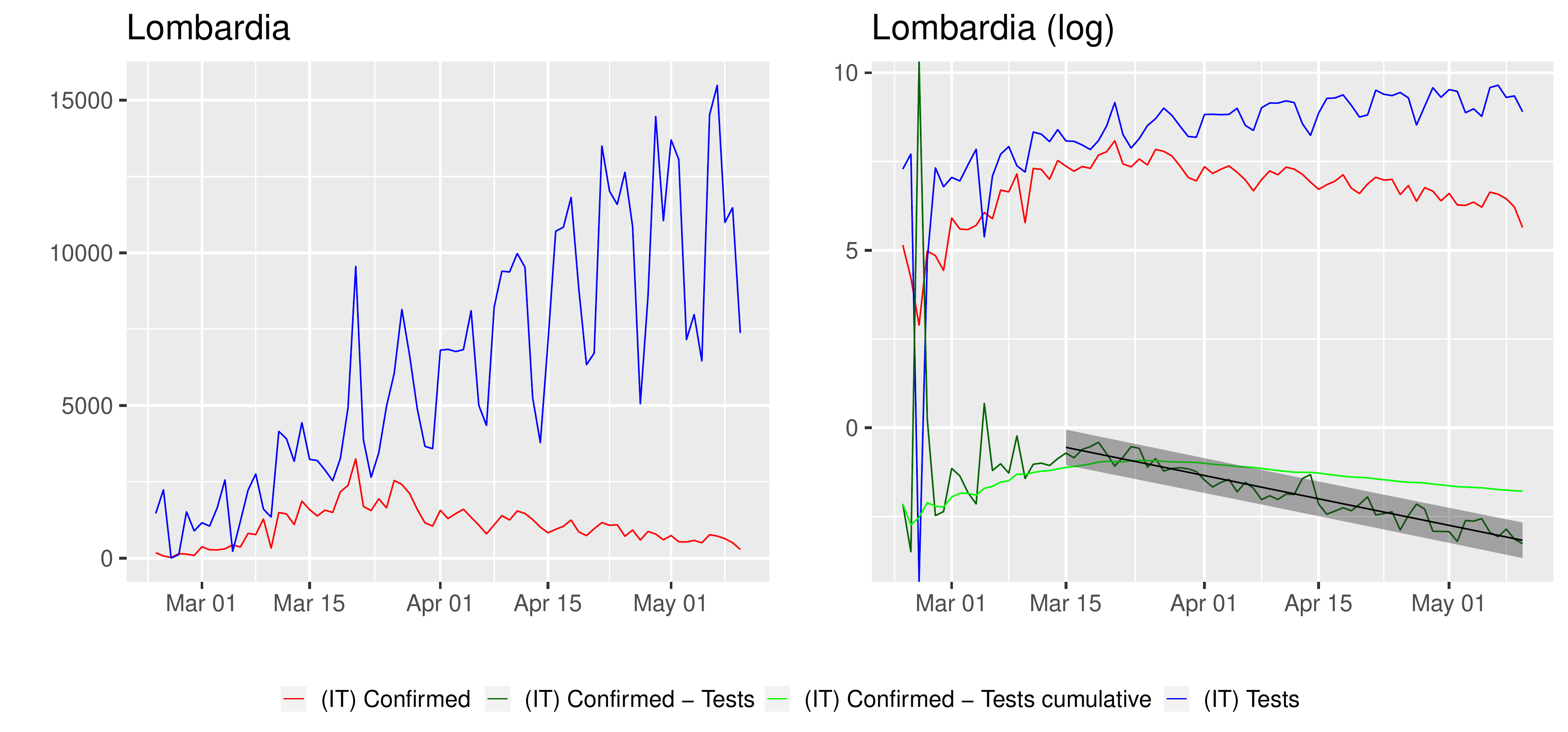}
\end{center}
\caption{Comparison of curves for Lombardia region. Left: $y$--axis on normal scale, right: on logarithmic
scale. 
Regression line shown for $\log(\mathrm{daily~confirmed})-\log(\mathrm{daily~tested})\sim time$
with $95\%$ prediction band. Slope of regression with $95\%$ confidence interval:
$a_{\mathrm{f}}=-0.047 (-0.051,-0.043)$, this corresponds to a half--life (in days) of $14.829 (13.712,16.144)$. 
The slope of the regression
$\log(\mathrm{daily~confirmed})\sim time$ is 
$a_{\mathrm{raw}}=-0.025 (-0.029,-0.021)$ corresponding to a half--life (in days) of $27.656 (23.961,32.698)$.
The slope of the regression
$\log(\mathrm{daily~tests})\sim time$ is 
$0.022 (0.016,0.030)$ corresponding to a doubling time (in days) of $31.970 (23.280,42.506)$.
Ratio of slopes for $a_{\mathrm{f}}/a_{\mathrm{raw}}=1.865$, with corresponding half--lives' ratio: $0.536$.
The slope of the regression
$\log(\mathrm{cumulative~confirmed})-\log(\mathrm{cumulative~tested})\sim time$ is 
$-0.016 (-0.018,-0.015)$ corresponding to a half--life (in days) of $42.525 (39.386,46.208)$.
}\label{figLombardyApp}
\end{figure}

\begin{figure}[!ht]
\begin{center}
\includegraphics[width=0.98\textwidth]{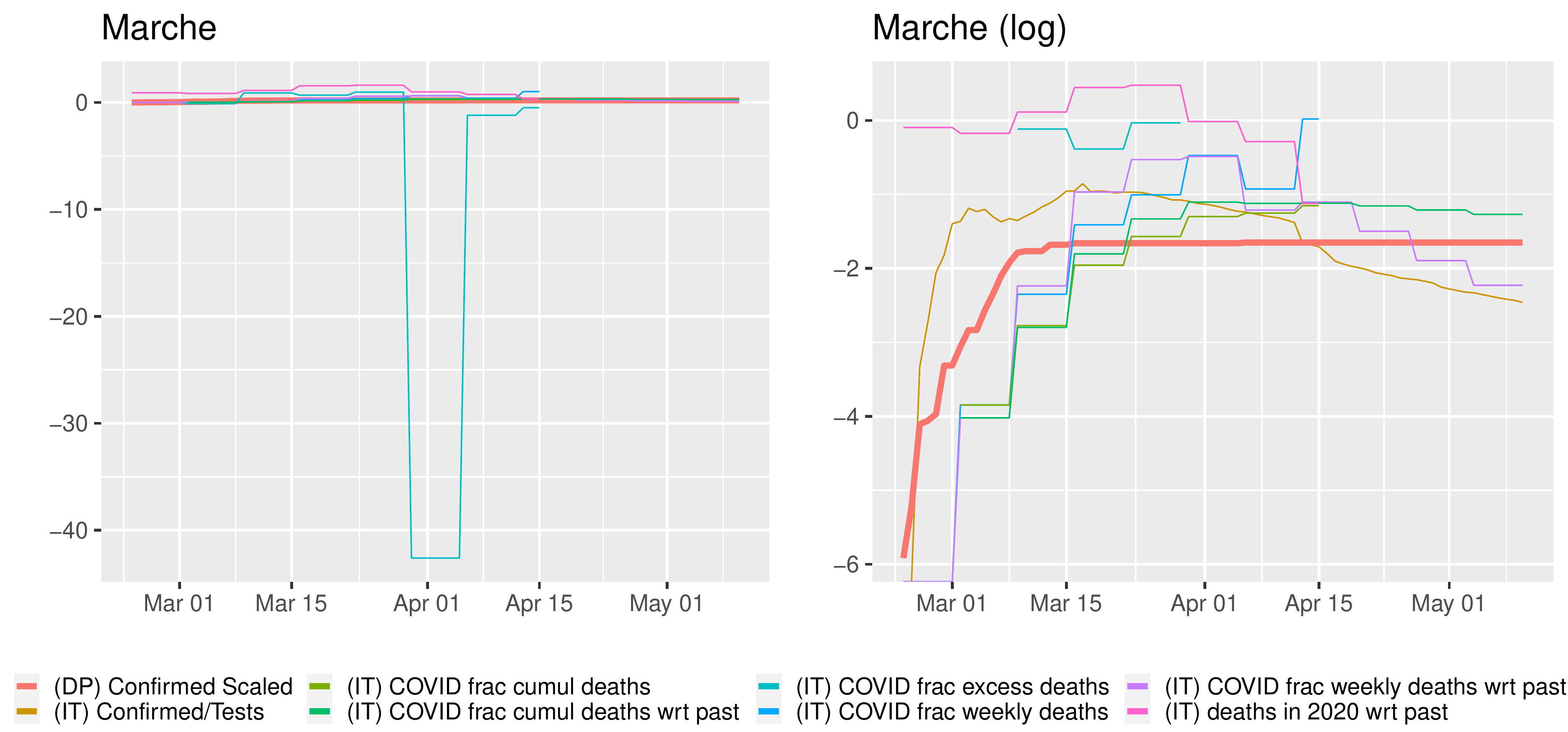}
\\
\includegraphics[width=0.98\textwidth]{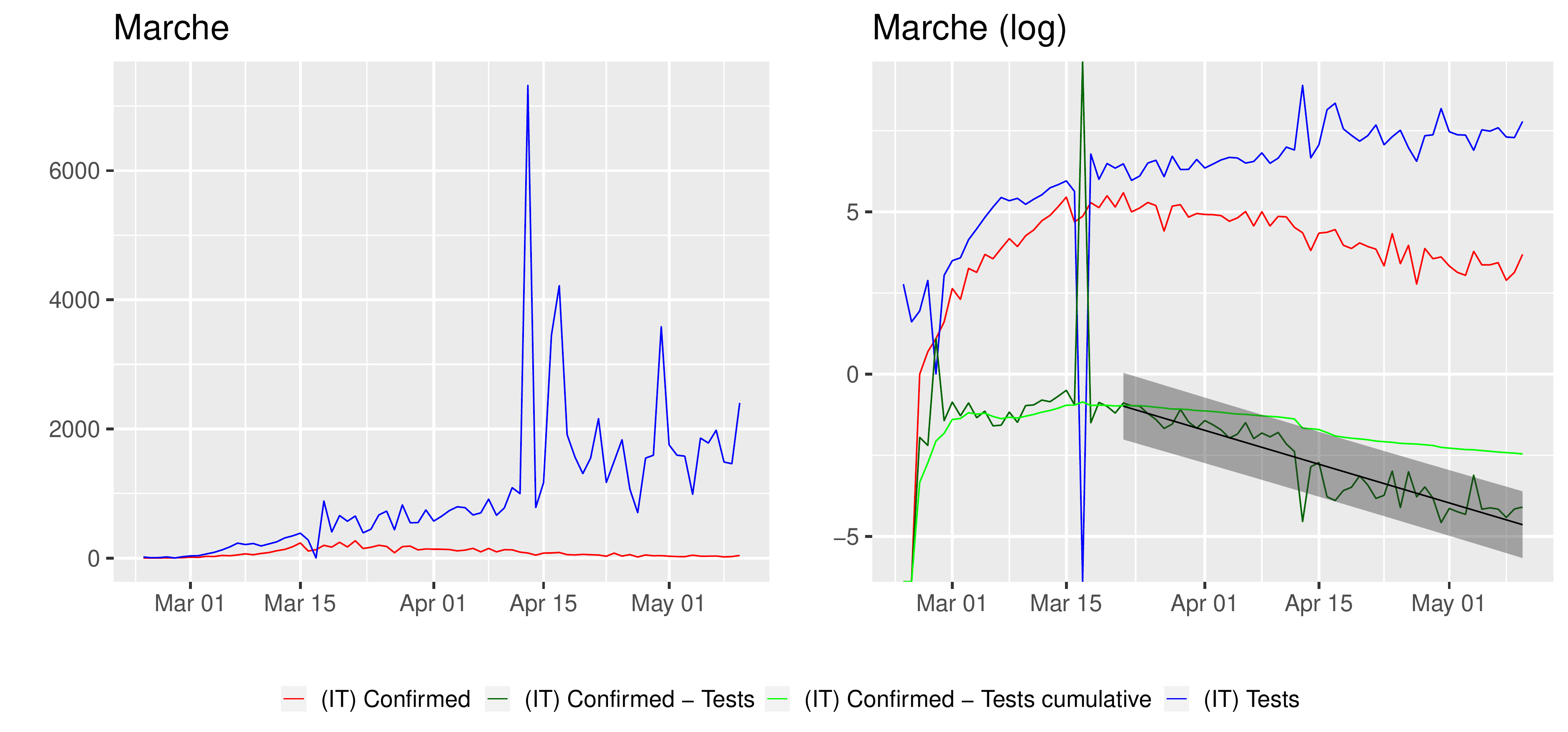}
\end{center}
\caption{Comparison of curves for Marche region. Left: $y$--axis on normal scale, right: on logarithmic
scale. 
Regression line shown for $\log(\mathrm{daily~confirmed})-\log(\mathrm{daily~tested})\sim time$
with $95\%$ prediction band. Slope of regression with $95\%$ confidence interval: 
$a_{\mathrm{f}}=-0.074 (-0.084,-0.065)$, this corresponds to a half--life (in days) of $9.304 (8.256,10.657)$. 
The slope of the regression
$\log(\mathrm{daily~confirmed})\sim time$ is 
$a_{\mathrm{raw}}=-0.046 (-0.052,-0.040)$ corresponding to a half--life (in days) of $14.990 (13.307,17.161)$.
The slope of the regression
$\log(\mathrm{daily~tests})\sim time$ is 
$0.028 (0.019,0.020)$ corresponding to a doubling time (in days) of $24.531 (35.397,35.914)$.
Ratio of slopes for $a_{\mathrm{f}}/a_{\mathrm{raw}}=1.611$, with corresponding half--lives' ratio: $0.621$.
The slope of the regression
$\log(\mathrm{cumulative~confirmed})-\log(\mathrm{cumulative~tested})\sim time$ is 
$-0.036 (-0.037,-0.034)$ corresponding to a half--life (in days) of $19.519 (18.515,20.639)$.
}\label{figMarche}
\end{figure}

\begin{figure}[!ht]
\begin{center}
\includegraphics[width=0.98\textwidth]{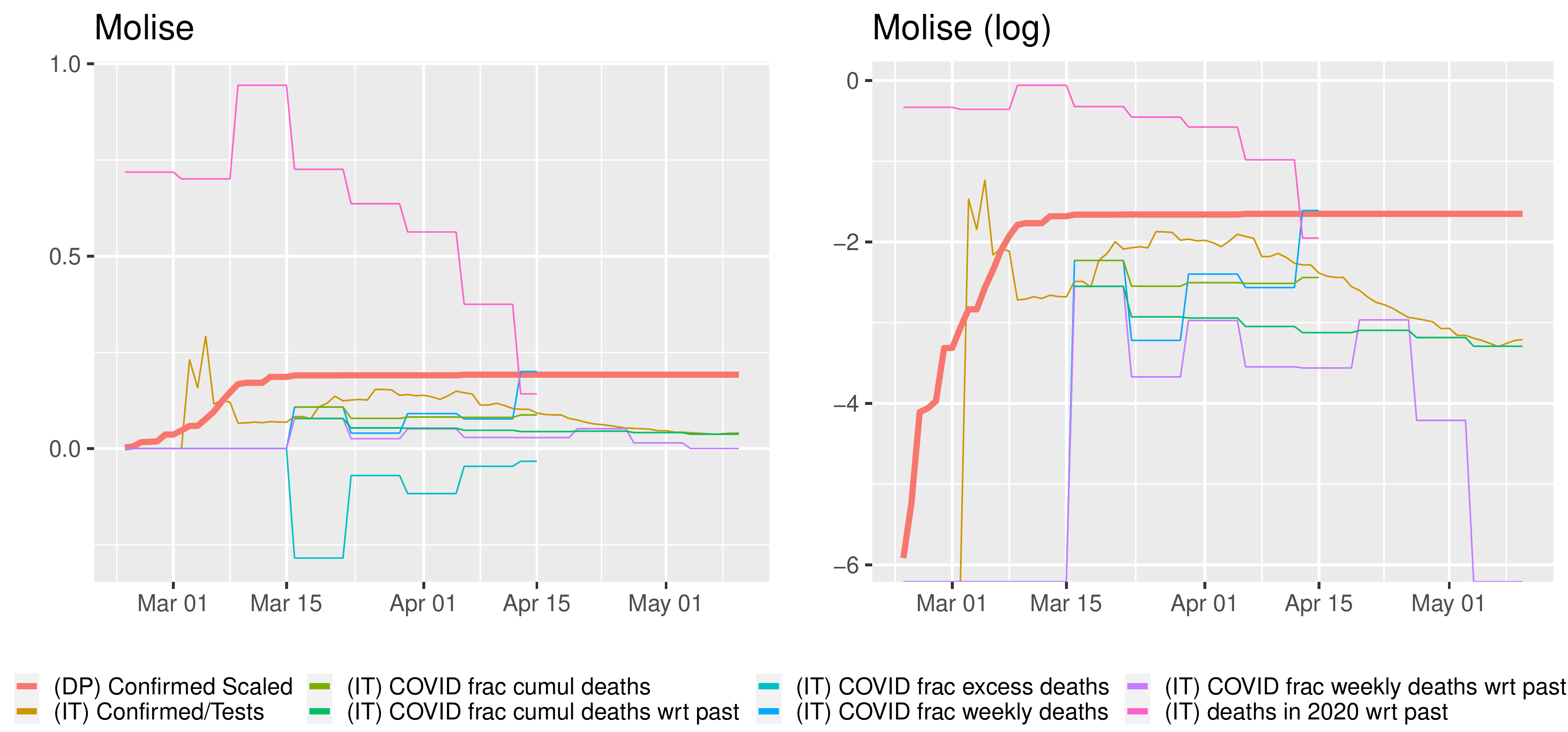}
\\
\includegraphics[width=0.98\textwidth]{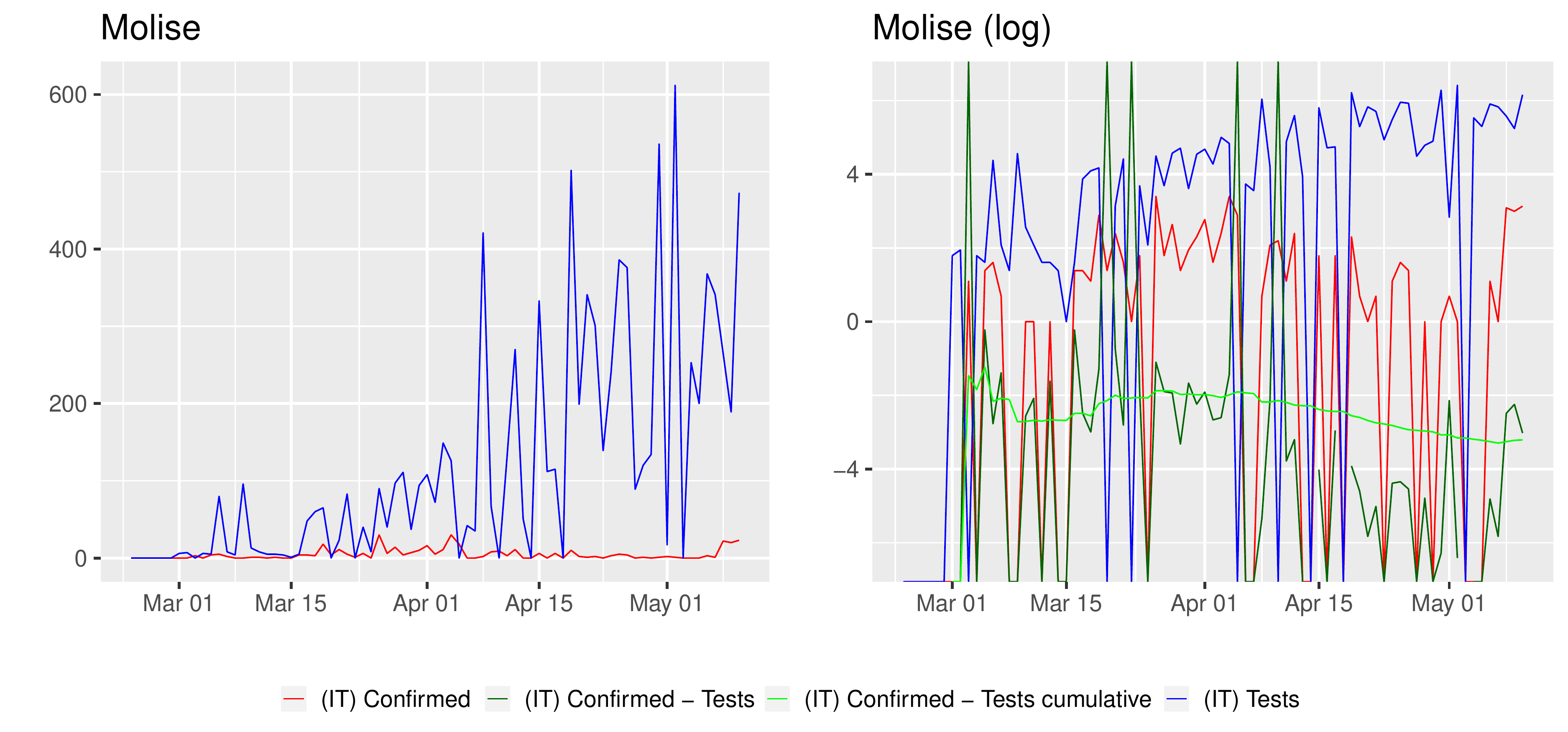}
\end{center}
\caption{Comparison of curves for Molise region. Left: $y$--axis on normal scale, right: on logarithmic
scale. 
}\label{figMolise}
\end{figure}

\begin{figure}[!ht]
\begin{center}
\includegraphics[width=0.98\textwidth]{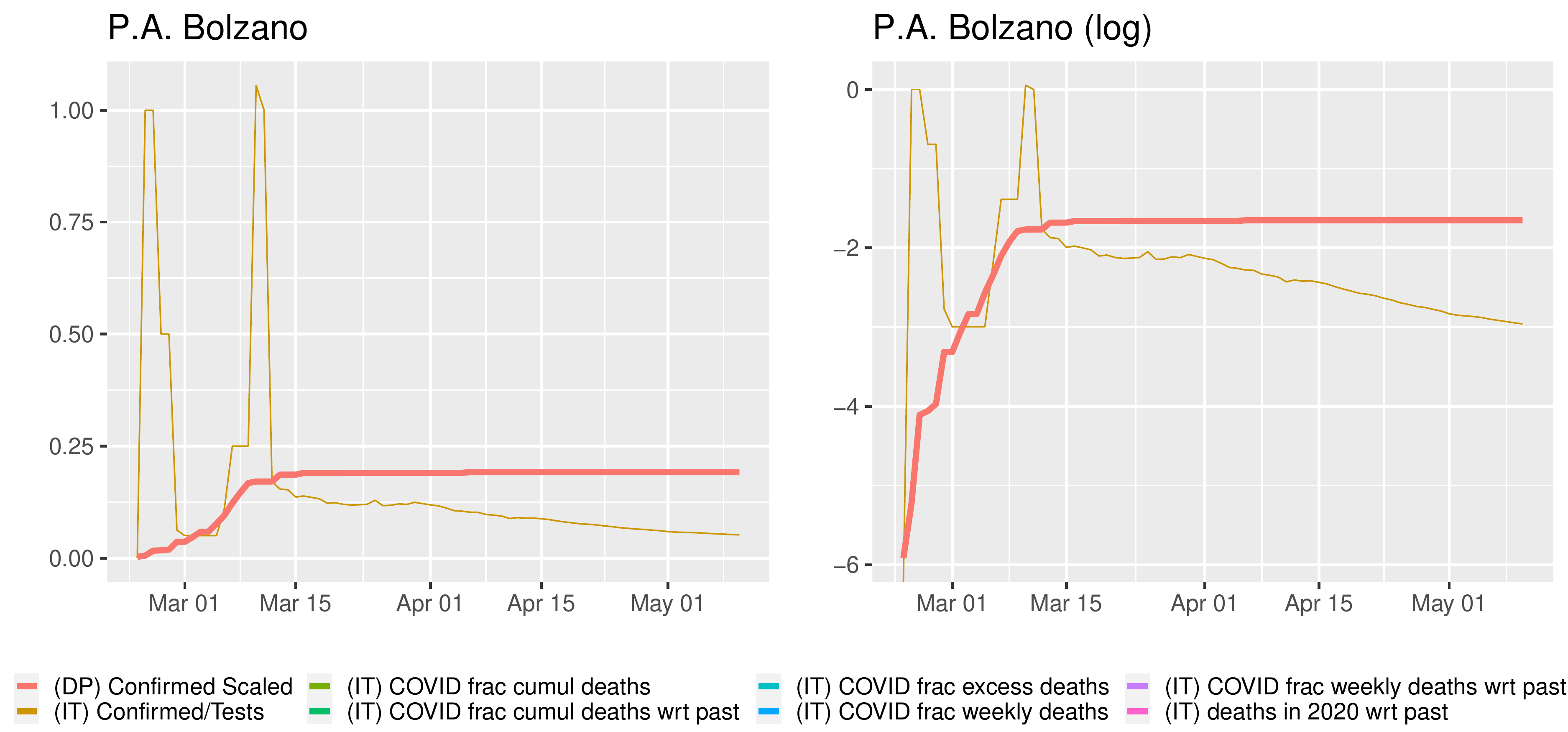}
\\
\includegraphics[width=0.98\textwidth]{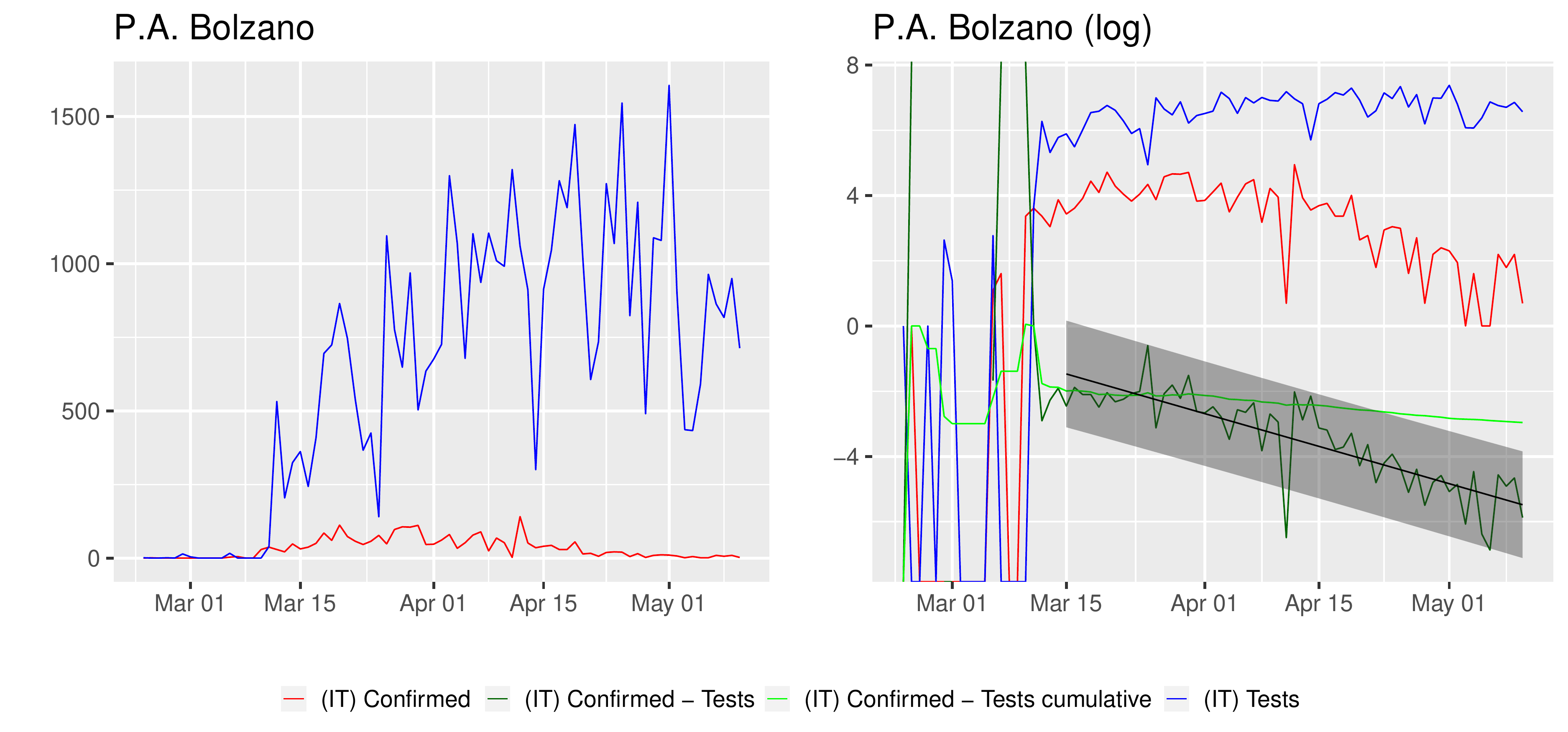}
\end{center}
\caption{Comparison of curves for P.A. Bolzano region. Left: $y$--axis on normal scale, right: on logarithmic
scale. Top row: scaling with respect to population death tolls not presented as
P. A. Bolzano is merged with P. A. Trento in deaths date provided by ISTAT.
Regression line shown for $\log(\mathrm{daily~confirmed})-\log(\mathrm{daily~tested})\sim time$
with $95\%$ prediction band. Slope of regression with $95\%$ confidence interval: 
$a_{\mathrm{f}}=-0.072 (-0.084,-0.059)$, this corresponds to a half--life (in days) of $9.681 (8.246,11.720)$. 
The slope of the regression
$\log(\mathrm{daily~confirmed})\sim time$ is 
$a_{\mathrm{raw}}= -0.061 (-0.074,-0.047)$ corresponding to a half--life (in days) of $11.385 (9.321,14.623)$.
The slope of the regression
$\log(\mathrm{daily~tests})\sim time$ is 
$ 0.011 (0.004,0.040)$ corresponding to a doubling time (in days) of $64.662 (17.302,191.080)$.
Ratio of slopes for $a_{\mathrm{f}}/a_{\mathrm{raw}}=1.176$, with corresponding half--lives' ratio: $0.850$.
The slope of the regression
$\log(\mathrm{cumulative~confirmed})-\log(\mathrm{cumulative~tested})\sim time$ is 
$ -0.018 (-0.019,-0.018)$ corresponding to a half--life (in days) of $37.589 (35.994,39.332)$.
}\label{figP.A.Bolzano}
\end{figure}

\begin{figure}[!ht]
\begin{center}
\includegraphics[width=0.98\textwidth]{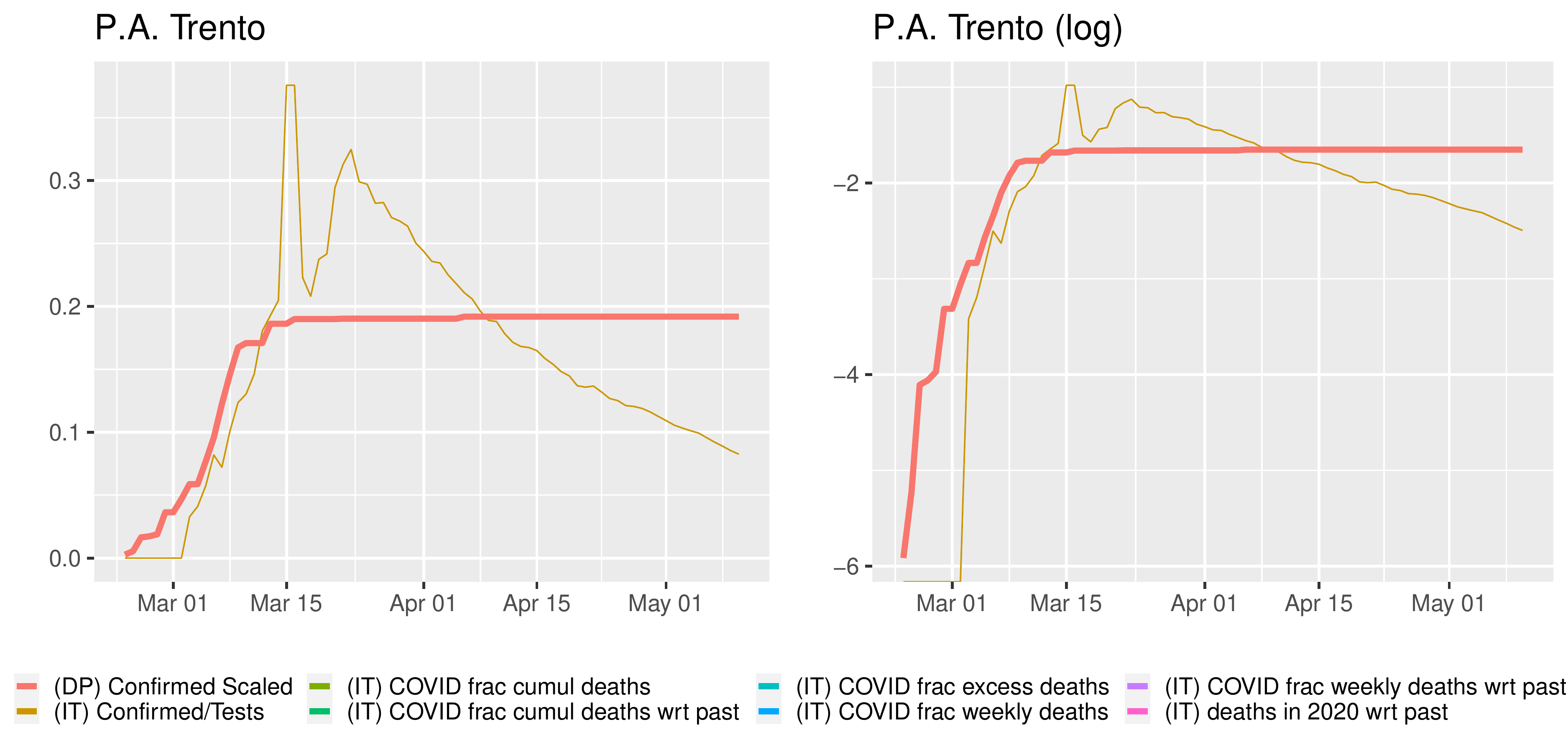}
\\
\includegraphics[width=0.98\textwidth]{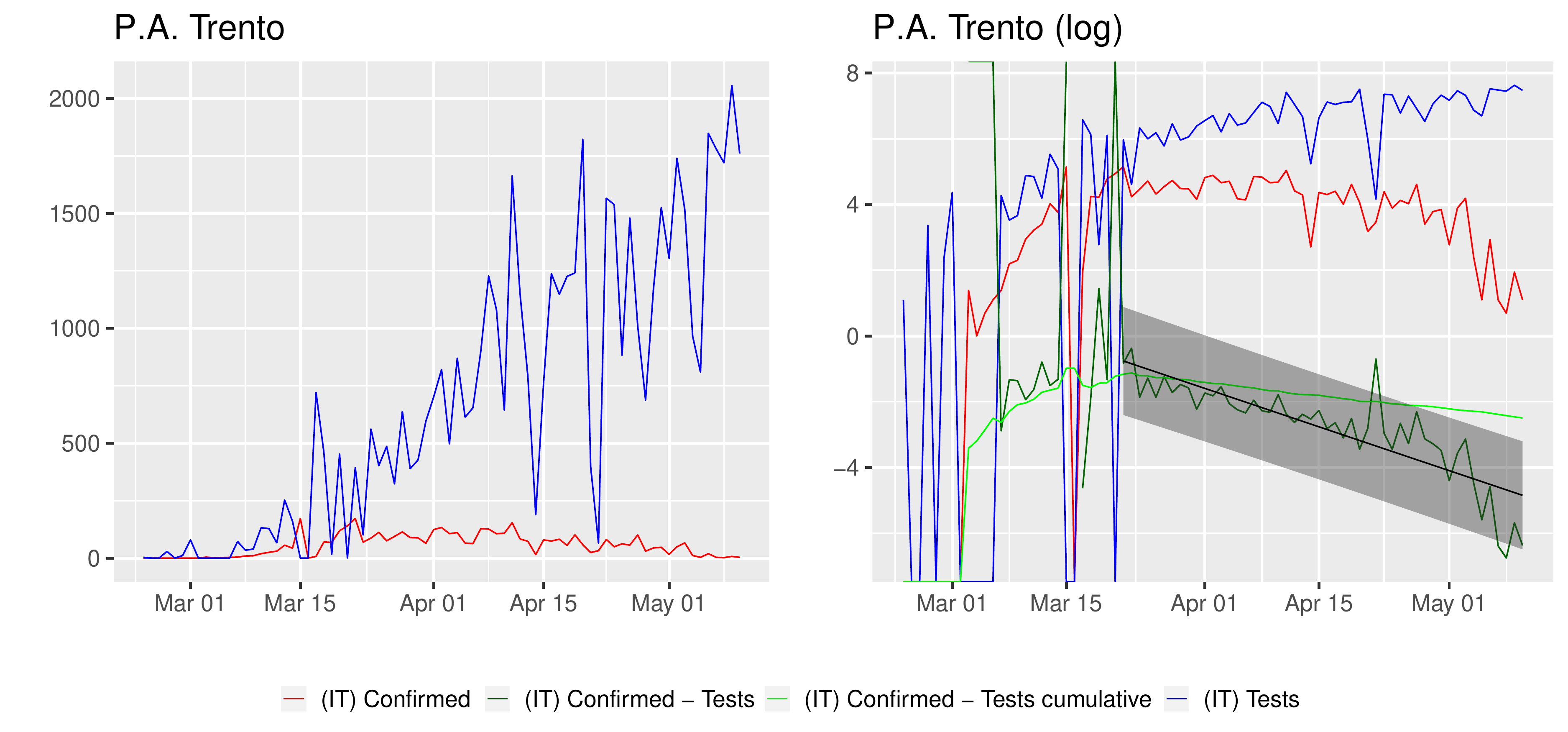}
\end{center}
\caption{Comparison of curves for P.A. Trento region. Left: $y$--axis on normal scale, right: on logarithmic
scale. Top row: scaling with respect to population death tolls not presented as
P. A. Bolzano is merged with P. A. Trento in deaths date provided by ISTAT.
Regression line shown for $\log(\mathrm{daily~confirmed})-\log(\mathrm{daily~tested})\sim time$
with $95\%$ prediction band. Slope of regression with $95\%$ confidence interval:
$a_{\mathrm{f}}= -0.083 (-0.099,-0.068)$, this corresponds to a doubling time (in days) of $8.308 (7.033,10.148)$. 
The slope of the regression
$\log(\mathrm{daily~confirmed})\sim time$ is 
$a_{\mathrm{raw}}=-0.055 (-0.069,-0.040)$ corresponding to a half--life (in days) of $12.714 (10.010,17.420)$.
The slope of the regression
$\log(\mathrm{daily~tests})\sim time$ is 
$0.029 (0.017,0.031)$ corresponding to a doubling time (in days) of $23.975 (22.069,39.784)$.
Ratio of slopes for $a_{\mathrm{f}}/a_{\mathrm{raw}}=1.530$, with corresponding half--lives' ratio: $0.653$.
The slope of the regression
$\log(\mathrm{cumulative~confirmed})-\log(\mathrm{cumulative~tested})\sim time$ is 
$-0.027 (-0.028,-0.027)$ corresponding to a half--life (in days) of $25.537 (25.139,25.948)$.
}\label{figP.A.Trento}
\end{figure}

\begin{figure}[!ht]
\begin{center}
\includegraphics[width=0.98\textwidth]{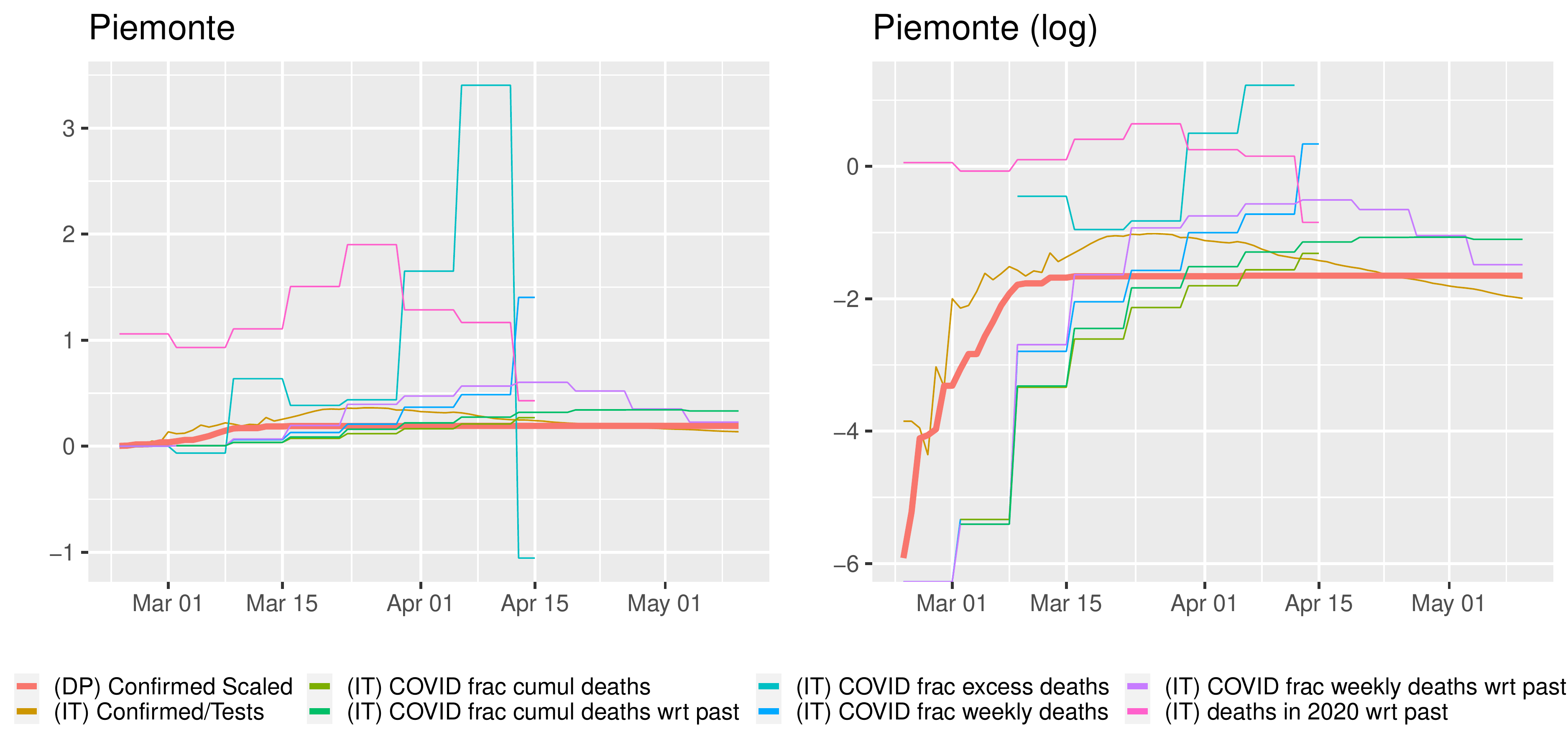}
\\
\includegraphics[width=0.98\textwidth]{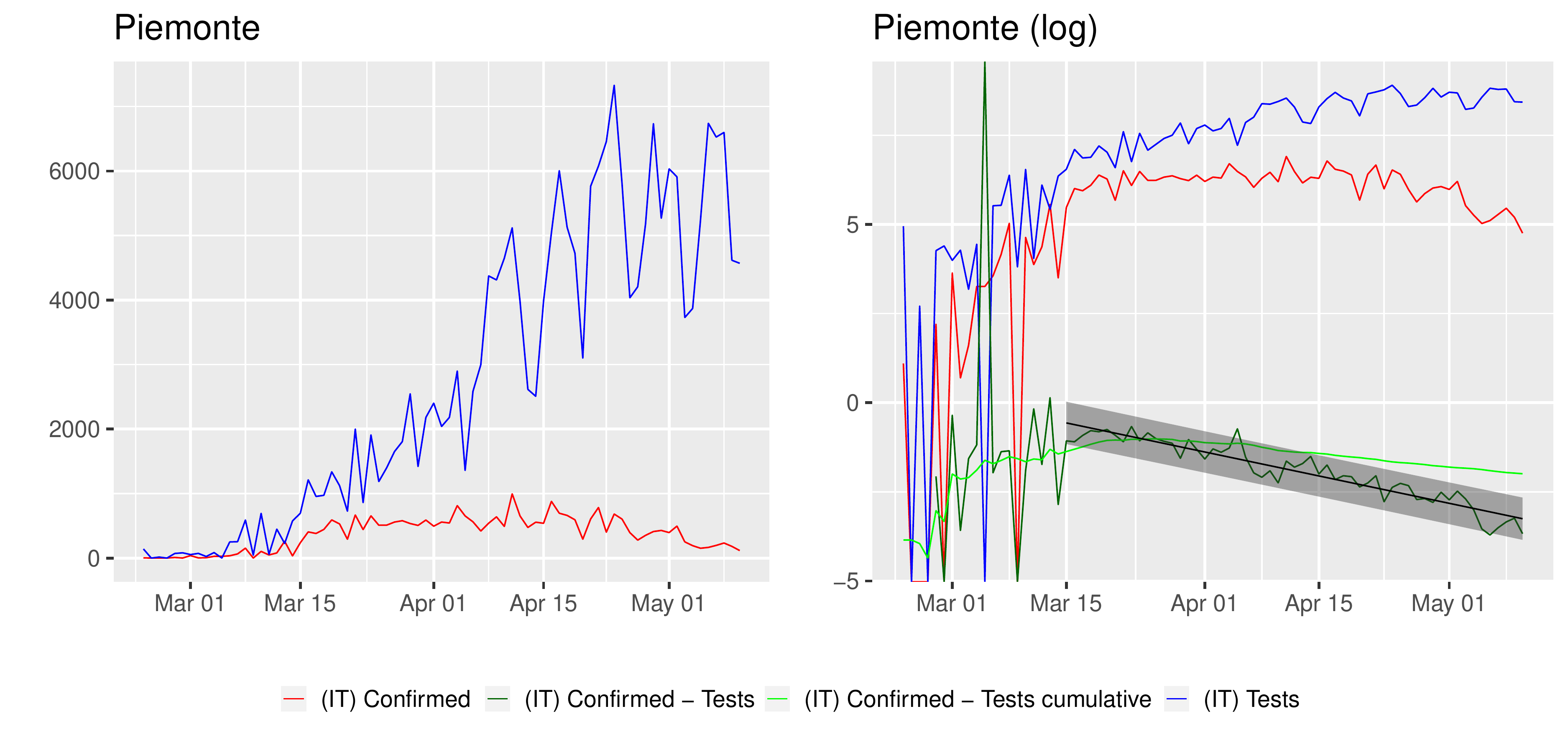}
\end{center}
\caption{Comparison of curves for Piemonte region. Left: $y$--axis on normal scale, right: on logarithmic scale. 
Regression line shown for $\log(\mathrm{daily~confirmed})-\log(\mathrm{daily~tested})\sim time$
with $95\%$ prediction band. Slope of regression with $95\%$ confidence interval: 
$a_{\mathrm{f}}=-0.048 (-0.052,-0.043)$, this corresponds to a half--life (in days) of $14.451 (13.207,15.954)$.
The slope of the regression
$\log(\mathrm{daily~confirmed})\sim time$ is 
$a_{\mathrm{raw}}=-0.013 (-0.019,-0.006)$ corresponding to a half--life (in days) of $53.716 (35.637,109.020)$.
The slope of the regression
$\log(\mathrm{daily~tests})\sim time$ is 
$0.035 (0.030,0.052)$ corresponding to a doubling time (in days) of $19.769 (13.415,23.179)$.
Ratio of slopes for $a_{\mathrm{f}}/a_{\mathrm{raw}}=3.717$, with corresponding half--lives' ratio: $0.269$.
The slope of the regression
$\log(\mathrm{cumulative~confirmed})-\log(\mathrm{cumulative~tested})\sim time$ is 
$-0.018 (-0.019,-0.016)$ corresponding to a half--life (in days) of $39.243 (35.548,43.796)$.
}\label{figPiemonte}
\end{figure}

\begin{figure}[!ht]
\begin{center}
\includegraphics[width=0.98\textwidth]{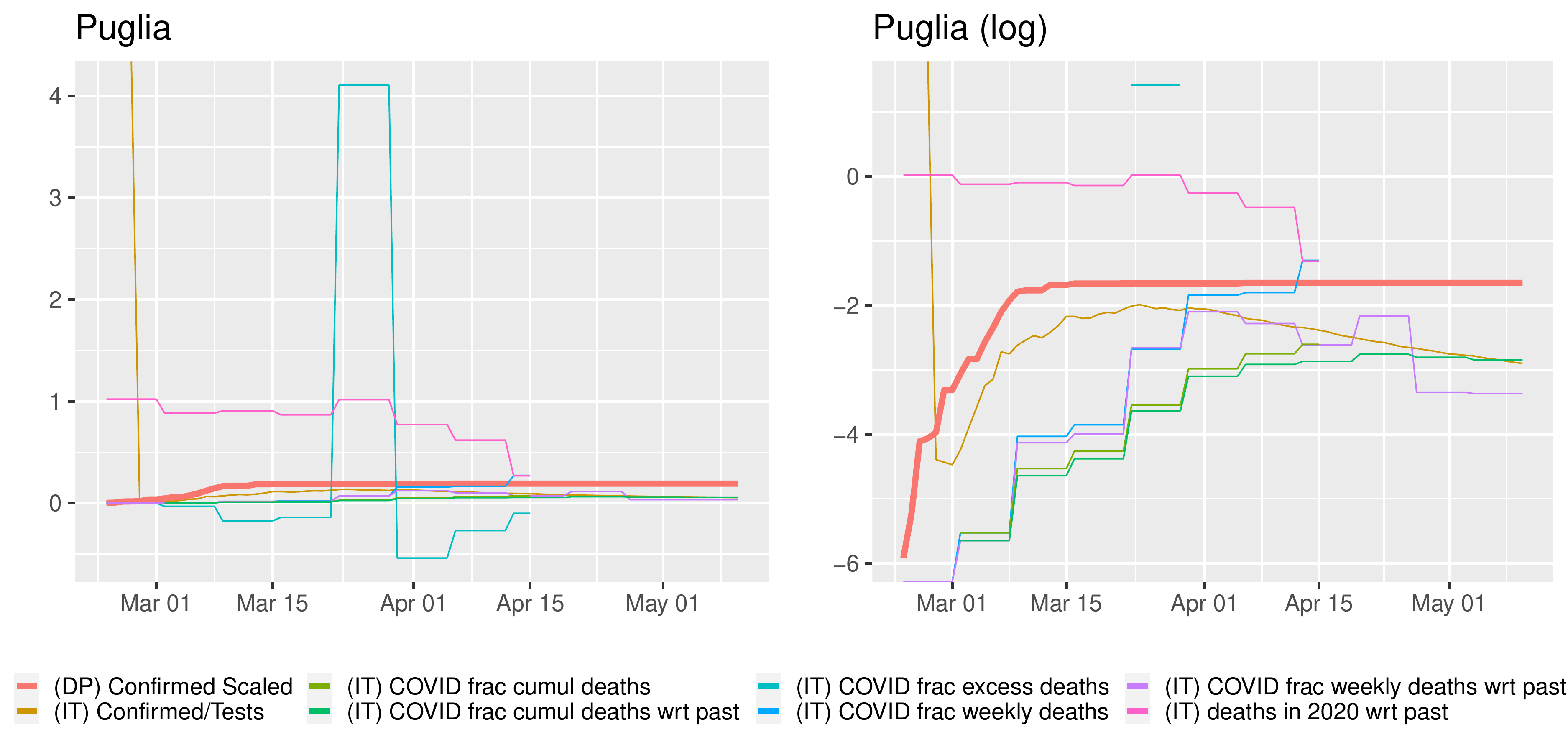}
\\
\includegraphics[width=0.98\textwidth]{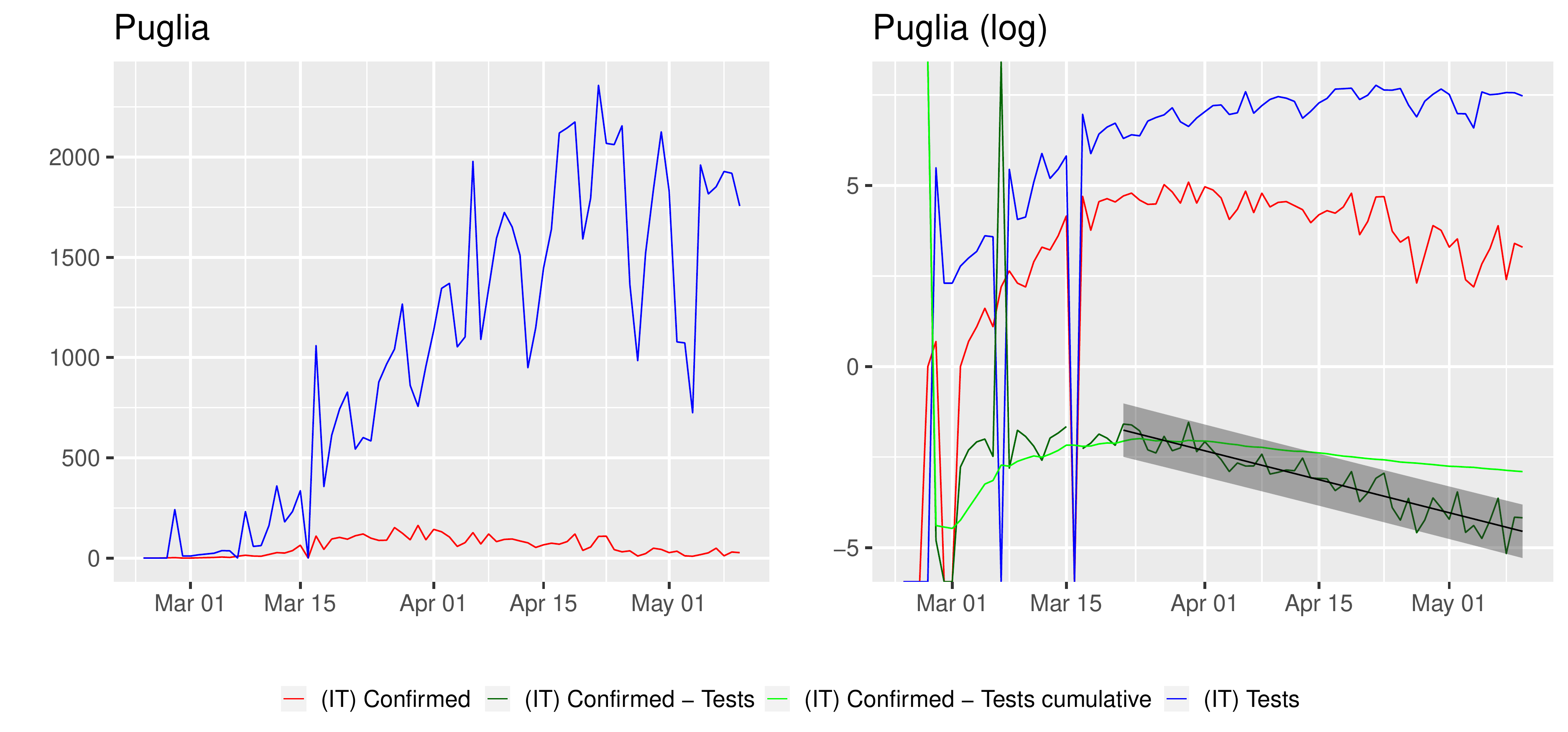}
\end{center}
\caption{Comparison of curves for Puglia region. Left: $y$--axis on normal scale, right: on logarithmic scale. 
Regression line shown for $\log(\mathrm{daily~confirmed})-\log(\mathrm{daily~tested})\sim time$
with $95\%$ prediction band. Slope of regression with $95\%$ confidence interval: 
$a_{\mathrm{f}}=-0.057 (-0.064,-0.050)$, this corresponds to a half--life (in days) of $12.157 (10.864,13.800)$. 
The slope of the regression
$\log(\mathrm{daily~confirmed})\sim time$ is 
$a_{\mathrm{raw}}=-0.041 (-0.050,-0.032)$ corresponding to a half--life (in days) of $16.877 (13.840,21.622)$.
The slope of the regression
$\log(\mathrm{daily~tests})\sim time$ is 
$0.016 (0.010,0.033)$ corresponding to a doubling time (in days) of $43.470 (20.965,68.915)$.
Ratio of slopes for $a_{\mathrm{f}}/a_{\mathrm{raw}}=1.388$, with corresponding half--lives' ratio: $0.720$.
The slope of the regression
$\log(\mathrm{cumulative~confirmed})-\log(\mathrm{cumulative~tested})\sim time$ is 
$-0.020 (-0.021,-0.019)$ corresponding to a half--life (in days) of $34.395 (33.273,35.594)$.
}\label{figPuglia}
\end{figure}

\begin{figure}[!ht]
\begin{center}
\includegraphics[width=0.98\textwidth]{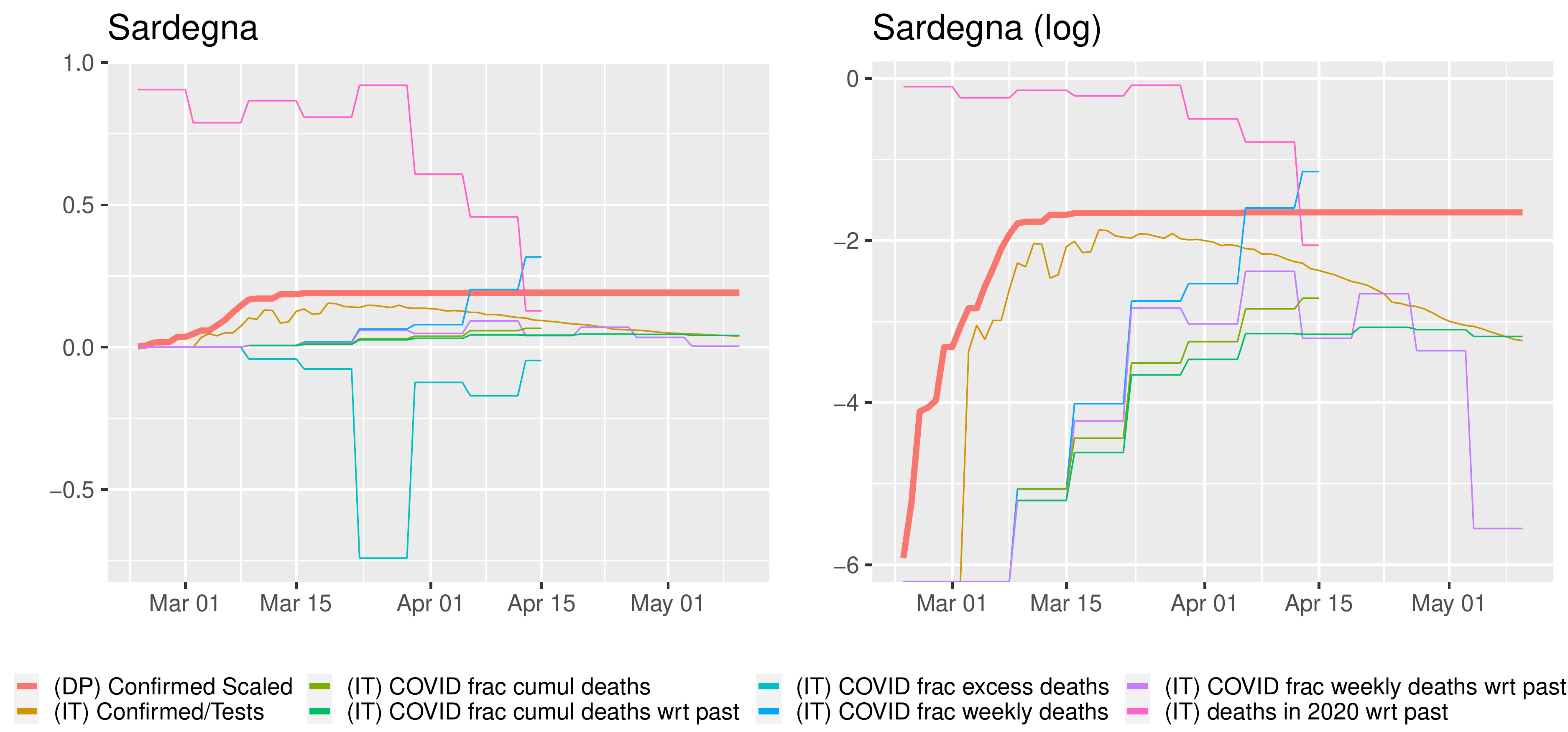}
\\
\includegraphics[width=0.98\textwidth]{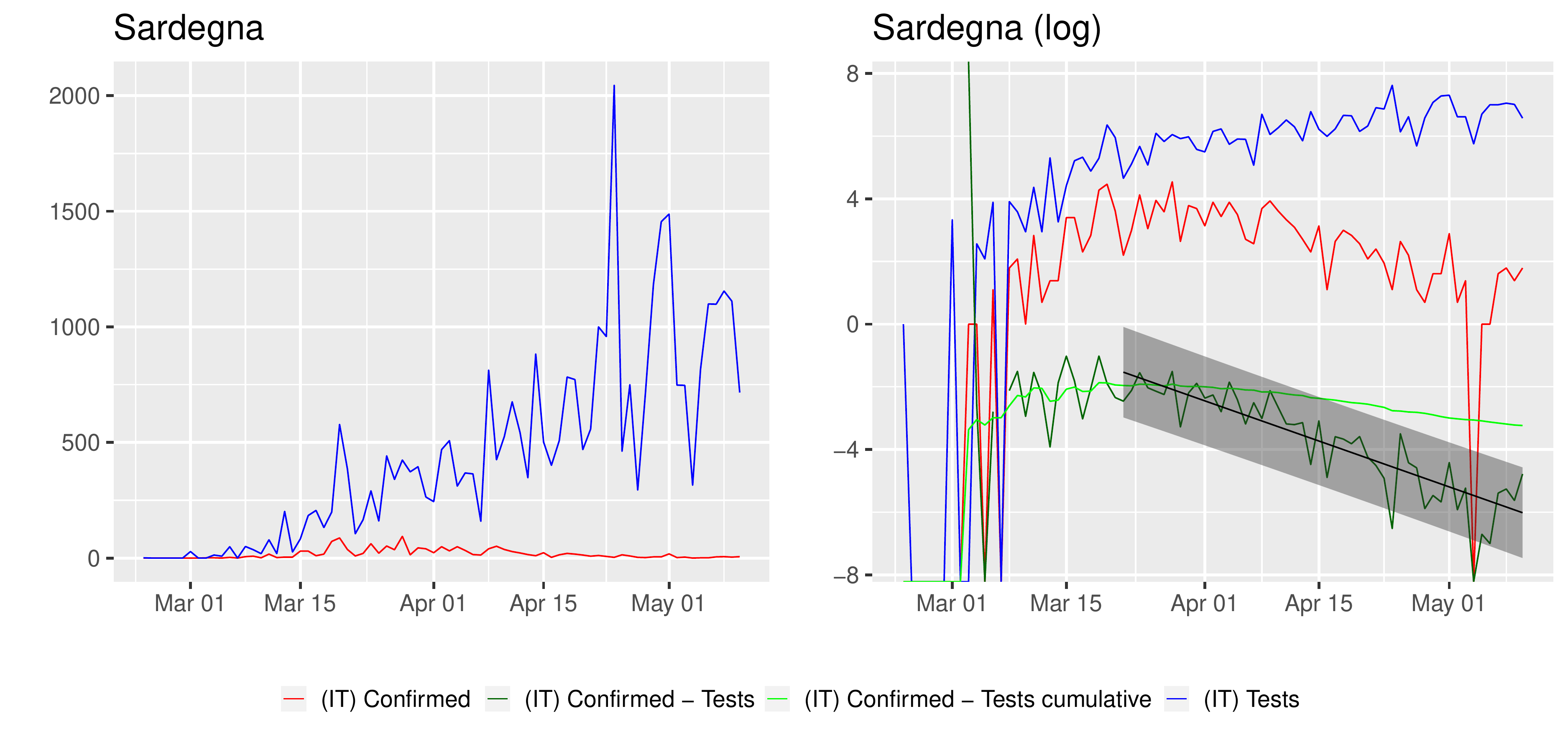}
\end{center}
\caption{Comparison of curves for Sardegna region. Left: $y$--axis on normal scale, right: on logarithmic
scale. 
Regression line shown for $\log(\mathrm{daily~confirmed})-\log(\mathrm{daily~tested})\sim time$
with $95\%$ prediction band. Slope of regression with $95\%$ confidence interval: 
$a_{\mathrm{f}}=-0.092 (-0.105,-0.078)$, this corresponds to a half--life (in days) of $7.573 (6.599,8.884)$. 
The slope of the regression
$\log(\mathrm{daily~confirmed})\sim time$ is 
$a_{\mathrm{raw}}=-0.058 (-0.072,-0.044)$ corresponding to a half--life (in days) of $11.902 (9.603,15.649)$.
The slope of the regression
$\log(\mathrm{daily~tests})\sim time$ is 
$0.031 (0.023,0.026)$ corresponding to a doubling time (in days) of $22.193 (26.372,30.382)$.
Ratio of slopes for $a_{\mathrm{f}}/a_{\mathrm{raw}}=1.572$, with corresponding half--lives' ratio: $0.636$.
The slope of the regression
$\log(\mathrm{cumulative~confirmed})-\log(\mathrm{cumulative~tested})\sim time$ is 
$-0.030 (-0.031,-0.028)$ corresponding to a half--life (in days) of $23.310 (22.147,24.602)$.
}\label{figSardegna}
\end{figure}

\begin{figure}[!ht]
\begin{center}
\includegraphics[width=0.98\textwidth]{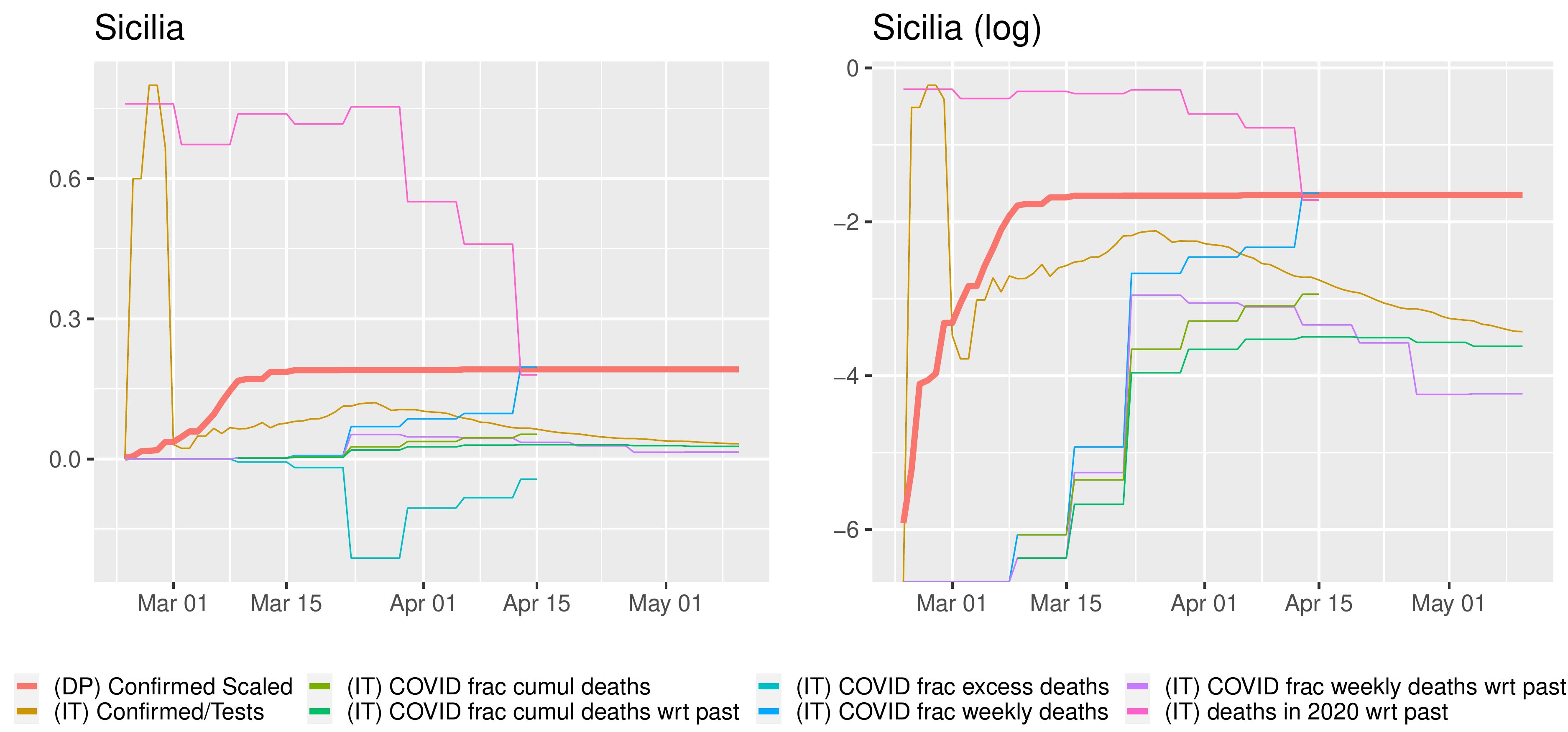}
\\
\includegraphics[width=0.98\textwidth]{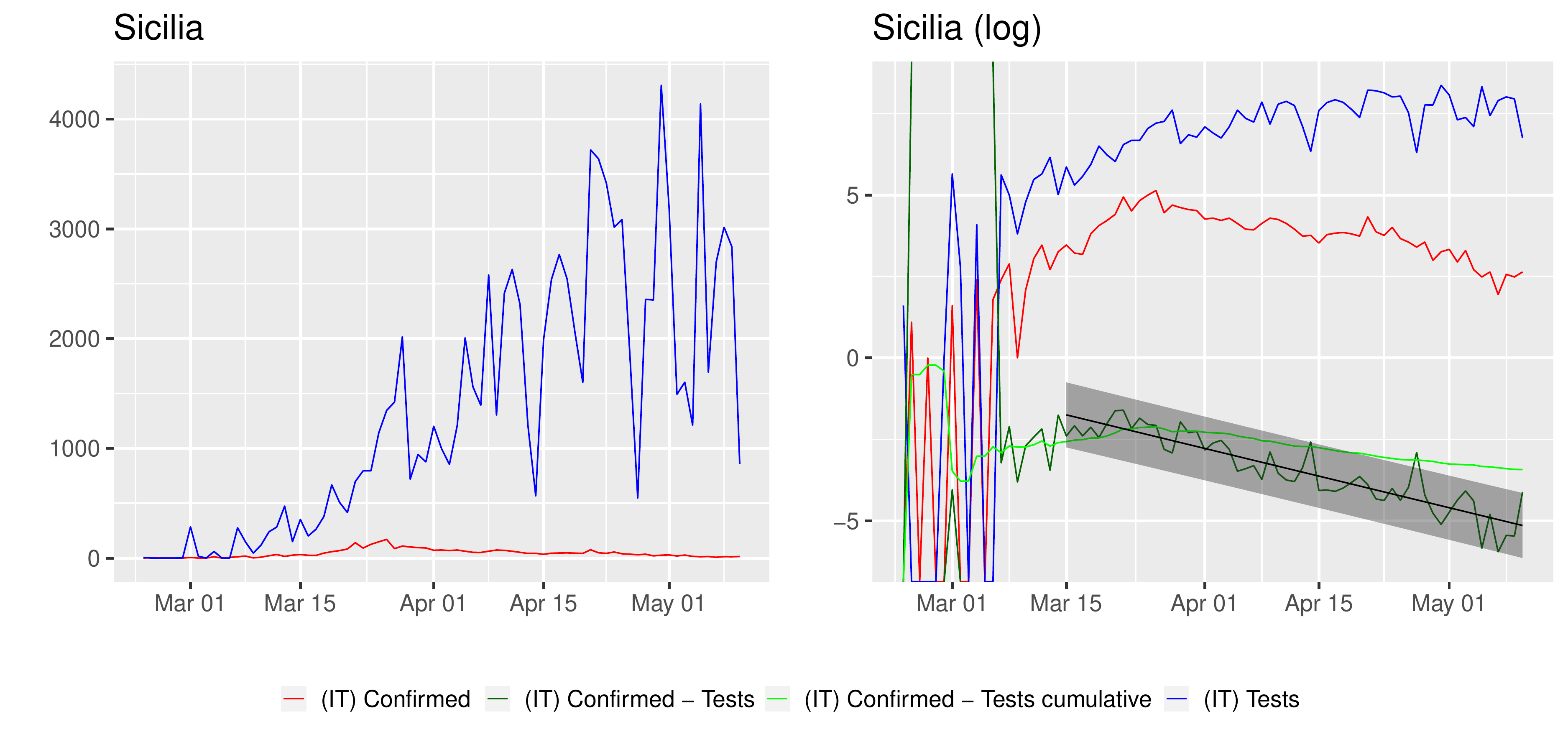}
\end{center}
\caption{Comparison of curves for Sicilia region. Left: $y$--axis on normal scale, right: on logarithmic
scale. 
Regression line shown for $\log(\mathrm{daily~confirmed})-\log(\mathrm{daily~tested})\sim time$
with $95\%$ prediction band. Slope of regression with $95\%$ confidence interval: 
$a_{\mathrm{f}}=-0.061 (-0.068,-0.053)$, this corresponds to a half--life (in days) of $11.430 (10.156,13.069)$. 
The slope of the regression
$\log(\mathrm{daily~confirmed})\sim time$ is 
$a_{\mathrm{raw}}=-0.030 (-0.038,-0.022)$ corresponding to a half--life (in days) of $22.916 (18.243,30.808)$.
The slope of the regression
$\log(\mathrm{daily~tests})\sim time$ is 
$0.030 (0.022,0.047)$ corresponding to a doubling time (in days) of $22.803 (14.642,31.528)$.
Ratio of slopes for $a_{\mathrm{f}}/a_{\mathrm{raw}}=2.005$, with corresponding half--lives' ratio: $0.499$.
The slope of the regression
$\log(\mathrm{cumulative~confirmed})-\log(\mathrm{cumulative~tested})\sim time$ is 
$-0.023 (-0.026,-0.021)$ corresponding to a half--life (in days) of $29.589 (26.719,33.149)$.
}\label{figSicilia}
\end{figure}

\begin{figure}[!ht]
\begin{center}
\includegraphics[width=0.98\textwidth]{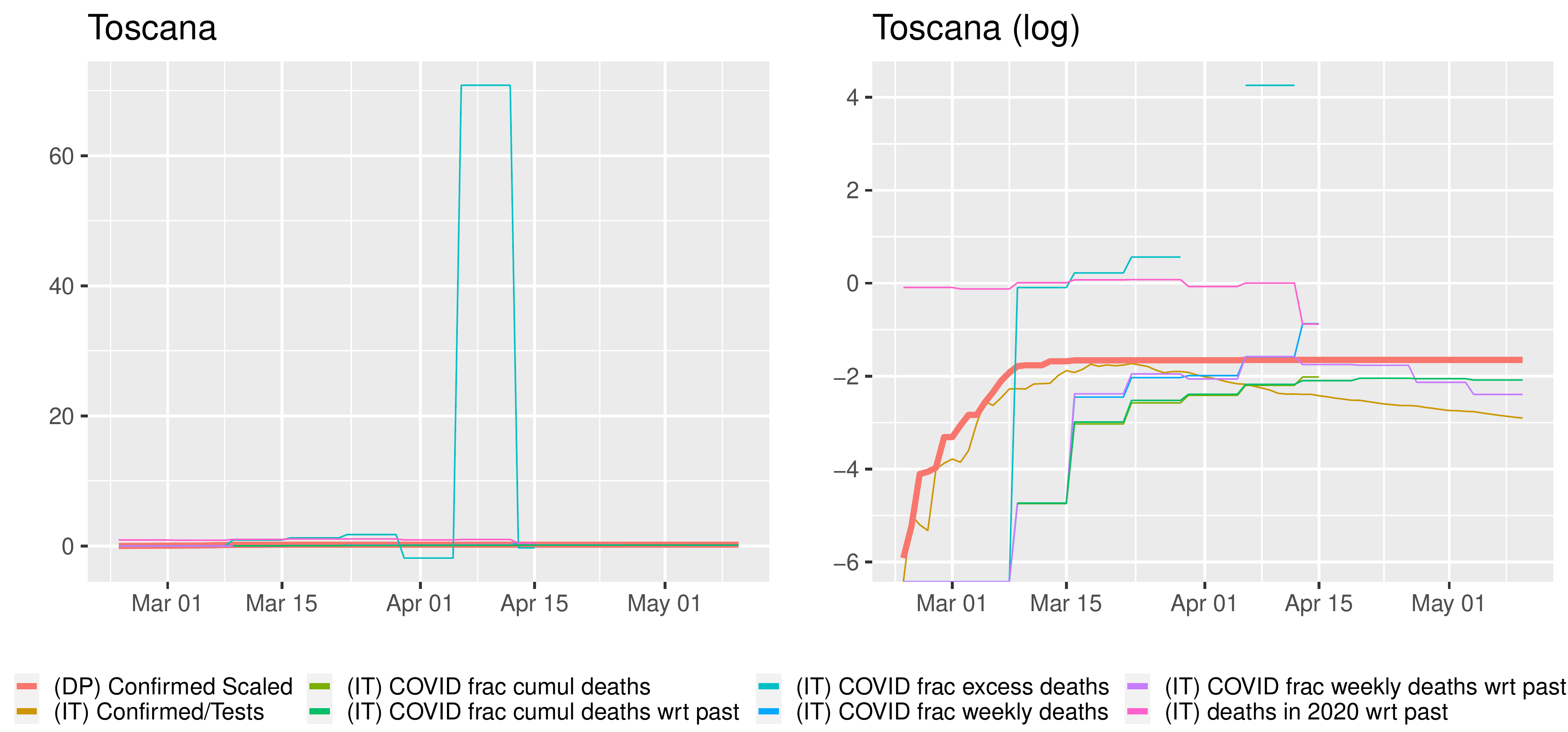}
\\
\includegraphics[width=0.98\textwidth]{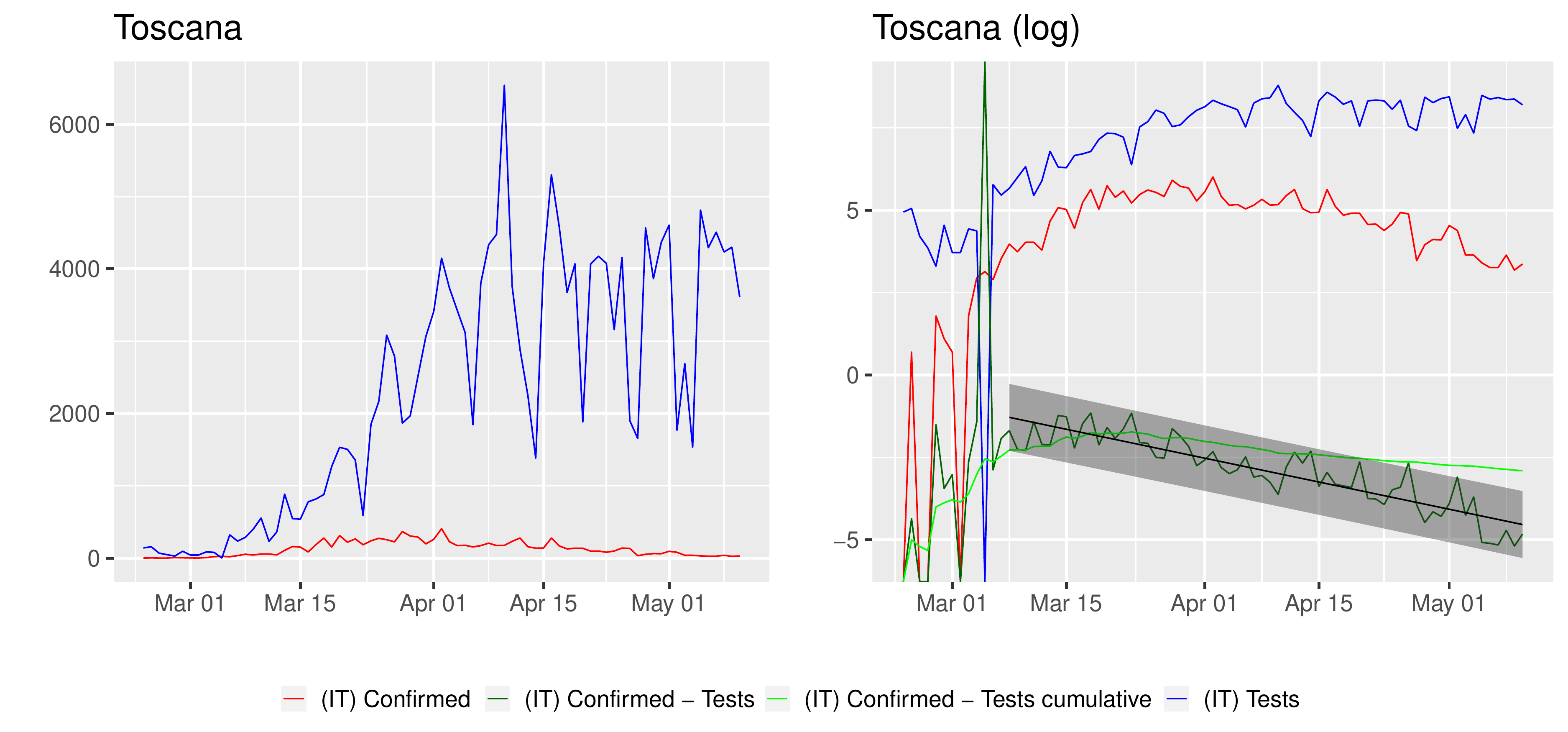}
\end{center}
\caption{Comparison of curves for Toscana region. Left: $y$--axis on normal scale, right: on logarithmic
scale. 
Regression line shown for $\log(\mathrm{daily~confirmed})-\log(\mathrm{daily~tested})\sim time$
with $95\%$ prediction band. Slope of regression with $95\%$ confidence interval: 
$a_{\mathrm{f}}=-0.052 (-0.058,-0.045)$, this corresponds to a half--life (in days) of $13.430 (11.920,15.379)$. 
The slope of the regression
$\log(\mathrm{daily~confirmed})\sim time$ is 
$a_{\mathrm{raw}}=-0.020 (-0.029,-0.011)$ corresponding to a half--life (in days) of $34.753 (23.956,63.267)$.
The slope of the regression
$\log(\mathrm{daily~tests})\sim time$ is 
$0.032 (0.024,0.059)$ corresponding to a doubling time (in days) of $21.890 (11.681,28.539)$.
Ratio of slopes for $a_{\mathrm{f}}/a_{\mathrm{raw}}=2.588$, with corresponding half--lives' ratio: $0.386$.
The slope of the regression
$\log(\mathrm{cumulative~confirmed})-\log(\mathrm{cumulative~tested})\sim time$ is 
$-0.017 (-0.019,-0.015)$ corresponding to a half--life (in days) of $40.971 (36.036,47.472)$.
}\label{figToscana}
\end{figure}

\begin{figure}[!ht]
\begin{center}
\includegraphics[width=0.98\textwidth]{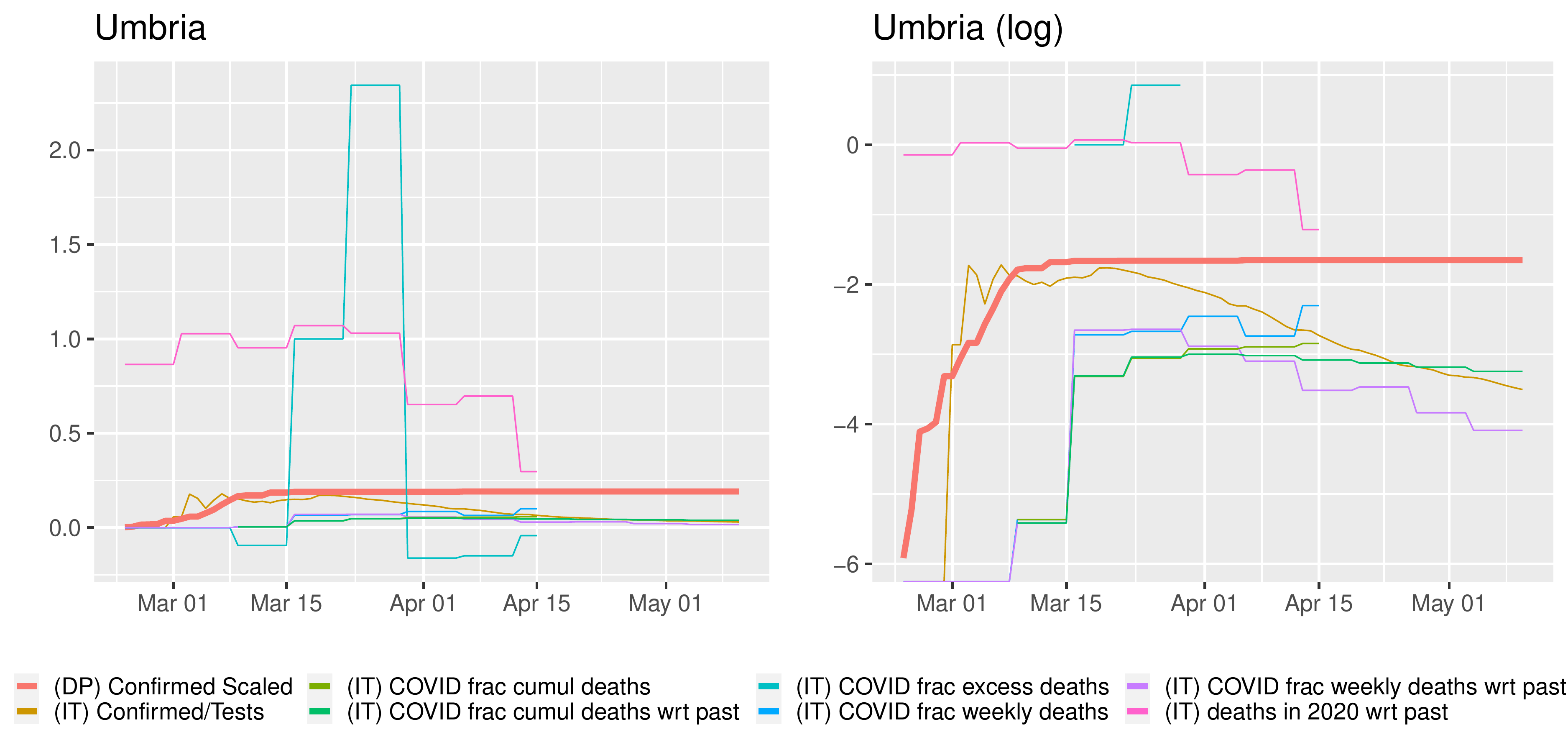}
\\
\includegraphics[width=0.98\textwidth]{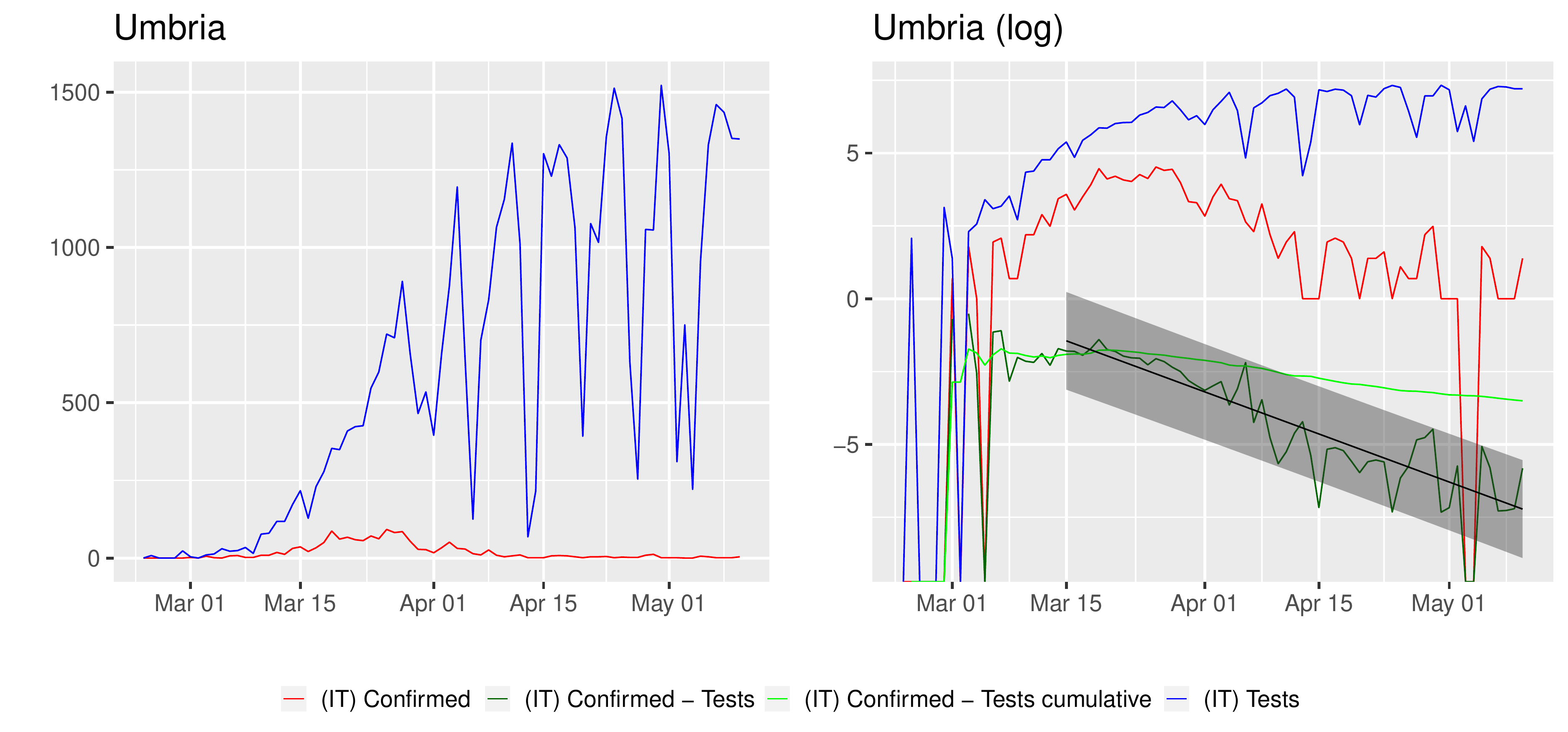}
\end{center}
\caption{Comparison of curves for Umbria region. Left: $y$--axis on normal scale, right: on logarithmic
scale. 
Regression line shown for $\log(\mathrm{daily~confirmed})-\log(\mathrm{daily~tested})\sim time$
with $95\%$ prediction band. Slope of regression with $95\%$ confidence interval: 
$a_{\mathrm{f}}=-0.103 (-0.116,-0.090)$, this corresponds to a half--life (in days) of $ 6.720 (5.959,7.704)$. 
The slope of the regression
$\log(\mathrm{daily~confirmed})\sim time$ is 
$-0.079 (-0.093,-0.065)$ corresponding to a half--life (in days) of $8.810 (7.478,10.720)$.
The slope of the regression
$\log(\mathrm{daily~tests})\sim time$ is 
$0.022 (0.011,0.033)$ corresponding to a doubling time (in days) of $32.197 (21.311,61.463)$.
Ratio of slopes for $a_{\mathrm{f}}/a_{\mathrm{raw}}=1.311$, with corresponding half--lives' ratio: $0.763$.
The slope of the regression
$\log(\mathrm{cumulative~confirmed})-\log(\mathrm{cumulative~tested})\sim time$ is 
$ -0.035 (-0.036,-0.033)$ corresponding to a half--life (in days) of $19.892 (19.118,20.731)$.
}\label{figUmbria}
\end{figure}

\begin{figure}[!ht]
\begin{center}
\includegraphics[width=0.98\textwidth]{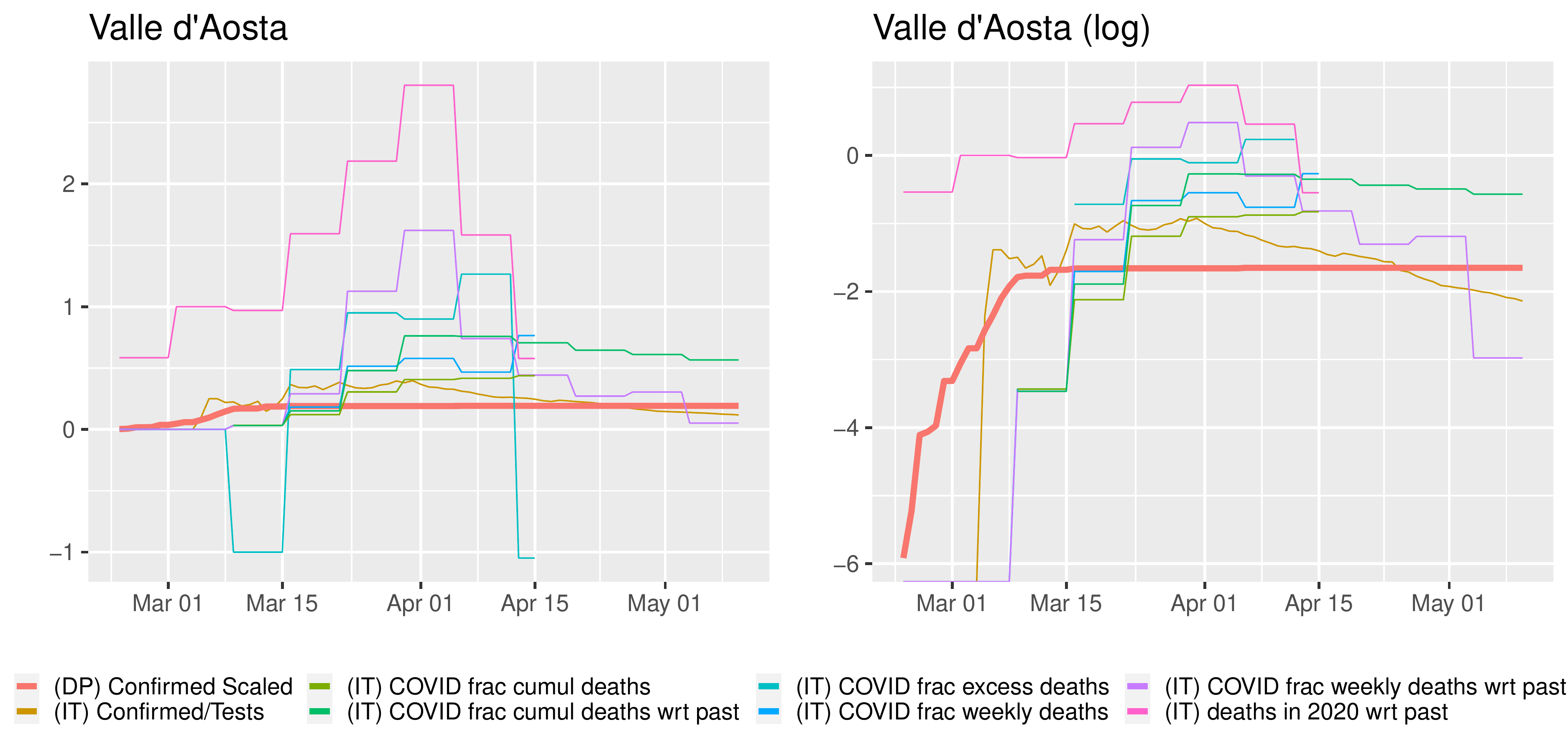}
\\
\includegraphics[width=0.98\textwidth]{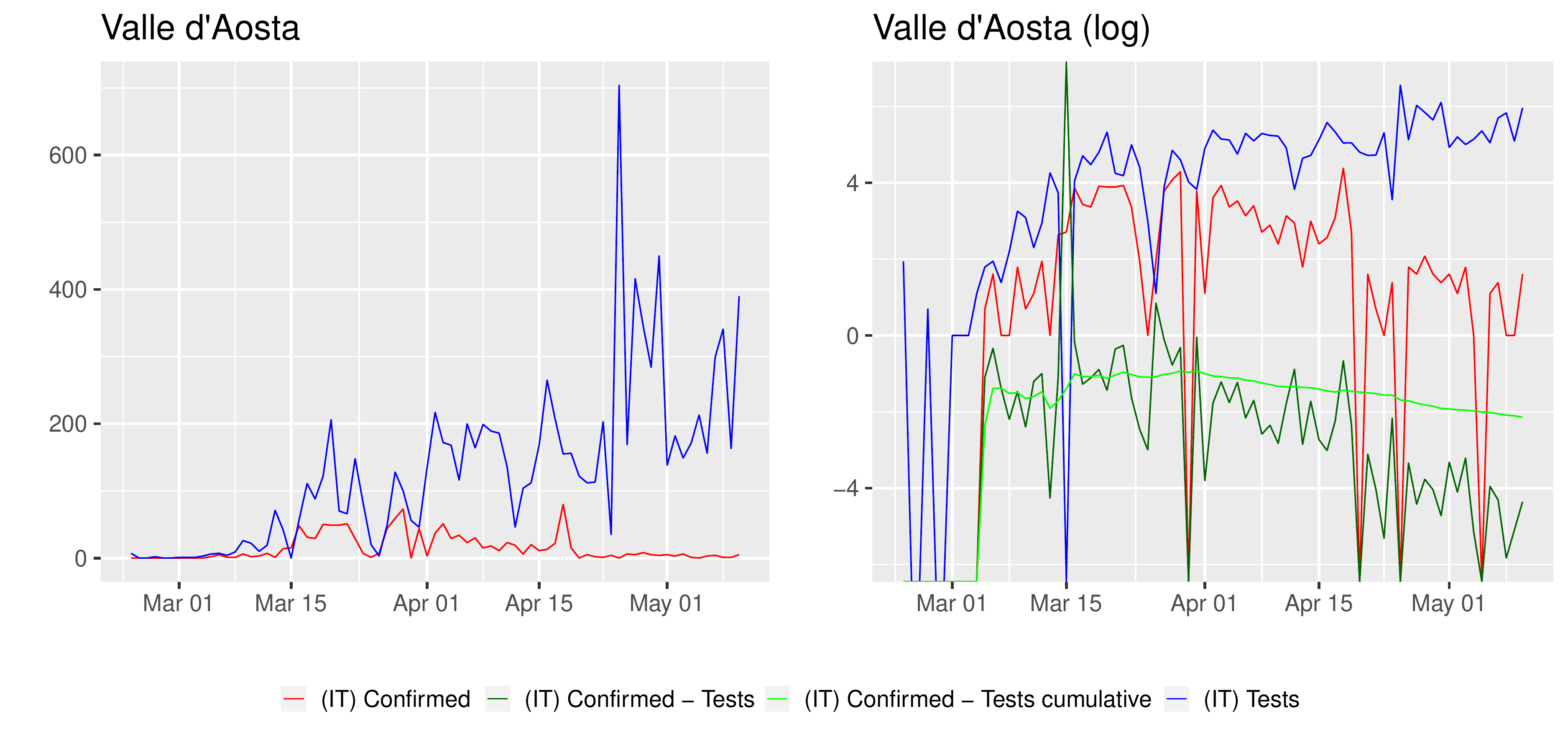}
\end{center}
\caption{Comparison of curves for Valle d'Aosta region. Left: $y$--axis on normal scale, right: on logarithmic scale. 
}\label{figValledAosta}
\end{figure}

\begin{figure}[!ht]
\begin{center}
\includegraphics[width=0.98\textwidth]{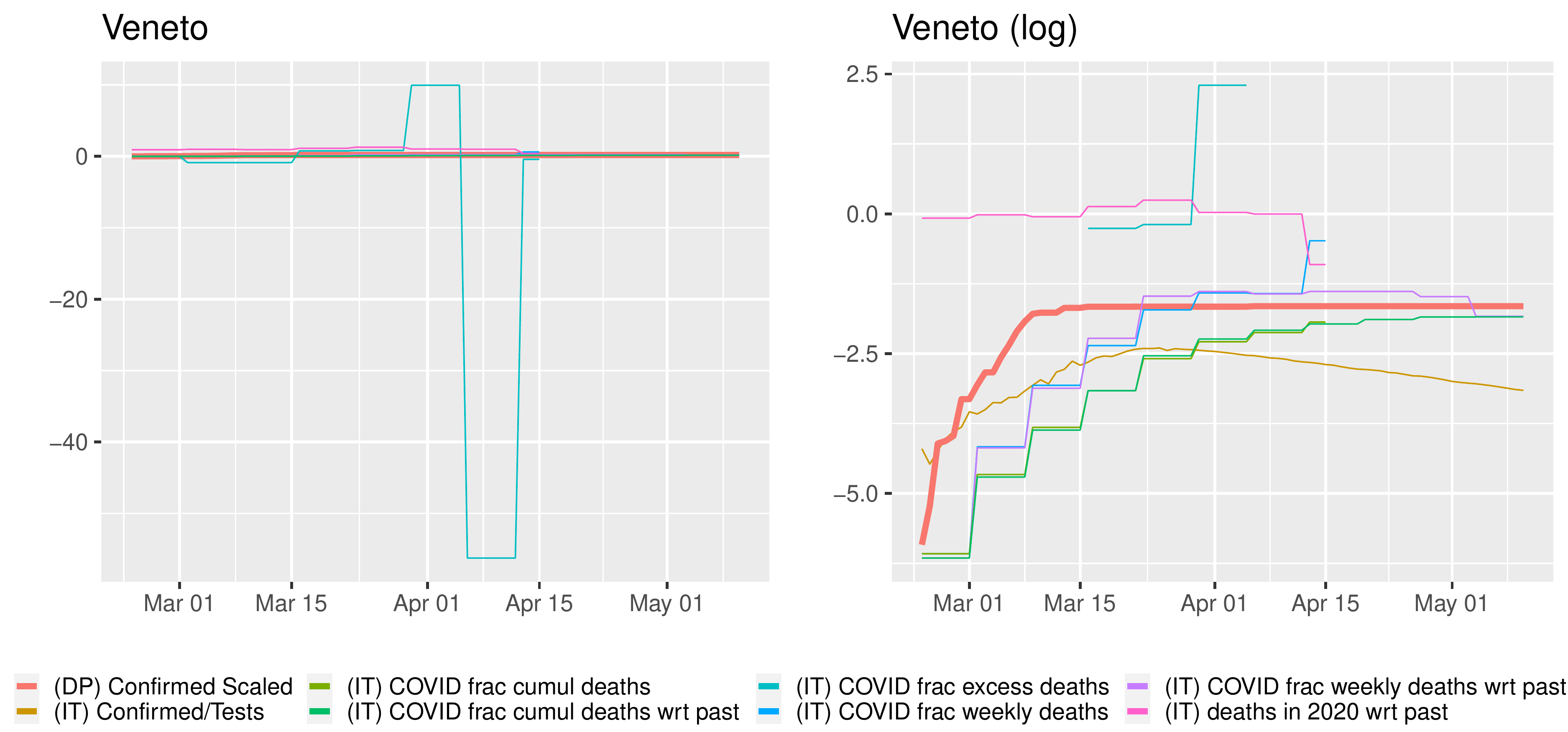}
\includegraphics[width=0.98\textwidth]{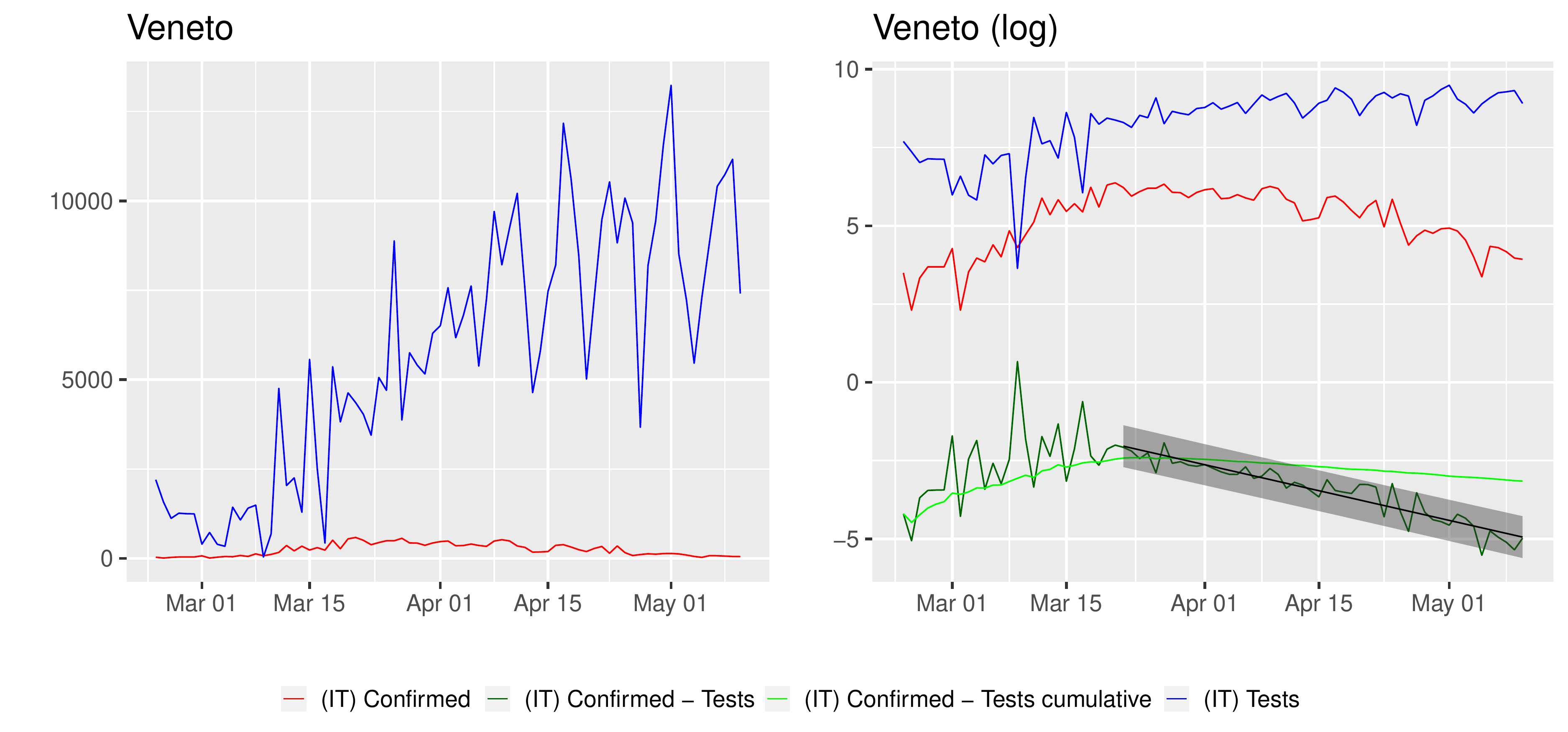}
\end{center}
\caption{Comparison of curves for Veneto region. Left: $y$--axis on normal scale, right: on logarithmic scale. 
Regression line shown for $\log(\mathrm{daily~confirmed})-\log(\mathrm{daily~tested})\sim time$
with $95\%$ prediction band. Slope of regression with $95\%$ confidence interval: 
$a_{\mathrm{f}}=-0.059 (-0.065,-0.053)$,
this corresponds to a half--life (in days) of $11.693 (10.592,13.049)$.
The slope of the regression
$\log(\mathrm{daily~confirmed})\sim time$ is 
$a_{\mathrm{raw}}=-0.047 (-0.054,-0.040)$ corresponding to a half--life (in days) of $14.843 (12.879,17.513)$.
The slope of the regression
$\log(\mathrm{daily~tests})\sim time$ is 
$0.013 (0.007,0.026)$ corresponding to a doubling time (in days) of $55.100 (26.467,95.637)$.
Ratio of slopes for $a_{\mathrm{f}}/a_{\mathrm{raw}}=1.269$, with corresponding half--lives' ratio: $0.788$.
The slope of the regression
$\log(\mathrm{cumulative~confirmed})-\log(\mathrm{cumulative~tested})\sim time$ is 
$-0.016 (-0.017,-0.016)$, corresponding to a half--life (in days) of $42.325 (40.947,43.799)$.
}\label{figVenetoApp}
\end{figure}

Furthermore, these slopes are steeper
than the slopes of the logarithm of the raw confirmed case counts on time.
With the exceptions of Lazio, P. A. Bolzano the $95\%$ confidence intervals for these
two slopes do not overlap, or overlap very slightly. The ratios of the two slopes
lie between $1.176$ (P. A. Bolzano) and $3.717$ (Piemonte). We report these ratios
alongside the slope estimates in the captions of Figs. \ref{figAbruzzo}---\ref{figVenetoApp}.
This means that 
the number of confirmed cases  will be confounded by the number of performed
tests and cannot be analyzed without them as a point of reference.

Hence, the raw confirmed case counts
are not representative of the virus' infection dynamics. The logarithm of the  fraction  of confirmed cases to total tests is modelled well by a linear function with an increasing number of daily tests being performed and has a steeper slope than the logarithm of the 
confirmed case counts. Drawing conclusions from
raw confirmed case data would seem to be mixing--in the study of the sampling effort (it is important to stress that we do not make any statements here concerning the interpretation of the confirmed cases to tests fraction).
Therefore, calibrating model parameters for this virus's dynamics should not be done based solely
on confirmed case counts, but maybe rather also on case fatalities or hospitalization data (given that classification
protocols are taken into account) as, e.g., \citet{TBri202004medRxiv,APugSSot2020arXiv, GVat2020arXiv} do. In fact, already \citet{SFlaetal2020}, critised 
\citep[as,][later also did following them]{APugSSot2020arXiv}
looking at case counts and postulated a focus on the ``observed deaths'' while \citet{GVat2020arXiv}
writes that ``the cumulative number of deaths can be regarded as a master variable''.
\citet{TBri202003medRxiv} developed an estimation methods based on the cumulative
reported number of case fatalities.

On the other hand, we also looked at the ratio of case fatalities
to the number of deceased per day. This has the analytical advantage,
of referring to something certain and well measured, detailed
records are collected (sooner or later) on the exact number of deceased in a given
time period. Here, there is hardly any chance of missing asymptomatic (of being dead) people.
\textit{If the assumption},
mentioned in the Introduction, that a significant proportion of the tests
are serological is true, then the ratio of case fatalities to 
all deceased should be telling us something about the cumulative proportion
of infected individuals. Our graphs (especially on the log--scale)
do not contradict this,
while the cumulative proportion of confirmed cases changes very slowly,
the ratio of case fatalities to total deceased per day
seems to look like an epidemic growth curve. 
Since Italy has very high quality data on the case fatalities,
this data could be further studied to assess the dynamics of the pandemic
(e.g. \citet{AMor2020arXiv} uses the raw death counts for assessing the dynamics of the pandemic, albeit at the country level).
This seems to be supported by that if one compares the curves to a potential
``gold standard''---the cumulative fraction of confirmed cases on
the \textit{Diamond Princess}, then
the case fatalities ratio seems to shadow this curve 
(on the initial part when the epidemic was taking place
on the cruise ship
and for some regions like Emilia--Romagna or Lombardia) but exceeds it.
One could hope that once all curves 
would flatten at the same level, then the epidemic will reach
the plateau. Unfortunately, at the level of some 
(e.g. Emilia--Romagna or Lombardia)
of the regions,
the scaled case fatalities grew and 
exceeded both the \textit{Diamond Princess} and cumulative fraction of confirmed
cases. 

\begin{figure}[!ht]
\begin{center}
\includegraphics[width=0.98\textwidth]{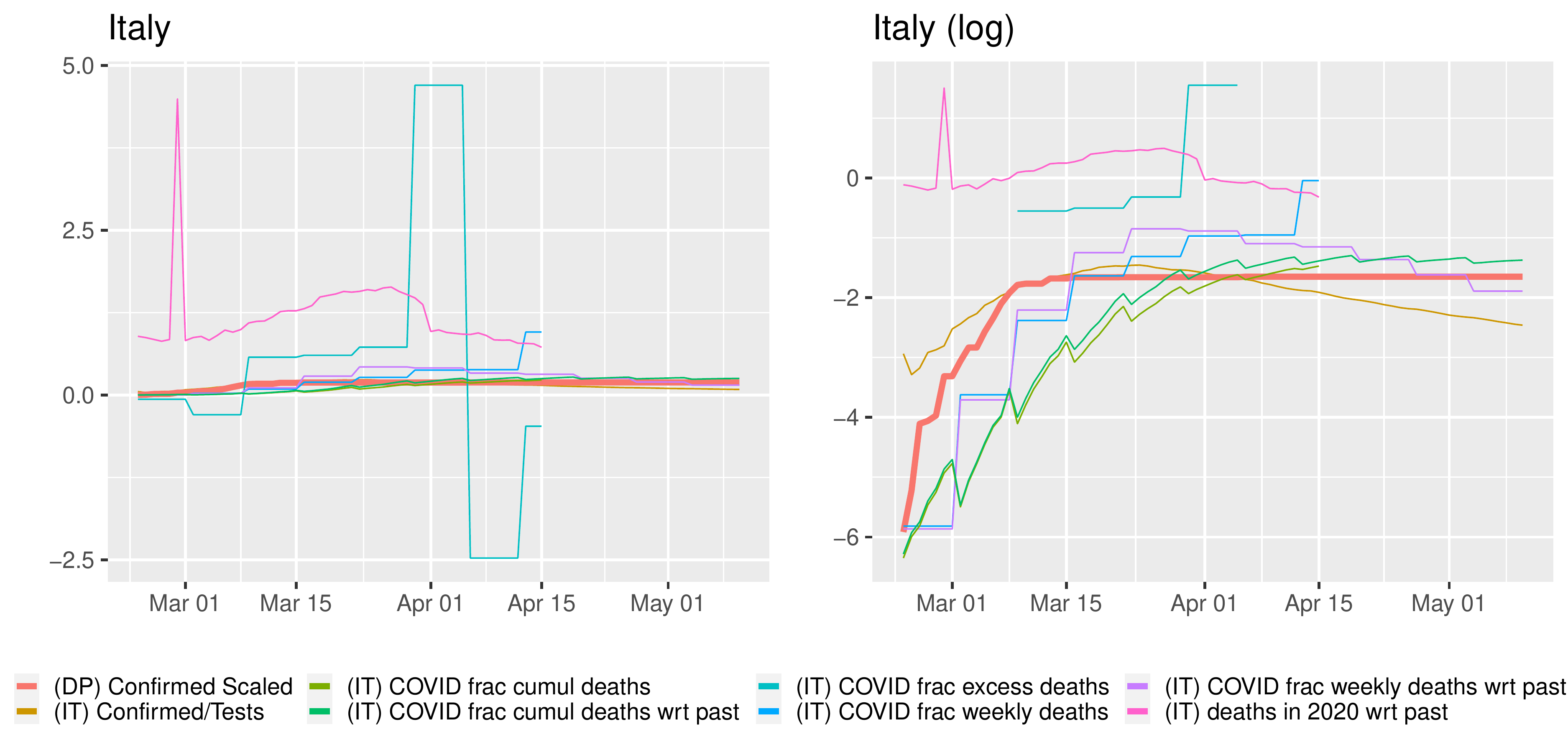}
\\
\includegraphics[width=0.98\textwidth]{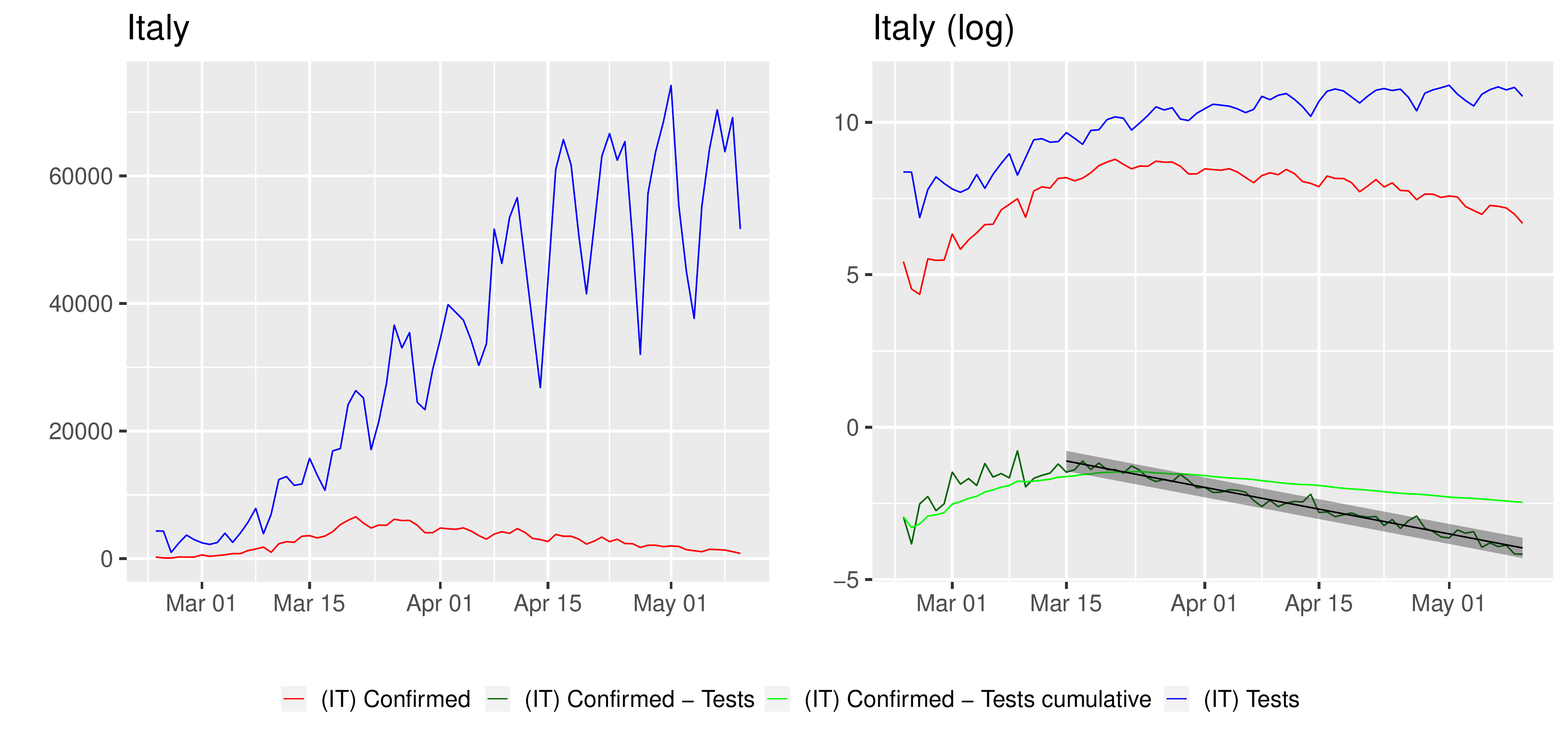}
\end{center}
\caption{Comparison of curves for the whole of Italy. Left: $y$--axis on normal scale, right: on logarithmic
scale. 
Regression line shown for $\log(\mathrm{daily~confirmed})-\log(\mathrm{daily~tested})\sim time$
with $95\%$ prediction band. Slope of regression with $95\%$ confidence interval:
$a_{\mathrm{f}}=-0.051 (-0.053,-0.048)$ corresponding to a half--life (in days) of $ 13.617 (12.972,14.330)$.
The slope of the regression
$\log(\mathrm{daily~confirmed})\sim time$ is 
$a_{\mathrm{raw}}=-0.027 (-0.031,-0.023)$ corresponding to a half--life (in days) of $25.631 (22.348,30.046)$.
The slope of the regression
$\log(\mathrm{daily~tests})\sim time$ is 
$0.024 (0.020,0.036)$ corresponding to a doubling time (in days) of $29.052 (19.224,35.217)$.
Ratio of slopes for $a_{\mathrm{f}}/a_{\mathrm{raw}}=1.882$, with corresponding half--lives' ratio: $0.531$.
The slope of the regression
$\log(\mathrm{cumulative~confirmed})-\log(\mathrm{cumulative~tested})\sim time$ is 
$-0.019 (-0.020,-0.018)$,this corresponds to a half--life (in days) of $35.882 (33.884,38.130)$. 
}\label{figItaly}
\end{figure}

\begin{figure}[!ht]
\begin{center}
\includegraphics[width=0.98\textwidth]{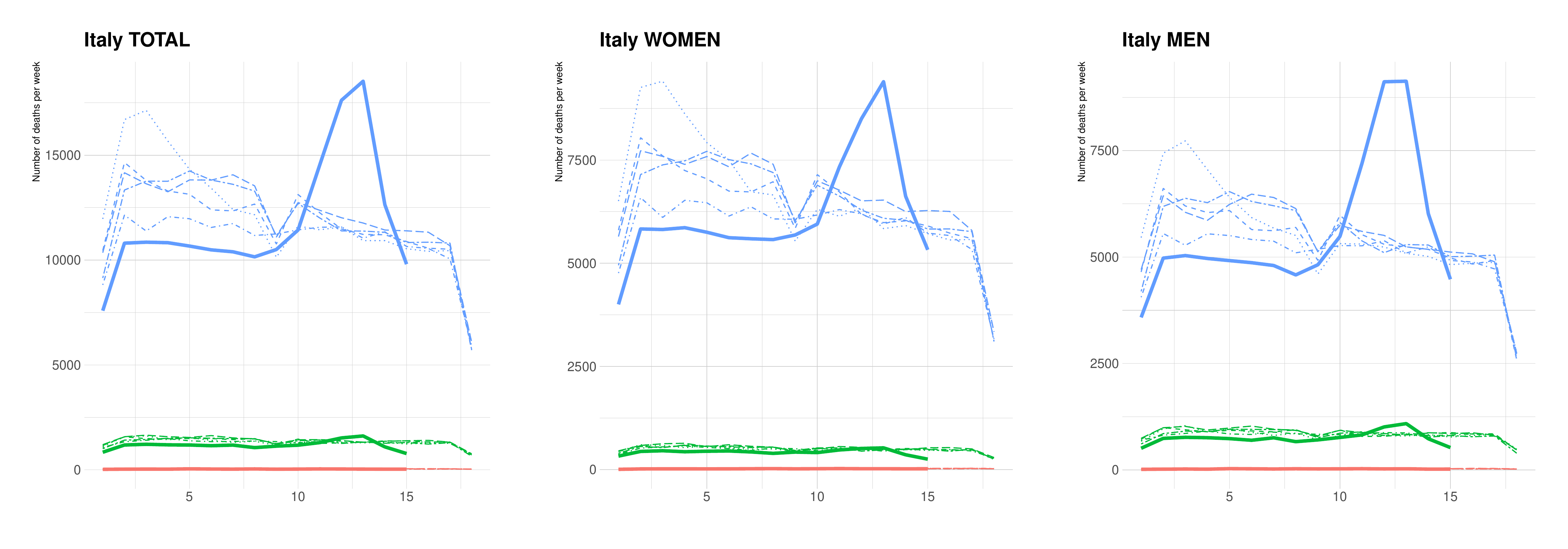} \\
\includegraphics[width=0.98\textwidth]{graphs/ITraw_mortality/Legend_FigRawMortality.pdf}
\end{center}
\caption{Weekly raw death toll comparison in different age groups between $2020$ and $2015$--$2019$
for whole of Italy. 
}\label{figRawDeathsItaly}
\end{figure}

We also compared the regional results to the same curves for the whole of 
Italy, Figure~\ref{figItaly}. On the one hand the same patterns 
are visible---the number of confirmed cases are related to the testing effort,
the case fatalities exceeding the \textit{Diamond Princess}'
cumulative confirmed cases and the confirmed cases fraction seems
to be stabilizing around the \textit{Diamond Princess}' and then dropping.
However, these graphs completely miss the regional variation. 
This is particularly
visible when looking at the total death tolls directly Figs. \ref{figRawDeathsItaly}
and \ref{figRawDeathsRegion}. Combined Italy shows a visible increase in the death toll
during the March--April period compared to previous years and the seasonal December/January peak.
However, this peak is driven by particular regions Emilia--Romagna, Lombardia and Piemonte
(Liguria, Marche, P.A. Bolzano combined with P.A. Trento and Valle d'Aosta also show
a big increase--but in raw numbers are much lesser than the other three). All the other regions'
peak is on the same level or lower than the December/January one and for some the 
death toll is on similar levels to the March--April one from previous years.
Furthermore, looking at epidemiological 
country level data would be especially misleading for Italy as Lombardy acted differently from Veneto in terms of their testing
strategies.

We believe that our presented view on the Italian regional data gives some
insights how the pandemic data reporting can be improved (if of course given the difficult situation it would be possible in practise). For the confirmed
cases count a break--down should be provided, how many of these
were medical personnel, how many had symptoms, how many were seriously
hospitalized, how many were tested for other reasons (e.g.
after contact). Similarly for the number of tests carried out and their
type (serological or not). 
The case fatalities counts, should also be put in perspective---with a report
of how many people died in total on the given day and how many deceased
were tested negatively. This would allow for estimating excess
mortality (crudely---compared to previous years' average or 
more exactly if number of deaths for the given time period are
available) and for correct scaling to compare to other
ratios. In fact in the time period $1^{\mathrm{st}}$ January--$15^{\mathrm{th}}$ April we are able to
visualize the excess mortality directly---the number of deceased (in each week) in $2020$
to the average from the past five years. 
The dataset is based on the $7904$ Italian municipalities.

To the best of our knowledge the presented here counts are at the moment the best
available data that can be used for scaling and putting the deceased counts in Italy in perspective.
The death counts seem to be collected in a consistent manner, both the number of case fatalities and the (used here) population death counts. This means that such
counts could be used as a proxy for monitoring the dynamics of the virus.

It is also a question whether the \textit{Diamond Princess}
can be considered as a gold--standard. Certainly at the beginning
it seems to behave like the other presented here curves. However,
the data very quickly ends, when the passengers were disembarked.
We do not know if it reached the plateau or would have still grown.
The confirmed case ratio seems to usually stay below/around this curve, slightly
go above and then drop. Scaled case fatality curves exceed the curve.

Finally, the counting methodology should be made readily available
for easy comparison between different countries. While of course each country is free
to follow their own protocol, without putting numbers into context one can
analyze data in an over--pessimistic or over--optimistic way.
The effect of different counting methods is pointed out by
\citet{MPicPAli2020arXiv,TheGuardian,FullFact}, when fitting parameters to the confirmed case counts
(in Lombardy, Bergamo and Brescia),
one has a change of coefficients following $10^{\mathrm{th}}$ March and $17^{\mathrm{th}}$ March, the latter can be possibly due to containment measures, but the former the authors are convinced is due to a change in the counting methodology. We have also abstained here from fitting any models to the data (the regression performed does not have as an aim modelling but formally testing what the respective curve could be telling us). It is known that due to different protocols between regions and changes in the protocols with time, the data is not homogeneous. In order to fit any model one would have to obtain documentation what were the measurement strategies for each region in the time periods.
In fact, when \citet{TAlbDFar2020arXiv} modelled the cumulative number of infections
in Italy through time (obtained using the \pkg{COVID19} package), they performed 
fits to date separately in different time intervals which corresponded to various
government introduced confinement measures.

\section{Source code and scripts}
The \pkg{COVID19} package is available from \url{https://cran.r-project.org/web/packages/COVID19/} .
\\ \noindent
The \proglang{R} script used to generate the graphics is available from
\url{https://github.com/krzbar/COVID19} .

\section{Acknowledgments}
We would like to thank Marco Picariello (MIUR-IISS) and Paola Aliani (Cognizant) for sharing their preliminary work on
data analysis of the Veneto region with us. We would like to thank Pierpaolo Malinverni (JRC) for his critical reading of an earlier version of the manuscript. All responsibility for the content of this document lies with the authors.
K.B. is supported by the
Swedish Research Council (Vetenskapsr\aa det) grant no. $2017$--$04951$.
M.O is supported by National Science Centre Poland grant no. $2014/14/E/NZ6/00182$.

\bibliography{ItalyCOVID19}

\bibliographystyle{chicago}

\clearpage

\section*{Appendix A: Curves for regions of Italy}

\clearpage

\clearpage

\clearpage

\clearpage

\clearpage

\clearpage

\clearpage

\clearpage

\clearpage

\clearpage

\clearpage
\section*{Appendix B: Death tolls for regions of Italy}

\begin{figure}[!htb]
\begin{center}
\includegraphics[width=0.98\textwidth]{graphs/ITraw_mortality/Legend_FigRawMortality.pdf} \\
\includegraphics[width=0.8\textwidth]{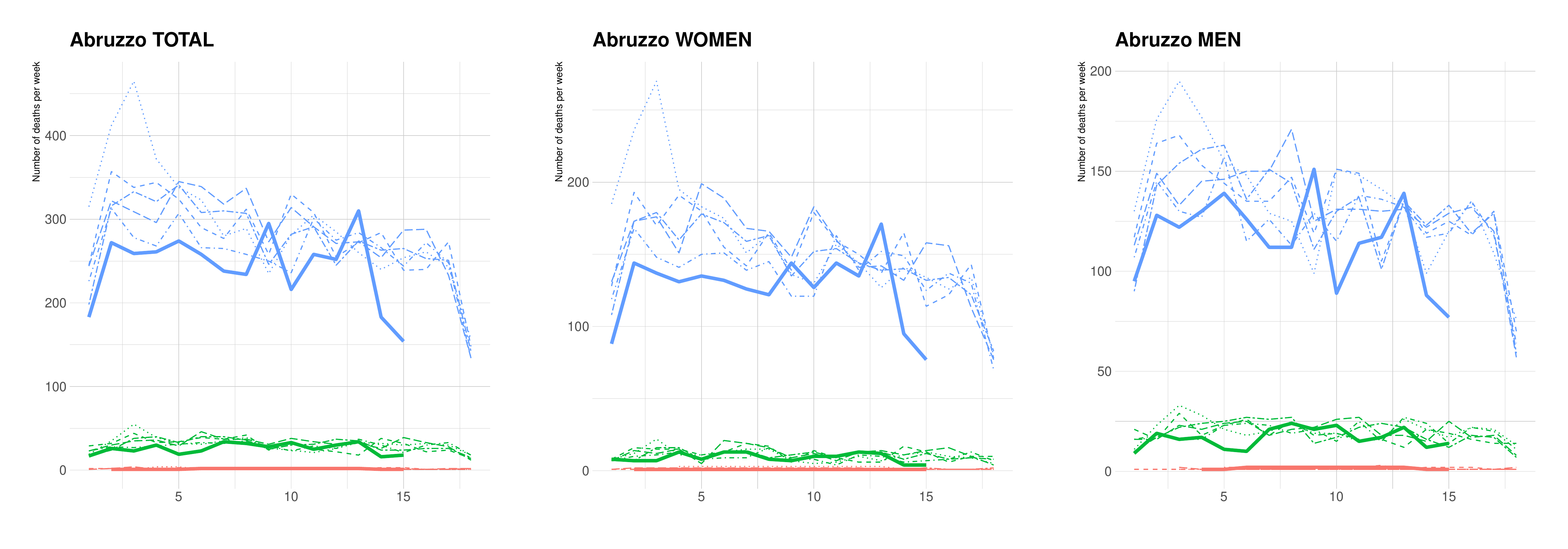}\\
\includegraphics[width=0.8\textwidth]{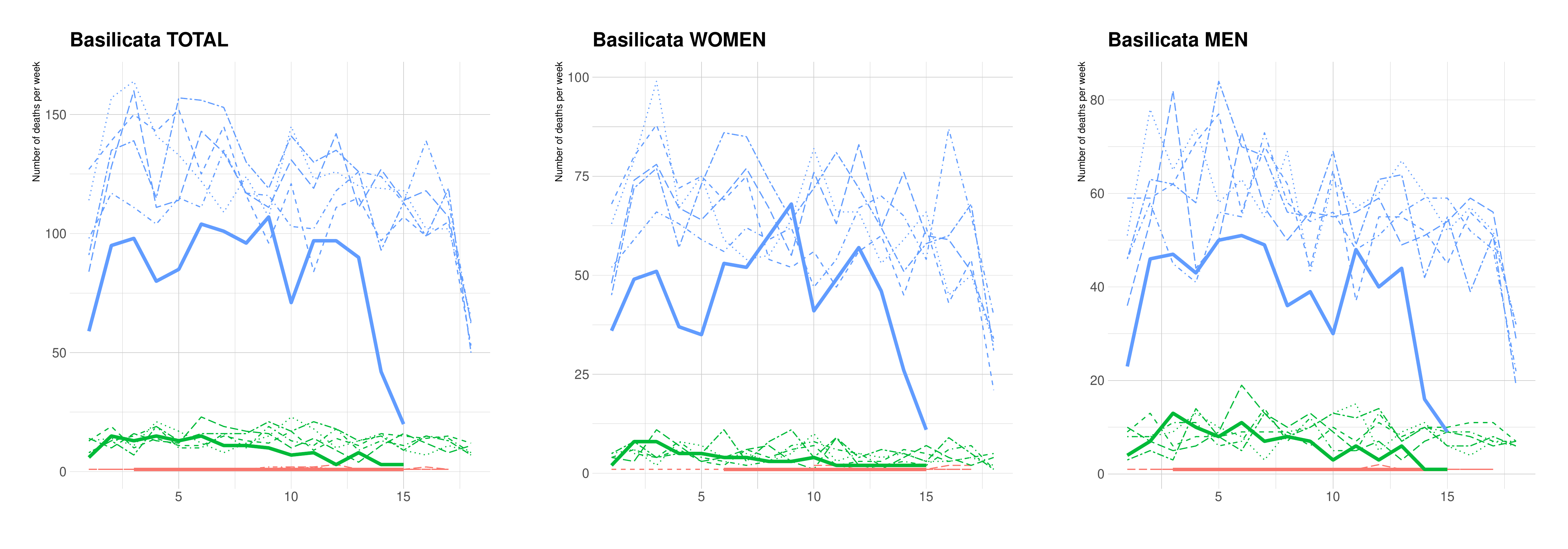}\\
\includegraphics[width=0.8\textwidth]{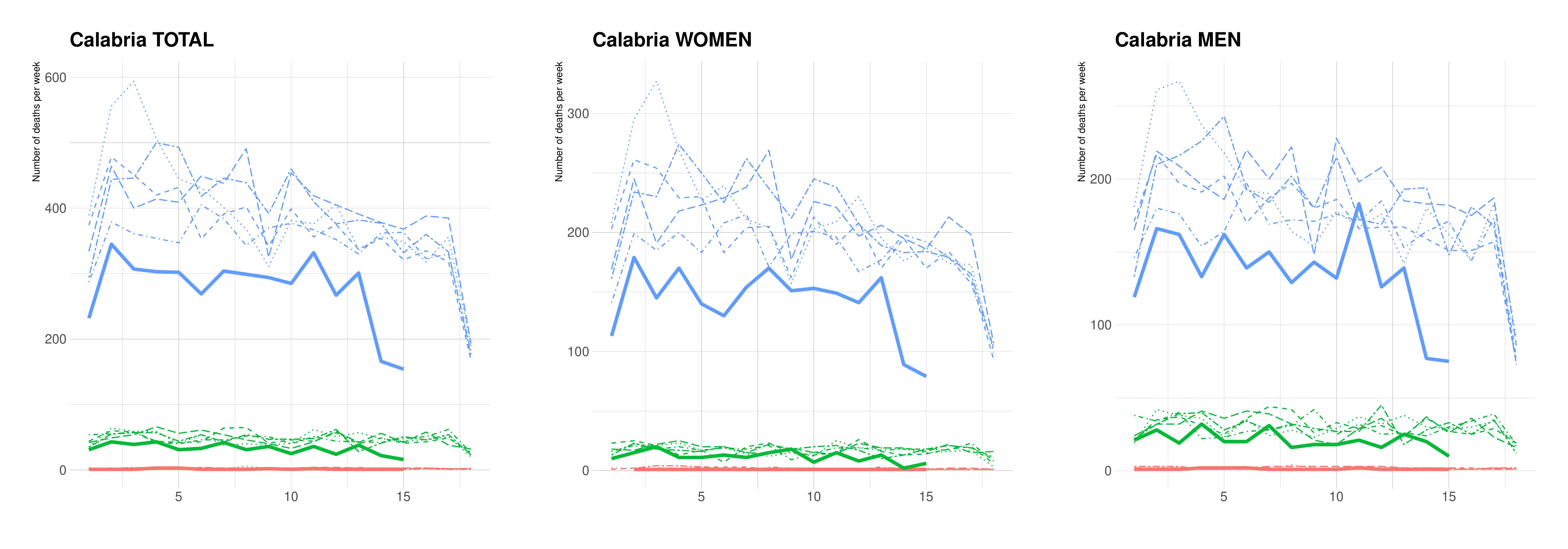}\\
\includegraphics[width=0.8\textwidth]{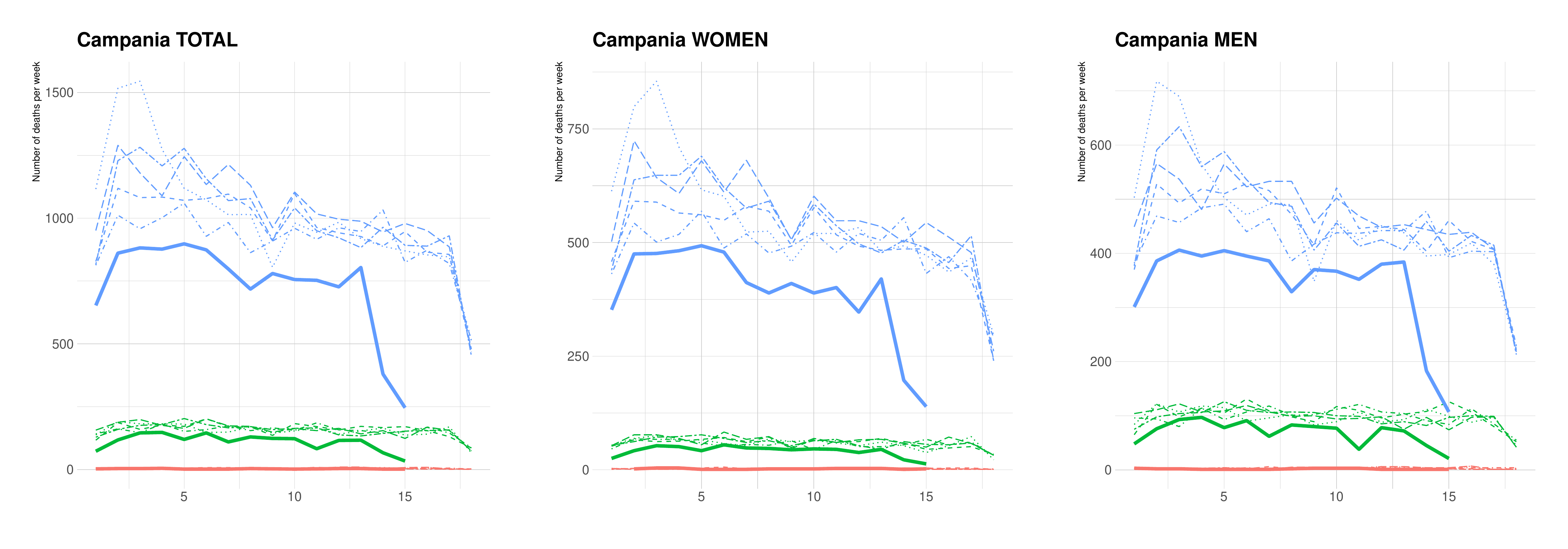}\\
\end{center}
\caption[]{}
\end{figure}
\begin{figure}[!htb]
\ContinuedFloat
\begin{center}
\includegraphics[width=0.98\textwidth]{graphs/ITraw_mortality/Legend_FigRawMortality.pdf} \\
\includegraphics[width=0.98\textwidth]{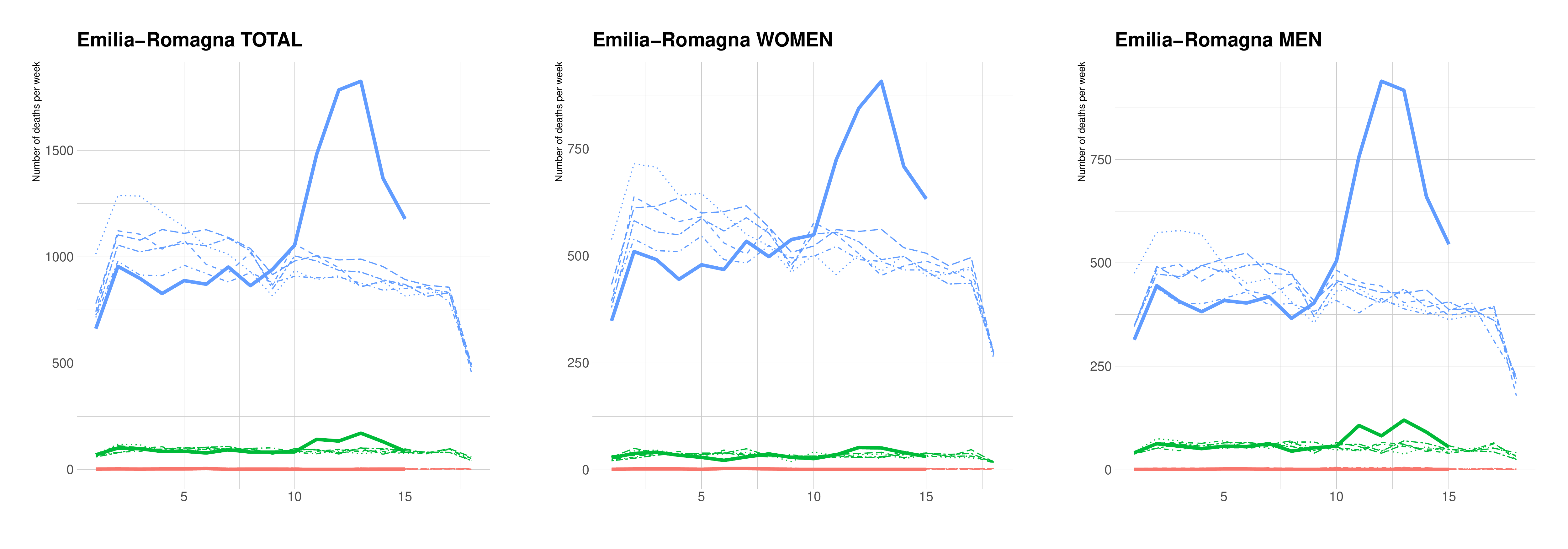}
\includegraphics[width=0.98\textwidth]{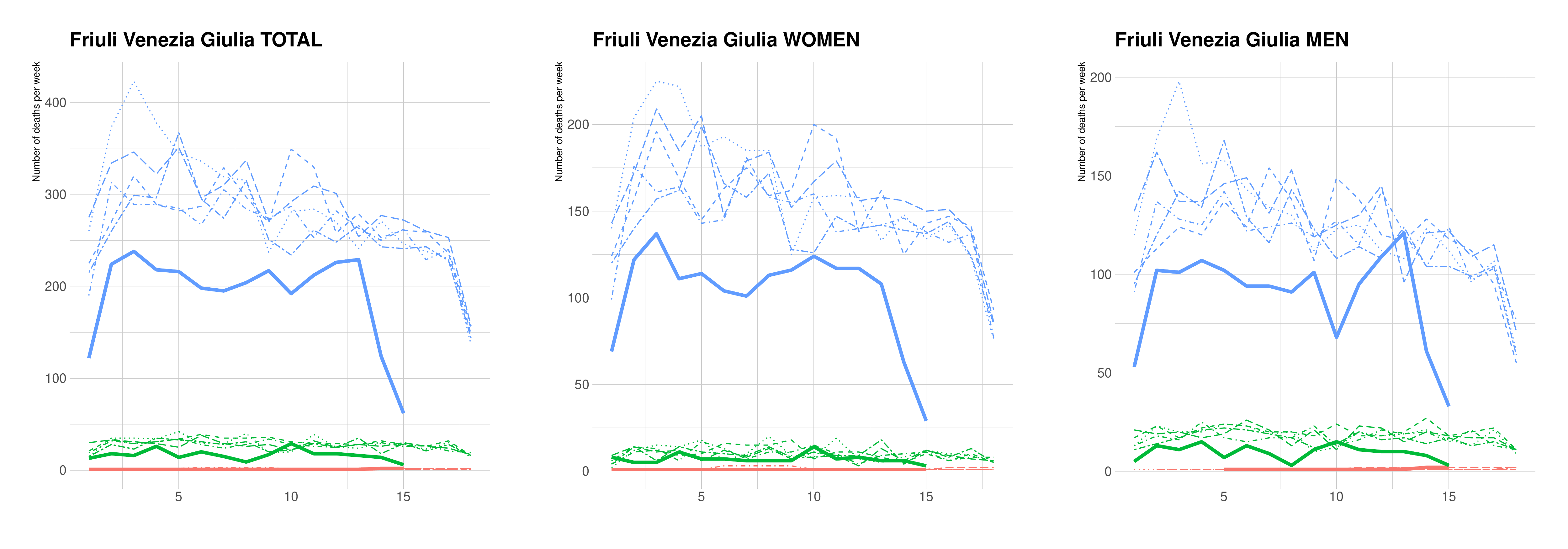}\\
\includegraphics[width=0.98\textwidth]{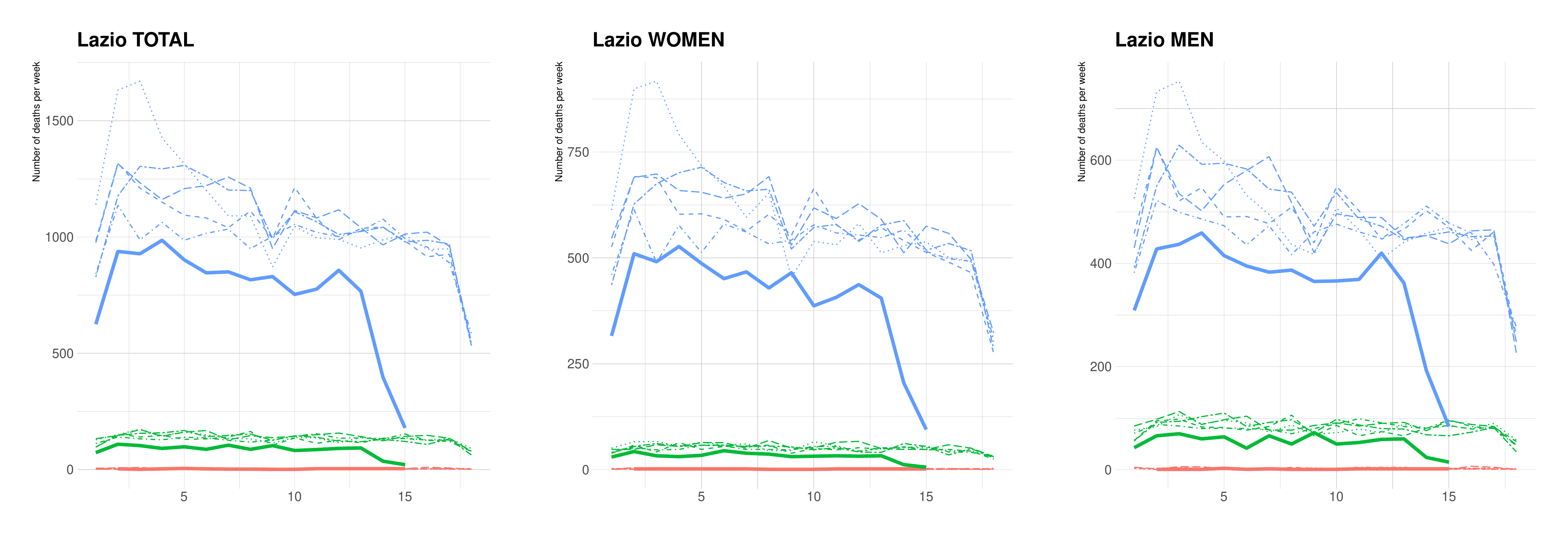}\\
\includegraphics[width=0.98\textwidth]{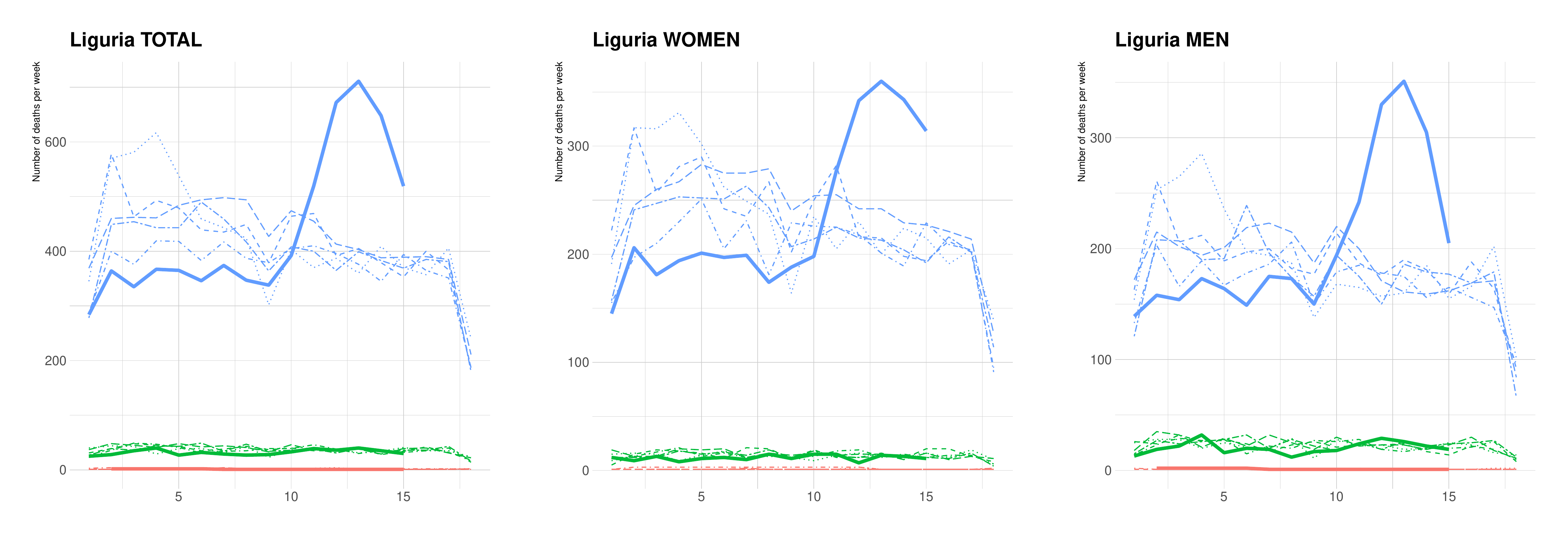}\\
\end{center}
\caption[]{}
\end{figure}
\begin{figure}[!htb]
\ContinuedFloat
\begin{center}
\includegraphics[width=0.98\textwidth]{graphs/ITraw_mortality/Legend_FigRawMortality.pdf} \\
\includegraphics[width=0.98\textwidth]{graphs/ITraw_mortality/Lombardia_rawmortality.pdf}\\
\includegraphics[width=0.98\textwidth]{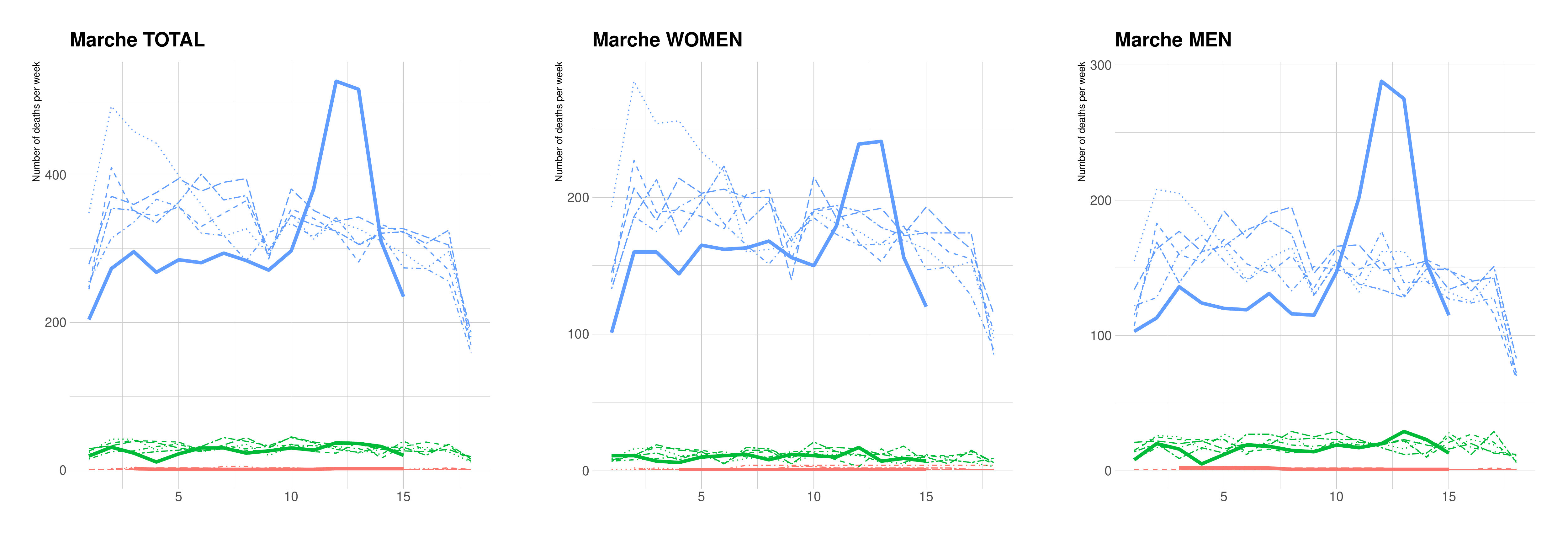}\\
\includegraphics[width=0.98\textwidth]{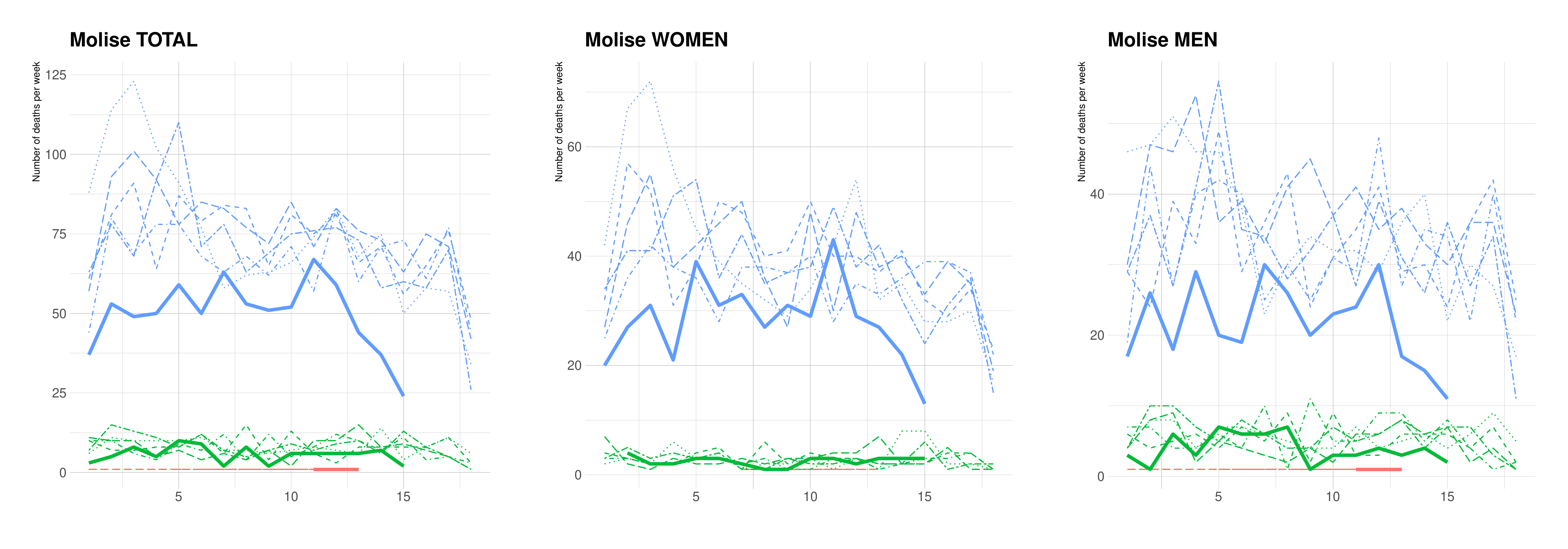}\\
\includegraphics[width=0.98\textwidth]{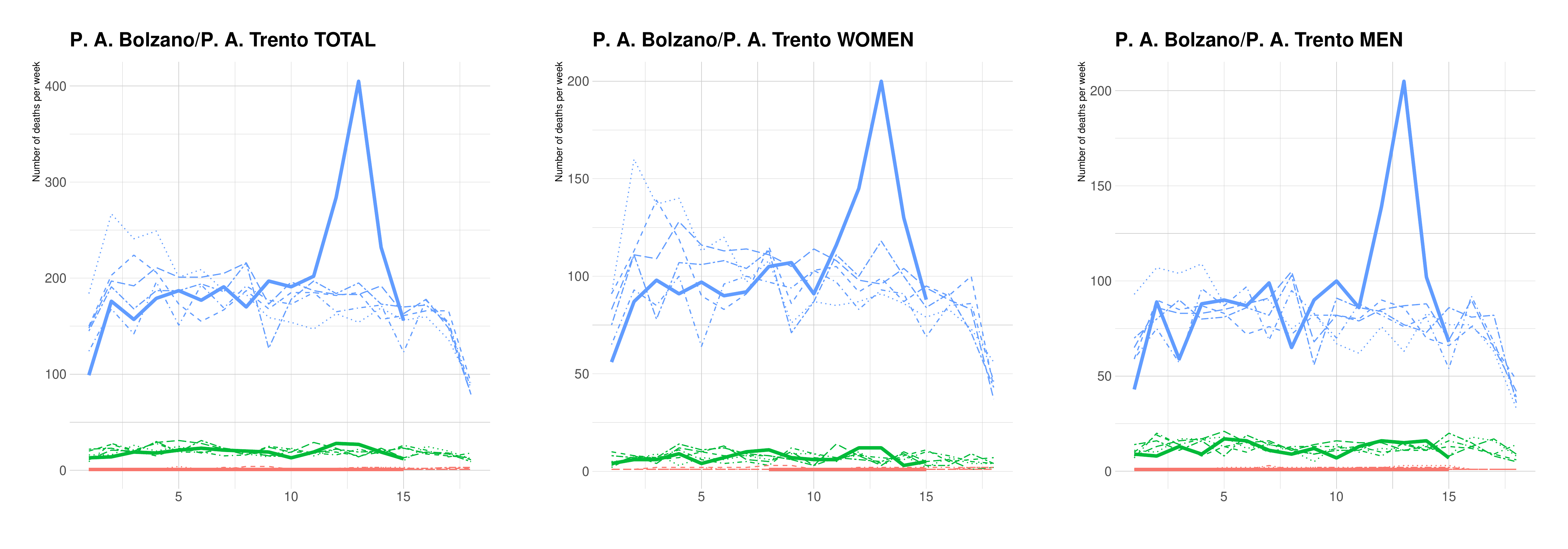}\\
\end{center}
\caption[]{}
\end{figure}
\begin{figure}[!htb]
\ContinuedFloat
\begin{center}
\includegraphics[width=0.98\textwidth]{graphs/ITraw_mortality/Legend_FigRawMortality.pdf} \\
\includegraphics[width=0.98\textwidth]{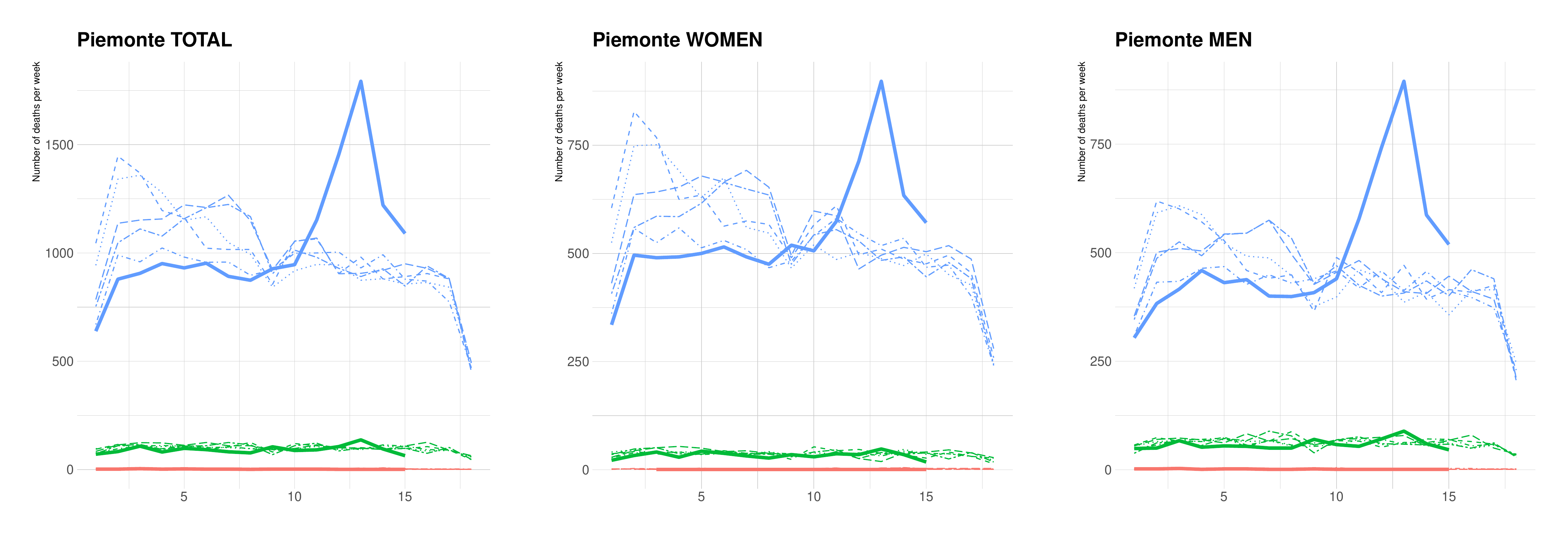}\\
\includegraphics[width=0.98\textwidth]{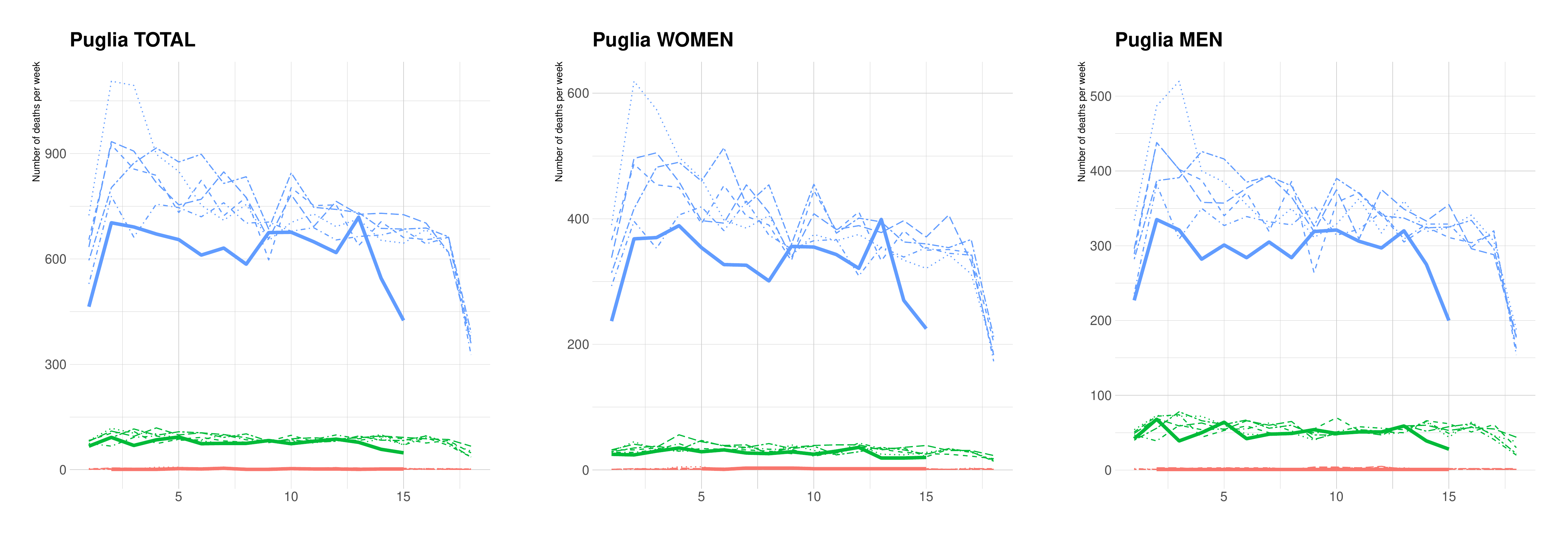}\\
\includegraphics[width=0.98\textwidth]{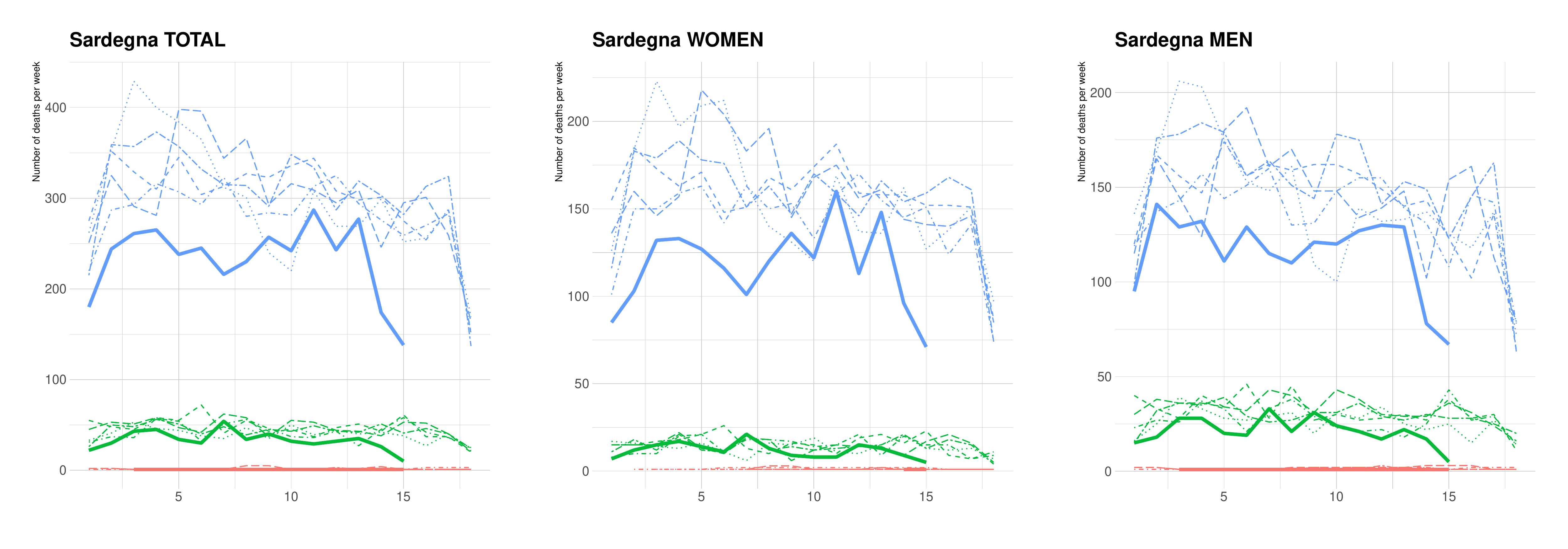}\\
\includegraphics[width=0.98\textwidth]{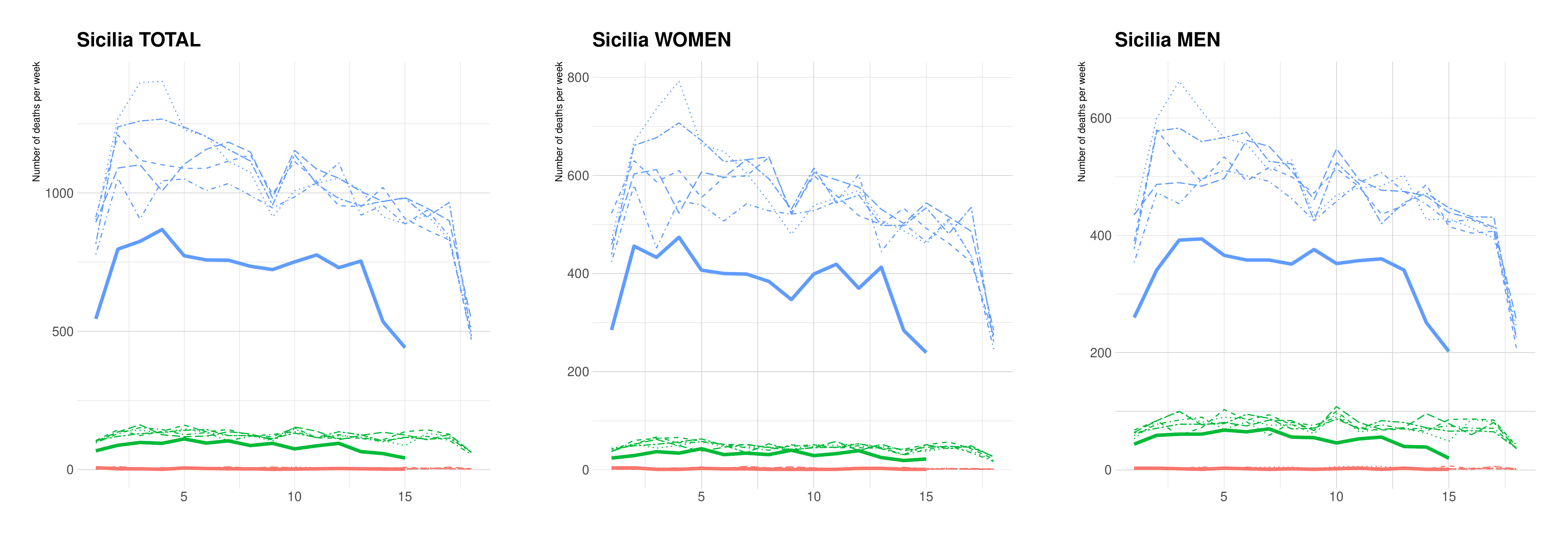}\\
\end{center}
\caption[]{}
\end{figure}
\begin{figure}[!htb]
\ContinuedFloat
\begin{center}
\includegraphics[width=0.98\textwidth]{graphs/ITraw_mortality/Legend_FigRawMortality.pdf} \\
\includegraphics[width=0.98\textwidth]{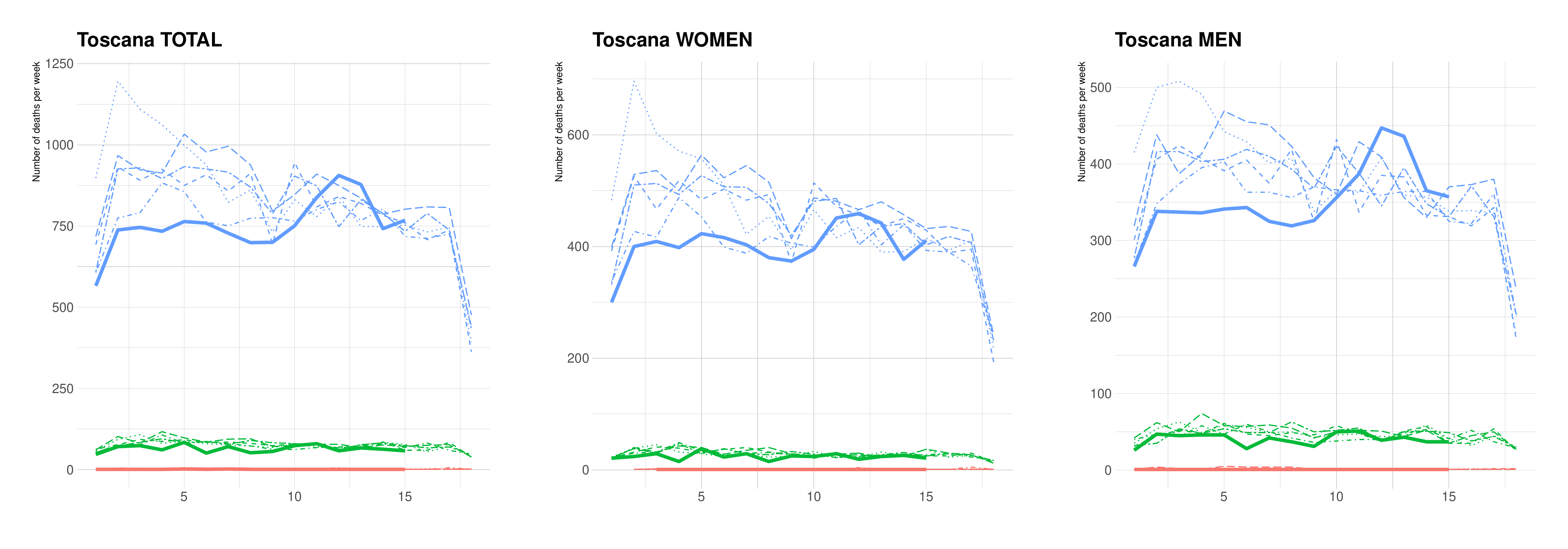}\\
\includegraphics[width=0.98\textwidth]{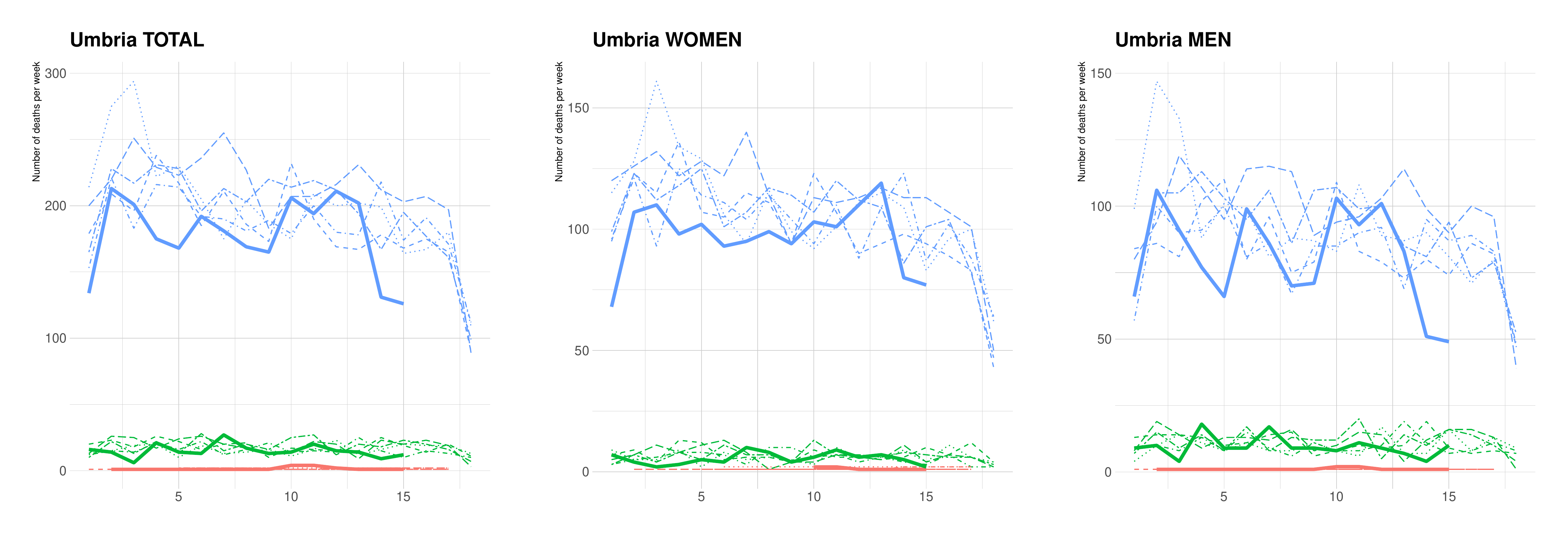}\\
\includegraphics[width=0.98\textwidth]{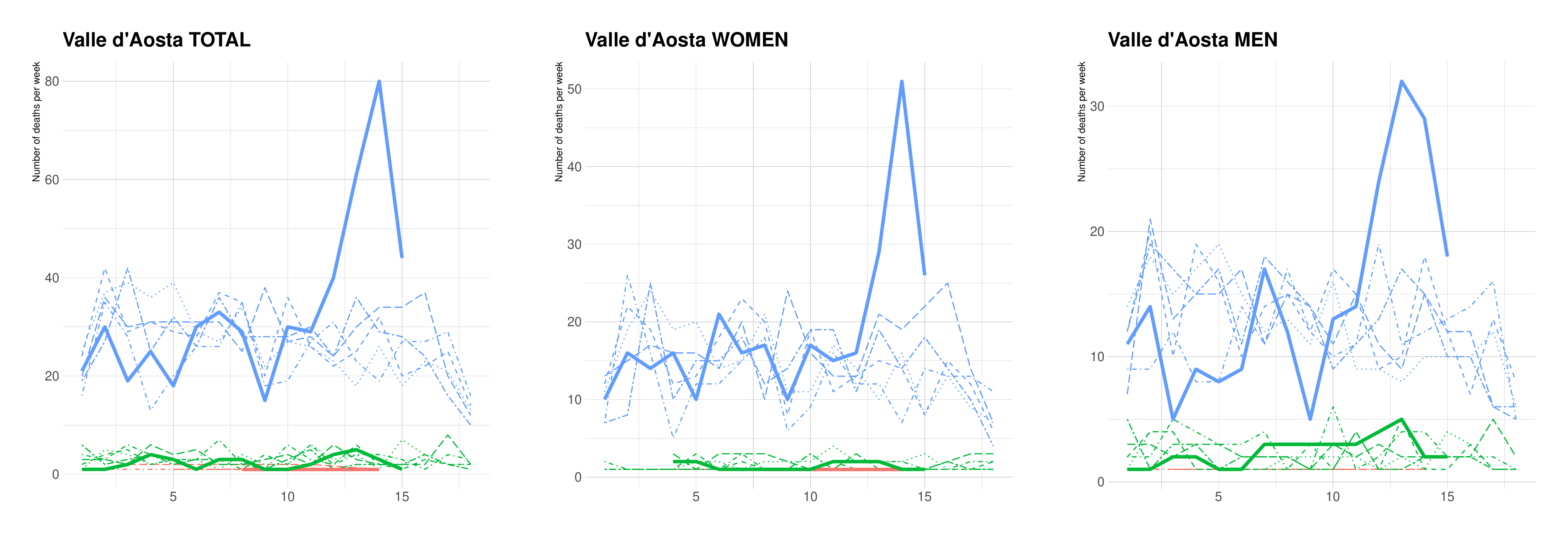}\\
\includegraphics[width=0.98\textwidth]{graphs/ITraw_mortality/Veneto_rawmortality.pdf}
\end{center}
\caption{Weekly raw death toll comparison in different age groups between $2020$ and $2015$--$2019$
for regions of Italy. 
}\label{figRawDeathsRegion}
\end{figure}

\end{document}